\documentclass[twocolumn]{aastex61}
\usepackage{color,url}

\begin{document}

\title{The Greater Taurus-Auriga Ecosystem I:\\ There Is A Distributed Older Population}

\author{Adam L. Kraus}
\affiliation{Department of Astronomy, The University of Texas at Austin, Austin, TX 78712, USA}

\author{Gregory J. Herczeg}
\affiliation{Kavli Institute for Astronomy and Astrophysics, Peking University, Yi He Yuan Lu 5, Haidian Qu, 100871 Beijing, China}

\author{Aaron C. Rizzuto}
\affiliation{Department of Astronomy, The University of Texas at Austin, Austin, TX 78712, USA}

\author{Andrew W. Mann}
\affiliation{Department of Astronomy, The University of Texas at Austin, Austin, TX 78712, USA}

\author{Catherine L. Slesnick}
\affiliation{Division of Physics, Mathematics, and Astronomy, California Institute of Technology, Pasadena, CA 91125, USA}
\affiliation{{Current Address: Agero, Inc., One Cabot Road, Medford, MA 02155}}

\author{John M. Carpenter}
\affiliation{Joint ALMA Observatory (JAO), Alonso de Cordova 3107, Vitacura-Santiago de Chile, Chile}

\author{Lynne A. Hillenbrand}
\affiliation{Division of Physics, Mathematics, and Astronomy, California Institute of Technology, Pasadena, CA 91125, USA}

\author{Eric E. Mamajek}
\affiliation{University of Rochester, Department of Physics and Astronomy, Rochester, NY 14627-0171, USA}
\affiliation{Jet Propulsion Laboratory, Pasadena, CA, USA}

\begin{abstract}

The Taurus-Auriga association and its associated molecular cloud are a benchmark population for studies of star and planet formation. The census of Taurus-Auriga has been assembled over seven decades and has inherited the biases, incompleteness, and systematic uncertainties of the input studies. The notably unusual shape of the inferred IMF and the existence of several isolated disk-bearing stars suggest that additional (likely disk-free) members might remain to be discovered. We therefore have begun a global reassessment of the membership of Taurus-Auriga that exploits new data and better definitions of youth and kinematic membership. As a first step, we reconsider the membership status of every disk-free star with spectral type later than F0, $3^h50^m < \alpha < 5^h40^m$, and $14\degr < \delta < 34\degr$ that we have found to be suggested as a candidate member in the literature. We combine data from the literature with Keck/HIRES and UH88/SNIFS spectra to test the membership of these candidates using their HR diagram positions, proper motions, radial velocities, H$\alpha$ emission, lithium absorption, and surface gravity diagnostics. Out of 396 candidate members, there are 218 confirmed or likely Taurus members, 160 confirmed or likely interlopers, and only 18 that still lack sufficient evidence to draw firm conclusions. A significant fraction of these stars ($81/218 = 37\%$) are not included in the most recent canonical member lists. Intriguingly, there are few additional members to the immediate vicinity of the molecular clouds, preserving the IMFs that have been deemed anomalous in past work. Many of the likely Taurus members are distributed broadly across the search area. When combined with the set of all known disk hosts, our updated census reveals two regimes: a high-density population with a high disk fraction (indicative of youth) that broadly traces the molecular clouds, and a low-density population with low disk fraction (hence likely older) that most likely represents previous generations of star formation. There is preliminary evidence of spatial and kinematic structure, including a possible link to the nearby 32 Ori association, though the lack of parallaxes prevents unambiguous interpretation of differences in sky position, RV, and proper motion. 

\end{abstract}


\section{Introduction}

\begin{figure*}
\hspace{-0.1in} \includegraphics[scale=0.6,trim={0 0 0 0},clip]{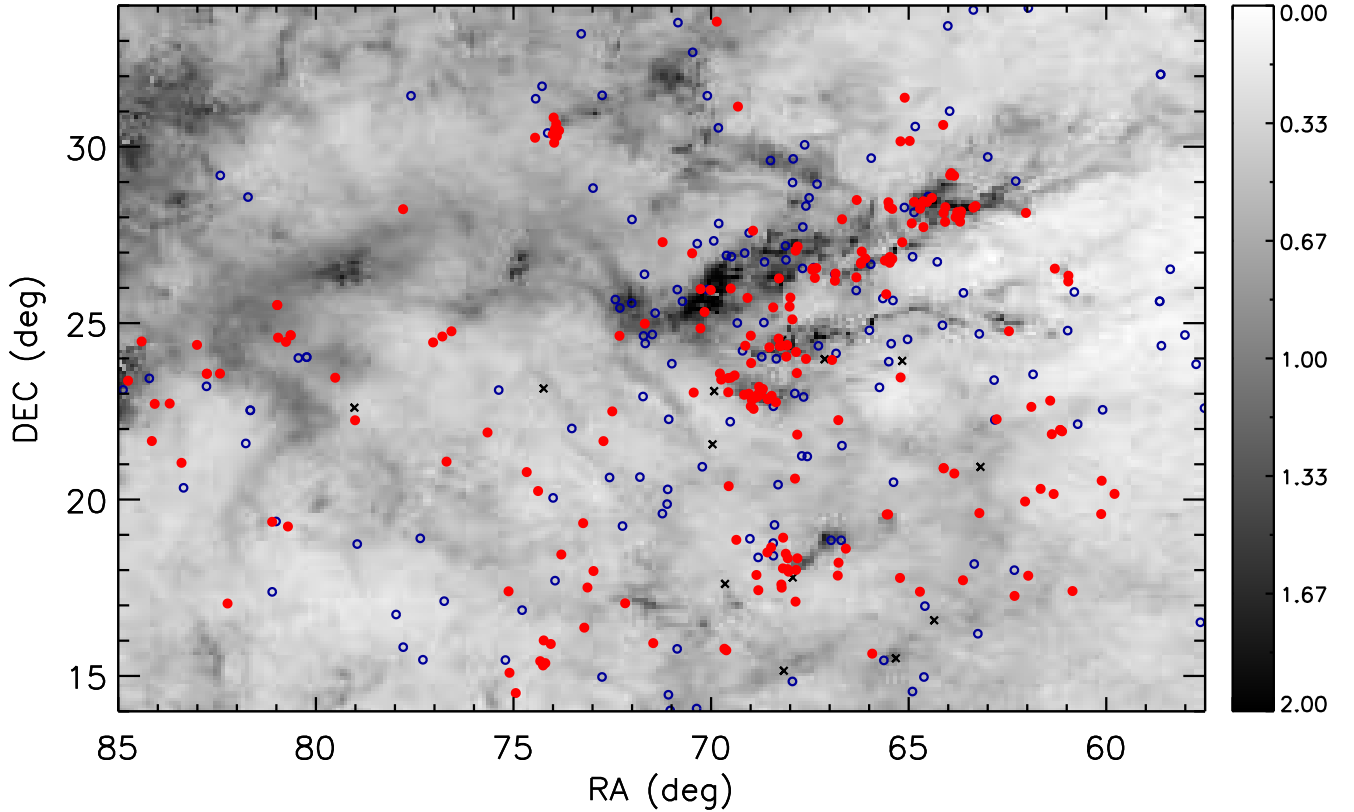}
\figcaption{\label{fig:map} A map of the candidate disk-free Taurus members that we consider in this study. The background image is an extinction map compiled by \citet{Schlafly:2014lr}. Objects that we assess as likely or confirmed members (``Y'' or ``Y?'' in Table~\ref{tab:diag}) are shown with filled red circles. Objects that we assess as likely or confirmed nonmembers (``N'' or ``N?'' in Table~\ref{tab:diag}) are shown with open blue circles. Objects which lack sufficient information for any definitely statement are shown with black crosses.}
\end{figure*}

The Taurus-Auriga star forming complex \citep{Kenyon:2008ij}, or more commonly ``Taurus'', is one of the most intensively studied regions of ongoing star formation in the Milky Way and an archetype for low-mass, distributed star formation. The close proximity ($d \sim 145$ pc; \citealt{Torres:2009ct}) and apparently young ages ($\tau \sim$ 0--5 Myr; \citealt{Kraus:2009fk}) make Taurus an ideal laboratory for observing individual young stars across a range of evolutionary stages, particularly for applications that are limited by flux or spatial resolution. However, the close proximity and distributed nature also mean that Taurus spans a large area on the sky (at least 15$\times$15 deg; \citealt{Kenyon:2008ij}), complicating the discovery of new members. The member census of Taurus has been slowly assembled since the 1940s, when the prototypical star T Tauri was recognized to be a newly-forming star \citep{Joy:1945sh}. In the ensuing decades, dozens of Class 0/I (envelope-bearing) protostars and several hundred Class II (disk-bearing) stars have been discovered (e.g., Hartmann \& Kenyon 1990; Kenyon \& Hartmann 1995; \citealt{Luhman:2010cr,Rebull:2010xf}). In the era of WISE, the census of Taurus members with associated circum(sub)stellar material is likely to be approaching completion \citep{Rebull:2011ve,Esplin:2014lr}.

The completeness of the census for Class III (disk-free) stars is much less clear; in the absence of circumstellar material to produce an infrared excess, disk-free stars are more difficult to distinguish from unassociated field stars. These stars have commonly been identified based on emission lines \citep{Herbig:1952rp,Herbig:1988no,Cohen:1979qm,Briceno:1993hc} or high-energy excesses in the X-ray \citep{Wichmann:1996fk,Scelsi:2008my} or UV \citep{Findeisen:2010mj,Gomez-de-Castro:2015kx} that denote the extreme stellar activity often found in young stellar coronae and chromospheres. However, those searches are potentially contaminated by the most active intermediate-age field stars and are subject to incompleteness for quiescent members that fall below survey limits. An alternate strategy is to search for members based on CMDs \citep{Briceno:1998pz,Luhman:2004bs,Slesnick:2006xr}; such surveys are less subject to incompleteness, but given the large survey area required, contamination is much more substantial. In either case, once a candidate is identified, its membership is often confirmed via spectroscopy to test for indicators of youth (such as lithium or low surface gravity) or for comovement in radial velocity. In sum, these strategies have identified $\sim$150 disk-free Taurus members to date. 

Many of the disk-free Taurus members trace the distribution of disk-bearing members, strongly suggesting that they formed during the same generation of star formation. However, many candidates have also been identified at larger distances from the classically accepted concentrations of young stars, leading to ongoing suggestions of a distributed (older) population (e.g., \citealt{Wichmann:1996fk,Li:1998fj,Slesnick:2006pi,Gomez-de-Castro:2015kx}) where disks have largely already dispersed. Many of these claims have been subsequently considered and taken as evidence of contamination by interloper field stars \citep{Briceno:1997qy}, and the distributed candidates do not appear in most of the canonical target lists used for studies of Taurus (e.g., \citealt{Kenyon:1995dg,Kenyon:2008ij,Luhman:2010cr}). However, a number of those candidates have been shown to comove with Taurus in proper motion \citep{Daemgen:2015uq} and radial velocity \citep{Cuong-Nguyen:2012ul}, and a number of low-mass candidates have low surface gravity that unambiguously indicates youth \citep{Slesnick:2006pi}. There also are a small number of comoving disk-hosting stars, such as StH$\alpha$34 \citep{White:2005gz,Hartmann:2005ve}, MWC 758 \citep{The:1994lr}, and even T Tauri itself, that lie at significant angular separation from the rest of the disk-hosting stars and are still included in canonical membership lists. StH$\alpha$34 also poses the additional mystery of being depleted of lithium (which indicates an age of $\ga$15--20 Myr given its M3 spectral type; \citealt{Binks:2013aa}), implying that star formation was occurring at the position and velocity of Taurus well before the current generation.

If Taurus did host a distributed population of comoving older members, built in previous generations of star formation, then that population would present both challenges and opportunities that have gone unrealized to date. Any older members that are included in the census of current star formation would dilute the apparent disk fraction and thereby bias the inferred disk lifetime (such as in calculations by \citealt{Haisch:2001om} and \citealt{Hillenbrand:2008pd}). Indeed, because disk lifetimes depend on stellar mass \citep{Carpenter:2006hf} and multiplicity \citep{Cieza:2009fr,Duchene:2010cj,Kraus:2012qe}, then measurements of the IMF and of multiple star formation must also be inextricably tied to the completeness of membership as a function of age and disk presence. Those older objects also represent an untapped resource for exoplanet searches (as recognized by \citealt{Daemgen:2015uq}) and studies of stellar rotation, multiplicity, and fundamental properties. 

The nature of this distributed population is also important for the existence and properties of ``post T Tauri'' stars (Herbig 1978): young stars that are no longer associated with molecular cloud material, or perhaps even any recognizable population, but have not yet reached the ZAMS. If star formation occurs at a constant rate over an extended interval, then star-forming regions should be surrounded by many such stars (e.g., Walter et al. 1988; Feigelson 1996), while star formation in a single burst or with a sharply increasing rate would tend to produce single-aged populations that evolve in unison (Palla \& Galli 1997). The mere existence of past generations of star formation would resolve the ongoing debate of molecular cloud lifetimes (e.g., \citealt{Hartmann:2001vn}; \citealt{Krumholz:2007rt}; \citealt{Murray:2011fr}; \citealt{Federrath:2015ys}), as kinematically related clouds must have existed since the first generation. Studies of nearby OB associations, most notably Scorpius-Centaurus, are revealing that age gradients and significant age spreads do indeed exist on scales of tens of parsecs (e.g., Pecaut \& Mamajek 2016).

In this paper, we present an initial re-examination of the disk-free stars in and around Taurus in order to distinguish bona fide members from unrelated field interlopers. In Section 2, we describe a comprehensive collection of all disk-free stars that have ever been suggested to be candidate Taurus members, along with other literature measurements that are relevant in determining their nature. In Section 3, we describe corresponding new measurements from our own intermediate-resolution optical spectroscopy with UH88/SNIFS and from archival high-resolution optical spectroscopy with Keck/HIRES. In Section 4, we describe conclusive indicators of youth (such as lithium and surface gravity) that we use to assess these objects, and in Section 5, we describe corresponding probabilistic indicators of youth (such as proper motions and radial velocities). In Section 6, we combine all of these tests into a holistic assessment of the nature of each object. Finally, in Section 7, we discuss the confirmation of a distributed population of Taurus members, and discuss the implications of that population for star and planet formation.

\section{Candidate Pre-Main Sequence Stars in Taurus-Auriga}

\subsection{Sample Construction}

Our sample consists of all objects that have been suggested as candidate disk-free members of Taurus by previous surveys, subject only to the limit that they must have a spectroscopically determined spectral type and were not immediately rejected with dispositive evidence of non-membership by that same survey. This set includes both those objects which are classically considered to be ``confirmed members'' (e.g., \citealt{Kenyon:1995dg,Luhman:2010cr,Rebull:2010xf}), as well as those which were suggested as candidates but are not typically included in the widely used Taurus census papers \citep{Wichmann:1996fk,Li:1998fj}. We tabulate the full list of objects in Table~\ref{tab:sample}, and plot their equatorial coordinate positions in Figure~\ref{fig:map}. We include both components of any pairs that are separated by $\rho > 10$\arcsec, but pairs of young stars closer than this limit are typically bound binaries \citep{Kraus:2008fr}, so we only consider the primary star in those cases. As we describe below, we also have compiled and tabulate the spectral type, extinction, $r'$ and $K_s$ magnitudes, and proper motion for each source. We also provide the discovery reference for the sources that have been identified to host binary companions.

Most of the canonical Taurus members in our sample were taken from the compilations by \citet{Rebull:2010xf} and \citet{Luhman:2010cr}, who analyzed Spitzer observations of Taurus-Auriga. From each census, we included all stars that they described as Class III sources. We also added the Class III sources reported as likely Taurus members by \citet{Kenyon:1995dg}, unless their status was subsequently amended to Class II or earlier.

A number of additional candidates have been identified from excess emission at UV or X-ray wavelengths, an indication of possible youth. We therefore consider the candidate Taurus members reported by \citet{Wichmann:1996fk}, which were selected from a compilation of optical counterparts of X-ray sources from the ROSAT All-Sky Survey (Neuhauser et al. 1995a), as well as 12 candidate Taurus members reported in a wider ROSAT survey by \citet{Li:1998fj} that were found to fall between 03$^h$50$^m < \alpha < $ 05$^h$40$^m$ and +15$\degr$ $< \delta <$ +31$\degr$. We do not yet consider the more widely separated ROSAT candidates reported by Neuhauser et al. (1995b), though, and instead defer an even wider search until the population within our existing search area is determined more robustly. We also include 2 sources of uncertain membership reported by \citet{Scelsi:2008my} as part of the XEST survey using XMM/Newton. Finally, we also add the 5 candidates identified by \citet{Findeisen:2010mj} as UV excess stars based on GALEX NUV and FUV fluxes.

Photometric surveys also have identified numerous candidates, typically with lower masses. We add the 29 candidate Taurus members that were identified by \citet{Slesnick:2006xr} to have low or intermediate gravity signatures and to fall between 03$^h$50$^m < \alpha < $ 05$^h$40$^m$. We also include 17 purported members from \citet{Esplin:2014lr} that were not flagged as disk hosts, as well as four candidate brown dwarfs from \citet{Reid:1999nq} and three candidate brown dwarfs from \citet{Gizis:1999yq}. Two intermediate-gravity objects which \citet{Luhman:2006dp} classified as young field objects are also included; one of the candidates from \citet{Gizis:1999yq} was similarly classified by \citet{Luhman:2006dp}. While this paper was under review, \citet{Luhman:2017lr} reported additional low-mass members; they are not included in our census, but remain as evidence that the Taurus census will continue to grow with the additional of deeper and broader surveys.

Finally, we also have added several other disk-free sources that previously were suggested to possibly be field stars, including V410 Anon 20 and V410 Anon 24 \citep{Strom:1994dq}; CIDA-13 \citep{Briceno:1999nm}; and NTTS 043220+1815, SAO 76411, and SAO 76428 \citep{Walter:1988ae}. These stars test whether our holistic approach to membership agrees with previous assessments in the literature (e.g., \citealt{Muzerolle:2003xd,Luhman:2009wd}.

We do not consider any Class II or earlier sources in our census, though they inform our definition of the kinematics that define ``Taurus''. The existence of a far-infrared excess due to circumstellar dust (if confirmed to also host gas, such as via spectroscopic signatures of accretion) constitutes prima facie evidence that an object is young ($\tau \la 20$ Myr), though some disk-bearing stars could be interlopers from the more distant Per OB2 association (e.g., Bally et al. 2008; Luhman et al. 2016). Furthermore, the optical veiling from accretion can complicate the interpretation of lithium absorption, and chromospheric H$\alpha$ emission from activity is typically superceded by much stronger emission from the magnetospheric accretion flows. Confirmation of young disk-bearing stars constitutes a fundamentally different observational problem, so we explicitly do not include any of the high-confidence Taurus members identified as Class I, Flat-Spectrum, or Class II by \citet{Kenyon:1995dg}, \citet{Luhman:2010cr}, or \citet{Rebull:2010xf}. As a corollary, we also do not include any binary companions to those stars unless they are sufficiently resolved as to be unclear whether their association is due to binarity or clustering; even if they are disk-free, the disk-bearing nature of a gravitationally bound binary companion is sufficient to confirm a star's youth.

\subsection{Photometry}

After constructing our sample of candidate disk-free Taurus members, we cross-referenced it with several literature catalogs to obtain optical and near-infrared photometry (to allow rejection of field stars that don't fall on the association sequence). We obtained $K_s$ photometry from 2MASS \citep{Skrutskie:2006nb}, which contained counterparts for every object. We then obtained $r'$ magnitudes for incomplete subsets of the sample from the APASS \citep{Henden:2012dq}, CMC15 \citep{Evans:2002sp}, and SDSS DR9 \citep{Ahn:2012ys} catalogs. For stars with magnitudes in multiple catalogs, we adopted the unweighted mean of all available magnitudes. We do not take a weighted mean because the intrinsic optical variability of young stars due to spots (e.g., \citealt{Cody:2014uq}) is generally larger than the photometric uncertainties ($\la 0.1$ mag for the catalogs we consider).

Finally, for the remaining sources that did not otherwise have $r'$ photometry available, we adopted $R$ magnitudes from the NOMAD survey \citep{Zacharias:2004zi} and from the QUEST survey of Taurus conducted by \citet{Slesnick:2006xr}. We cross-calibrated the NOMAD and QUEST photometry using sources which also had at least one measurements in the $r'$ surveys, and found that there was a systematic offset of $r'-R=+0.2$ mag for \citet{Slesnick:2006xr} and of $r'-R=+0.3$ for NOMAD, without evidence of a significant color term. These offsets are similar to the conversion from VEGAmag (for $R$) to ABmag (for $r'$), which is $+0.16$ mag at $\lambda = 6250$ \AA. After applying our calculated offsets to transform their $R$ magnitudes to $r'$ magnitudes, we find that the standard deviation in the color difference is $\sim$0.2 mag for \citet{Slesnick:2006xr} and $\sim$0.34 mag for NOMAD. This uncertainty was judged to be satisfactory for construction of color-magnitude diagrams and estimation of extinction. We did not find any $r'$ or $R$ photometry for 19 sources, all of which are ultracool objects (SpT $\ge$ M7.5) or early type stars in the high-extinction region surrounding V410 Tau.

\subsection{Proper Motions}

We obtained proper motions for our sample from two sources. For optically bright stars, we adopted proper motions from the UCAC4 catalog \citep{Zacharias:2012lr}; these proper motions had typical uncertainties of $\sigma_{\mu} \la 1$ mas/yr for stars brighter than $r' \sim 11$, growing to $\sigma_{\mu} \sim 3$ mas/yr at $r' \sim 14$. For fainter stars ($r' \ga 14$), we calculated new proper motions from a cross-match of USNO-B1.0 \citep{Monet:2003yt}, 2MASS \citep{Skrutskie:2006nb}, and SDSS \citep{Ahn:2012ys} using the methodology first described by \citet{Kraus:2007mz} and recently applied by \citet{Kraus:2014rt}. These proper motions have typical uncertainties of 3--4 mas/yr down to $r' \sim 18$, degrading below that level and not yielding any measurements for $r' \ga 20$. In the brightness range where candidates had proper motions available from UCAC4 and from our catalog cross-match, we adopted the proper motion with the smaller uncertainty.

\subsection{Spectral Types and Extinction}

By construction, all of the targets that we consider were assigned a spectral type by at least one previous study, and many have been assigned an extinction estimate. We have searched the literature to find all available measurements of the spectral type and extinction for each target, adopting the measurement that we deem to be most robust. Where possible, we have adopted spectral types for which extinction was simultaneously determined from the same spectrum. A simultaneous fit minimizes the correlation between the quantities, as otherwise extinction reddens the spectrum and mimicks the overall continuum shape of a cooler object.

For objects without extinction estimates, we have used the $r'-K$ color and the assigned spectral type to estimate the approximate extinction. We specifically assumed that each star had a photospheric $r'-K$ color matching the field color-SpT relations that we have previously constructed \citep{Kraus:2007mz}, and then inferred the extinction ($A_V$) from the color excess of the observation above that photospheric color, assuming the $R_V=3.1$ interstellar dust reddening law of \citet{Schlegel:1998yj}.

There are two possible sources of systematic uncertainty in this calculation: young stars might not follow the same SpT-color relation as field stars, and the dust in Taurus-Auriga might not follow the same reddening law as interstellar dust. Young star SpT-color relations are now available for some filter systems (e.g., Pecaut \& Mamajek 2013), but not for the SDSS optical filters or for any R filter, so the magnitude of this effect remains uncertain. However, most of the quantities calculated from this exercise are relatively modest ($A_V < 2$ mag) and we only use the extinction to modify $K$ magnitudes (where $A_K = 0.11 A_V$; \citealt{Schlegel:1998yj}), so these systematic uncertainties should not strongly influence our results. We do not expect any systematic errors due to the presence of optical excess (from accretion) or near-infrared excess (from inner disks) because our sample was constructed to only include objects which do not have a disk.

\subsection{H$\alpha$, Lithium, and RVs}

We sifted through the literature to find existing measurements for $EW[H\alpha]$, $EW[Li_{6708}]$, and $v_{rad}$; we list these measurements and the large number of corresponding references in Tables~\ref{tab:litha}, \ref{tab:litli} and \ref{tab:litrv}, respectively. In cases with multiple measurements, we adopted the value that was judged to be obtained with the highest S/N and the highest spectral resolution.

\section{New and Archival Observations and Analysis}

\subsection{Keck/HIRES High-Resolution Optical Spectroscopy}

High-resolution spectroscopic observations for 35 candidates from the surveys of \citet{Slesnick:2006xr} and \citet{Li:1998fj} were previously obtained with Keck/HIRES on 2006 Dec 12 and 2006 Dec 13 (PIs Carpenter and Slesnick). These observations are particularly useful in casting light on the nature of objects east of $\alpha = 5^h$ that \citet{Slesnick:2006xr} suggested might be a distributed population of intermediate-age objects. We downloaded those observations, extracted the spectra, and analyzed them to measure the radial velocity, H$\alpha$ equivalent width, and Li equivalent width for each object. Our analysis of the HIRES data is very similar to the methods described in \citet{Kraus:2011lr} and \citet{Kraus:2014rt}. We extracted and wavelength-calibrated each spectrum using the MAKEE pipeline\footnote{\url{http://spider.ipac.caltech.edu/staff/tab/makee}}, refining the wavelength solution by cross-correlating the 7600\AA\, telluric absorption band against that of the O7 spectral standard S Mon \citep{Morgan:1973vn}. We list the observations and their salient features (the epoch, integration time, and SNR) in Table~\ref{tab:hires}.

For each science spectrum, we measured the broadening function \citep{Rucinski:1999yr}\footnote{\url{http://www.astro.utoronto.ca/\~rucinski/SVDcookbook.html}} with respect to the closest match (in spectral type) among the spectral standard stars observed during the same observing run: HD 9986 (G2), HD 166 (K0), HD 79211 (K7), GJ 393 (M2), GJ 402 (M4), and GJ 406 (M6). We adopted the RVs reported for each standard by \citet{Chubak:2012lr}, which should have a systematic uncertainty of $<$100 m/s. We fit each broadening function with a Gaussian function to determine the absolute RV ($v_{rad}$) and the standard deviation of the line broadening ($\sigma_{v}$), which is a convolution of the rotational broadening and instrumental resolution. Based on our past use of this method for Keck/HIRES data, we estimate the uncertainty in the RV to be $\sim$0.5 km/s due to remaining wavelength calibration mismatch between science and standard spectra. To estimate $v \sin(i)$ from $\sigma_v$, we constructed a relation between the quantities by broadening each template spectrum by a range of values using the IDL task {\em lsf\_rotate} (Gray 1992; Hubeny \& Lanz 2011), and then measuring $\sigma_{v}$ for the broadened spectra using the corresponding original spectra as templates. Finally, we also measured the equivalent width of the H$\alpha$ and Li$_{6708}$ lines with respect to the surrounding continuum or pseudo-continuum using the IRAF task splot. We list all of these measurements in Table~\ref{tab:hires}.

\subsection{UH88/SNIFS Intermediate-Resolution Optical Spectroscopy}

The spectral properties for some candidate Taurus members were measured in the 1990s or earlier, so we obtained new intermediate-resolution optical spectra to update their spectral type, extinction, H$\alpha$ equivalent width, and surface gravity. We observed 32 candidates with the SuperNova Integral Field Spectrograph (SNIFS, \citealt{Aldering:2002rt}) on the University of Hawaii 2.2m telescope between November 2014 and January 2015 (PIs Herczeg and Mann). These observations are similar to those described by \citet{Mann:2015ys}, which describes some of the observations and reductions in more detail. To briefly summarize, SNIFS is an integral field spectrograph with an FOV of 6$\times$6\arcsec, feeding red (3200-5200\AA) and blue (5150-9700\AA) arms at a resolution of $R\simeq1000$. Standard data reduction was done by the SNF pipeline as described in \citet{Bacon:2001fr}. Additional flux calibration was applied to each spectrum using standard stars and/or a model of the atmosphere above Mauna Kea as described in \citet{Mann:2015ys}. We list the observations and their salient features (the epoch and integration time) in Table~\ref{tab:snifs}.

We have used the low-resolution optical spectra from SNIFS to calculate joint constraints on the spectral type and extinction of the candidate Taurus members. We compared each observed spectrum to a sequence of field dwarfs \citep{Pickles:1998zr,Bochanski:2007ky} that we artificially reddened using an
$R_V=3.1$ reddening law \citep{Savage:1979jk}, selecting the best-fit model. We describe this process in more detail in \citet{Rizzuto:2015fj} and \citet{Kraus:2015mz}, where it was applied to members of the Upper Scorpius OB association. Spectral types themselves are only defined by half-subclasses and differ systematically depending on the classifier, so we assess a final uncertainty of $\pm$0.5 subclass on M spectral types and a corresponding uncertainty of $\sim$0.3 mag on the extinction. The uncertainties for the small number of F--K stars are 2--3 subclasses.

From the SNIFS spectra, we also calculate the equivalent width of the H-$\alpha$ line for each of the candidate Taurus member. We fit a polynomial continuum or pseudo-continuum across the H-$\alpha$ line region using the surrounding wavelength range as reference, and then calculated the equivalent width compared to the fitted continuum in the line region. We also compute the gravity-sensitive sodium index (Na-8189; \citealt{Slesnick:2006xr}) for candidate members with spectral type later than K5. For stars later than approximately M2, the index robustly differentiates low-gravity giants, intermediate-gravity PMS dwarfs, and high-gravity main sequence field stars. The Na-8189 index measures the strength of the Na I doublet relative to the pseudo-continuum, and is calculated as the flux ratio between two 30 \AA\, bands, the first centered on the Na doublet at 8189 \AA\, and the second on the pseudo-continuum at 8150 \AA. We list all of these measurements in Table~\ref{tab:snifs}. 

Finally, we also have measured the Na-8189 index for Taurus candidates in our sample that were observed by \citet{Herczeg:2014oq}, and report the spectral indices in Table~\ref{tab:litna}.

\section{Conclusive Membership Tests from Stellar Properties}

\begin{figure*}
\epsscale{1.0}
\includegraphics[scale=0.9,trim={0 1.5cm 0 0},clip]{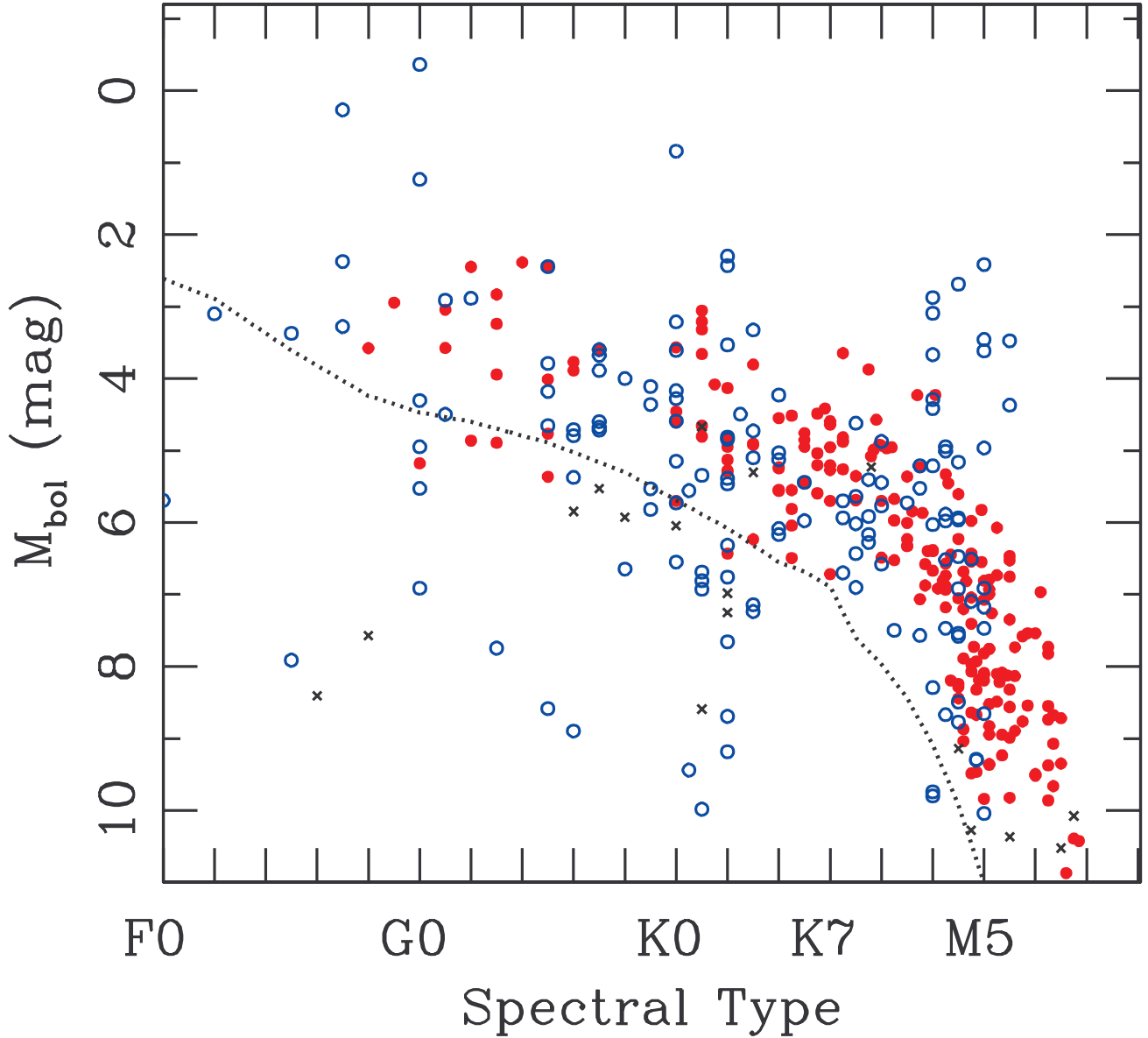}
\figcaption{\label{fig:hrd} An HR diagram for our sample of candidate disk-free Taurus members. The points are color coded as in Figure~\ref{fig:map}, but based on the membership assessment we {\it would have made} based only on the other tests that we use, without using the HR diagram position. The spectral types are taken from Table 1, and the bolometric absolute magnitudes are calculated as described in Section 4.1. The black dashed line denotes the field main sequence (Kraus \& Hillenbrand 2007), shifted to the distance of Taurus-Auriga ($d = 145$ pc). One way for a target to be assessed as a field interloper is to fall below this sequence.}
\end{figure*}

\begin{figure*}
\epsscale{1.0}
\includegraphics[scale=0.9,trim={0 6.5cm 0 0},clip]{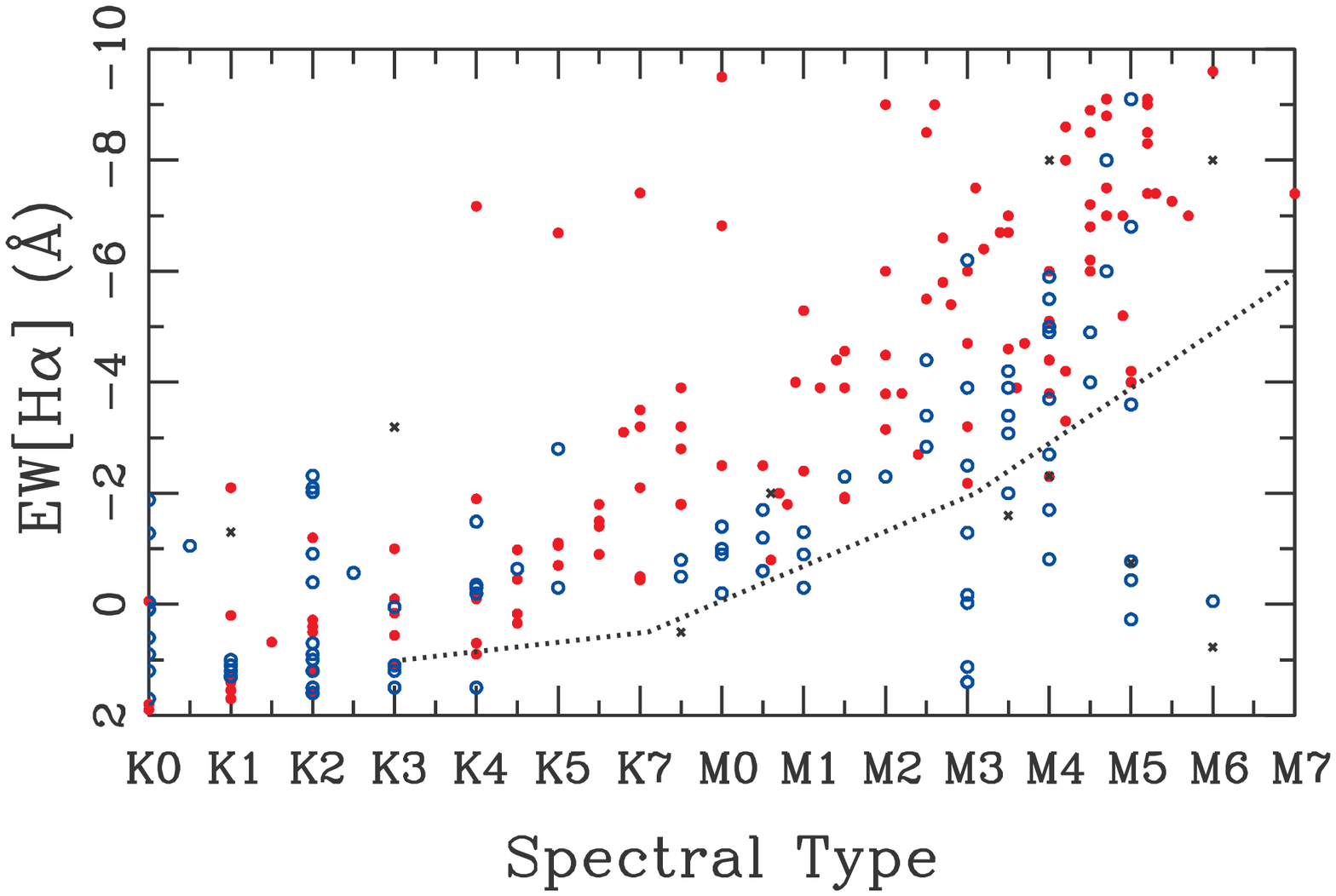}
\figcaption{\label{fig:halpha} H$\alpha$ equivalent width as a function of spectral type for our sample of candidate disk-free Taurus members. The points are color coded as in Figure~\ref{fig:map}, but based on the membership assessment we {\it would have made} based only on the other tests that we use, without using the H$\alpha$ equivalent width. The spectral types are taken from Table 1, and the H$\alpha$ equivalent widths are taken from Tables 2, 3, or 5. The black dashed line denotes the lower envelope of $EW[H\alpha]$ seen in the $\sim$40 Myr Tucana-Horologium moving group (Figure 6 of Kraus et al. 2014). One way for a target to be assessed as a field interloper is to fall below this lower envelope.}
\end{figure*}

\begin{figure*}
\epsscale{1.0}
\includegraphics[scale=0.9,trim={0 6.5cm 0 0},clip]{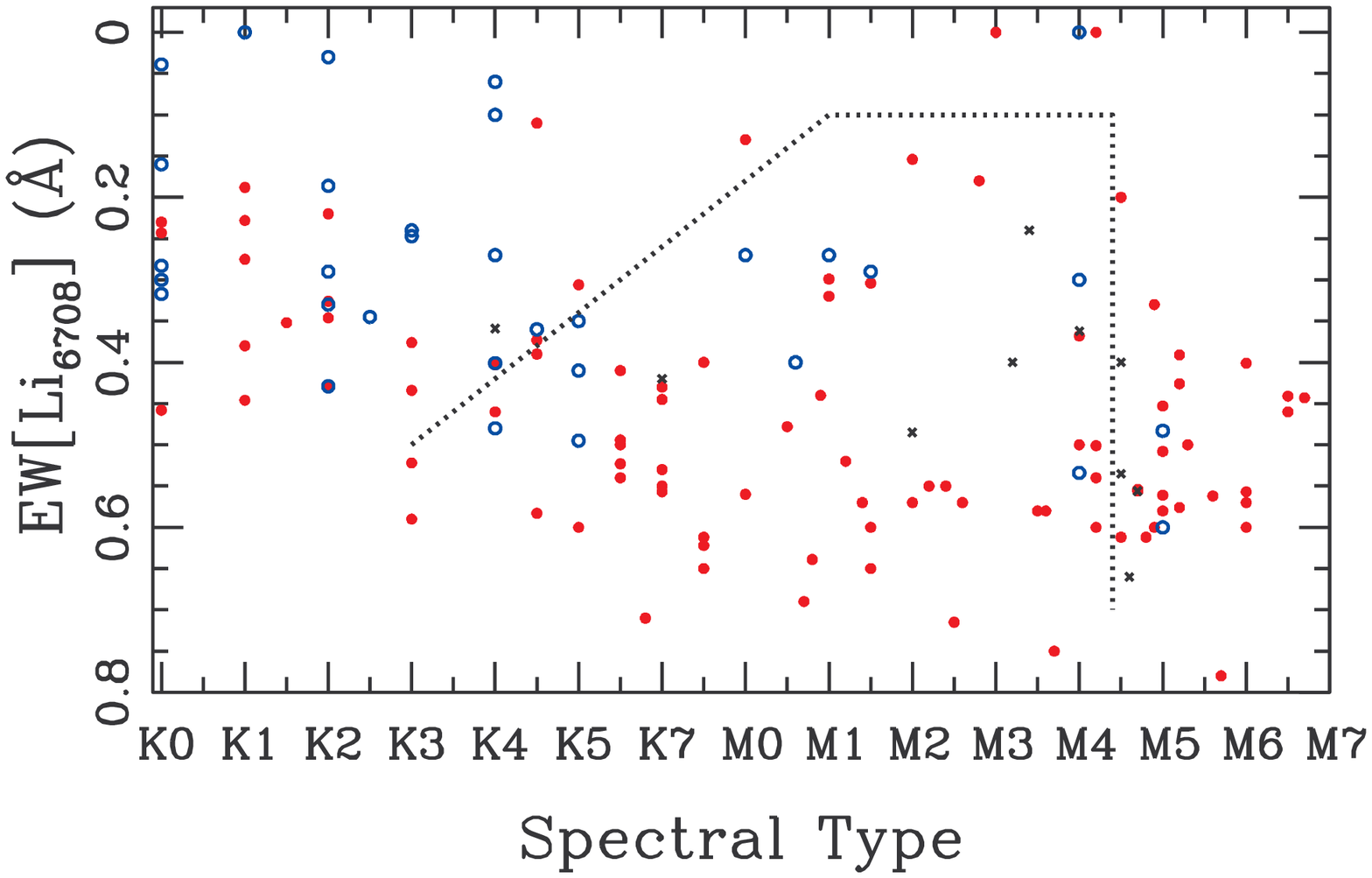}
\figcaption{\label{fig:lithium} Lithium equivalent width as a function of spectral type for our sample of candidate disk-free Taurus members. The points are color coded as in Figure~\ref{fig:map}, but based on the membership assessment we {\it would have made} based only on the other tests that we use, without using the lithium equivalent width. The spectral types are taken from Table 1, and the Li equivalent widths are taken from Tables 2 or 6. The black dashed line denotes the limiting envelope of $EW[Li]$ absorption seen in the $\sim$40 Myr Tucana-Horologium moving group (Figure 7 of Kraus et al. 2014). One way for a target to be assessed as a bona fide Taurus member is to have stronger lithium absorption than this envelope, denoting an age younger than Tuc-Hor.}
\end{figure*}

\begin{figure*}
\epsscale{1.0}
\includegraphics[scale=0.9,trim={0 6.5cm 0 0},clip]{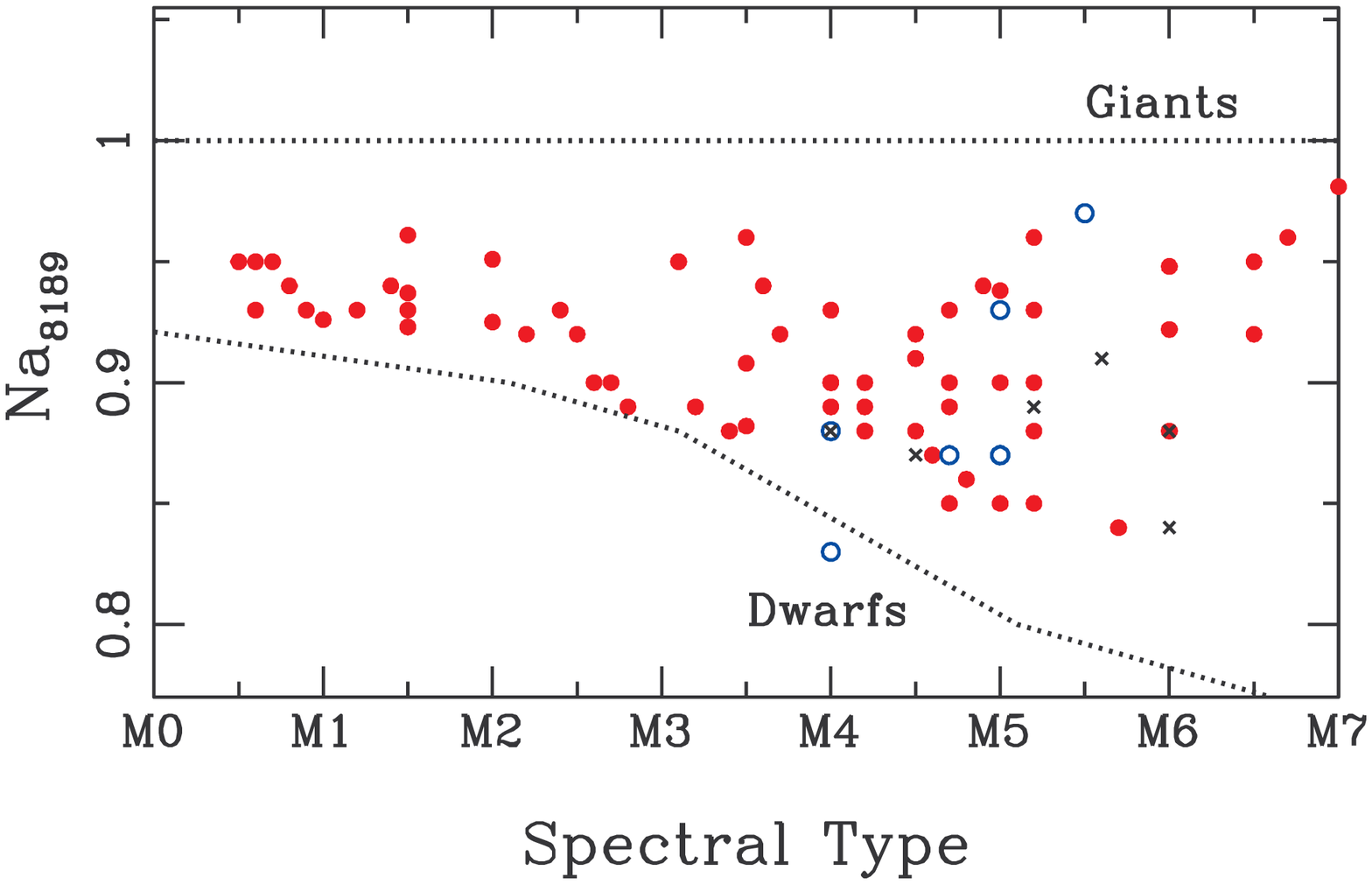}
\figcaption{\label{fig:sodium} Sodium doublet index (Na$_{8189}$) as a function of spectral type for our sample of candidate disk-free Taurus members. The points are color coded as in Figure~\ref{fig:map}, but based on the membership assessment we {\it would have made} based only on the other tests that we use, without using the surface gravity information. The spectral types are taken from Table 1, and the sodium doublet indix measurements are taken from Table~\ref{tab:snifs}, Table~\ref{tab:litna}, and the objects observed by \citet{Slesnick:2006xr} that are in our sample.  The black dashed lines indicate the approximate sequences expected for dwarfs (with deep absorption) and for giants (with no absorption), as defined by Slesnick et al. (2007). Pre-main-sequence objects should fall between the two sequences.}
\end{figure*}

The first steps of our census are to reject those candidates for which one or more features mark them conclusively as old field objects, and then to confirm those candidates which have unambiguous markers of youth. We are able to reject stars if they fall below the Taurus sequence on an HR diagram or if they have H$\alpha$ emission that falls below the lower envelope seen for other young populations. We can confirm objects if they have lithum absorption (as a function of spectral type) that exceeds the upper envelope of an intermediate-age population, or if they have surface gravities which are markedly lower than those of main-sequence objects. We ultimately reject 82 conclusive field interlopers, accept 159 confirmed very young objects ($\tau \la $40 Myr), and are left with 150 candidates to be considered with probabilistic kinematic tests (Section 5).

\subsection{Interlopers on the HR Diagram}

Given the spectral type, extinction, and magnitude measurements compiled in Section 2, we can place each object in our sample in the HR diagram and test for agreement with the age and distance of Taurus. Rather than invoking a temperature scale that is likely to be gravity-dependent (e.g., \citealt{Luhman:2003pb,Pecaut:2013zr}), we use the observed spectral type as a proxy for effective temperature. To compute luminosity (specifically $M_{bol}$), we combine the observed 2MASS $K_s$ magnitude with the dwarf $BC_K$ bolometric corrections of \citet{Kraus:2007mz}, subtract the extinction as converted from $A_V$ to $A_K$ using the relations of \citet{Schlegel:1998yj}, and add the mean distance modulus of Taurus ($d = 145$ pc or $DM=5.8$ mag; \citealt{Torres:2009ct}). This assumed distance might be spurious for field interlopers, but without individual distance measurements, we can only construct the test so as to disprove the hypothesis of Taurus membership, not to specifically address the distances to each target.

In Figure~\ref{fig:hrd}, we show the HR diagram that we have constructed for our sample. We also show the field main sequence \citep{Kraus:2007mz} as shifted to the distance of Taurus. By definition, any pre-main sequence star at that distance must sit above the field sequence. We therefore reject the 44 candidates falling below this limit as a conclusively identified field interlopers.

\subsection{Interlopers with Low H$\alpha$ Emission}


H$\alpha$ emission due to accretion from a disk has long been taken as a positive indicator of youth since Walker (1972) and even Ambartsumian (1947). However, H$\alpha$ emission from chromospheric activity is also a marker of likely youth even among stars that have dispersed their disks, as activity diagnostics have been demonstrated to decline with increasing age (e.g., Skumanich 1972). This test is not unambiguous; old stars can show significant H$\alpha$ emission if they are rapid rotators (such as in tidally locked short-period binaries). We nonetheless can use the absence of H$\alpha$ as an indicator of non-youth, and hence can use it to reject field interlopers from our sample.

In Figure~\ref{fig:halpha}, we plot the H$\alpha$ equivalent width measurements for our candidate disk-free Taurus members as a function of spectral type. We also show the lower envelope for K3--M6 stars that we defined for the 40 Myr Tuc-Hor moving group in \citet{Kraus:2014rt}, based on a similar envelope in IC 2602 and IC 2391 \citep{Stauffer:1997ly}. There are 29 candidate Taurus members in this spectral type range that fall below the Tuc-Hor threshold, and hence must be older than 40 Myr. We do not assess non-membership with H$\alpha$ for any stars with spectral types earlier than K3 or later than M7. Earlier stars do not show a clean separation from the field even in Tuc-Hor, while later-type young objects appear to have a turndown in H$\alpha$ emission line strength, perhaps as the cool atmospheres become increasingly neutral.

\subsection{Confirmation of Youth with Lithium}

Lithium provides a much less ambiguous indicator of youth, at least in the restricted spectral type range of late-K and M dwarfs. By the age of Tuc-Hor ($\sim$40 Myr), lithium is absent for spectral types of M0--M4 and notably depleted from primordial values for spectral types of K3--K7, but remains unburned for spectral types earlier than K3 or later than M4. We therefore have identified an upper envelope for lithium equivalent widths, calibrated with the Tuc-Hor sequence, that identifies objects with SpT = K3--M4 as being notably young ($\tau \la 40$ Myr).

In Figure~\ref{fig:lithium}, we plot the lithium equivalent widths for our candidate disk-free Taurus members as a function of spectral type. We also show the upper envelope for K3--M4 stars that was found for the 40 Myr Tuc-Hor moving group in \citet{Kraus:2014rt}. There are 71 candidate Taurus members in this spectral type range that exceed the Tuc-Hor threshold, indicating that they must be younger than 40 Myr. There are also seven objects in the M1--M4 range that are reported to have $EW[Li] = 0 $ m\AA, indicating that they are older than the lithium burning timescale at those temperatures ($\tau \sim 15$--20 Myr). However, these objects are not necessarily nonmembers; at least some members of Taurus (such as StHa34; \citealt{White:2005gz}) are lithium depleted while still hosting protoplanetary disks which mark them as unambiguously young. 

Lithium is not useful as an age indicator for F-G stars and most K stars because the lithium burning timescale is quite long. The upper envelope of the lithium sequence for the 120 Myr Pleiades open cluster \citep{King:2000cr} does not differ appreciably from the lithium sequence for Tuc-Hor, and both show a substantial spread in EW[Li] as a function of spectral type that might be tied to rotation \citep{Somers:2015ly}. Given that most of our sample members were selected to be active (and hence likely younger than a few hundred million years) based on X-ray or UV emission, then lithium existence among these $<$K3 stars is degenerate with activity and does not constitute a confirming observation. The same argument nominally applies for late-M stars, as the lithium depletion boundary even in the Pleiades only corresponds to a spectral type of $\sim$M5. Most of the late-type objects in our sample were chosen from their position in the HR diagram (without any activity-related criterion that would preselect for youth), so the presence of lithium should still be seen as highly suggestive of Taurus membership, but it is not used as a membership criterion in our analysis.

\subsection{Confirmation of Youth from Low Surface Gravity}

Low surface gravity also provides an unambiguous indicator of youth for objects that are still collapsing toward their final main-sequence radius, manifested in gravity-sensitive features (such as alkali absorption lines) with strengths intermediate between giants and dwarfs. Gravity diagnostics have been exploited by most surveys for M-type Taurus members over the past 15 years (e.g., \citealt{Luhman:1998oz,Slesnick:2006xr}) because they can be conducted at lower resolution than testing for lithium or measuring radial velocities. 

Most gravity-based assessments of youth are based on qualitative comparisons of spectra to standard stars, with only a few surveys conducting this comparison in a quantitative manner (e.g., \citealt{Lyo:2004dg}; \citealt{Slesnick:2006xr}). For consistency, we have followed the methods of \citet{Slesnick:2006xr} for SpT$\ge$M3 candidates that were observed with SNIFS. We have not attempted to reanalyze the qualitative gravity assessments out of the literature, instead adopting them to designate 91 objects as conclusively young due to low surface gravity. The quantitative approach used by \citet{Slesnick:2006xr} also allows for a more nuanced approach. They found a distributed population of mid-late M dwarfs surrounding Taurus with gravity signatures that were intermediate between young stars and dwarfs, consistent with ages younger than the Pleiades. These objects could represent previous generations of star formation in Taurus, so we include them in our sample, but do not regard their gravity assessment as confirming membership. \citet{Luhman:2009wd} also noted three other objects as having qualitatively intermediate gravity, so we also include them as candidates without treating the surface gravity as a conclusive sign of membership.

In Figure~\ref{fig:sodium}, we plot the Na$_{8189}$ index as a function of spectral type for the objects from \citet{Slesnick:2006xr} and for the objects that we have observed with SNIFS, as well as dwarf and giant sequences as outlined by \citet{Slesnick:2006xr}. All of the candidate Taurus members sit above the dwarf sequence, suggesting that they have not contracted to their final main-sequence radius. However, some could fall among the same intermediate-gravity population identified by \citet{Slesnick:2006xr}. To match their division between intermediate-gravity and low-gravity objects, we require candidates to have a Na$_{8189}$ index that is $\ge$0.05 above the dwarf sequence in order for low gravity to be taken as dispositive evidence of membership. For objects which sit above the dwarf sequence but do not reach this threshold, we take the measurement to be indeterminate of membership. One object sits below this sequence, and it appears to be a nonmember due to other criteria as well.

\section{Probabilistic Membership Tests from Kinematics}

\begin{figure*}
\epsscale{1.0}
\includegraphics[scale=0.9,trim={0 6.5cm 0 0},clip]{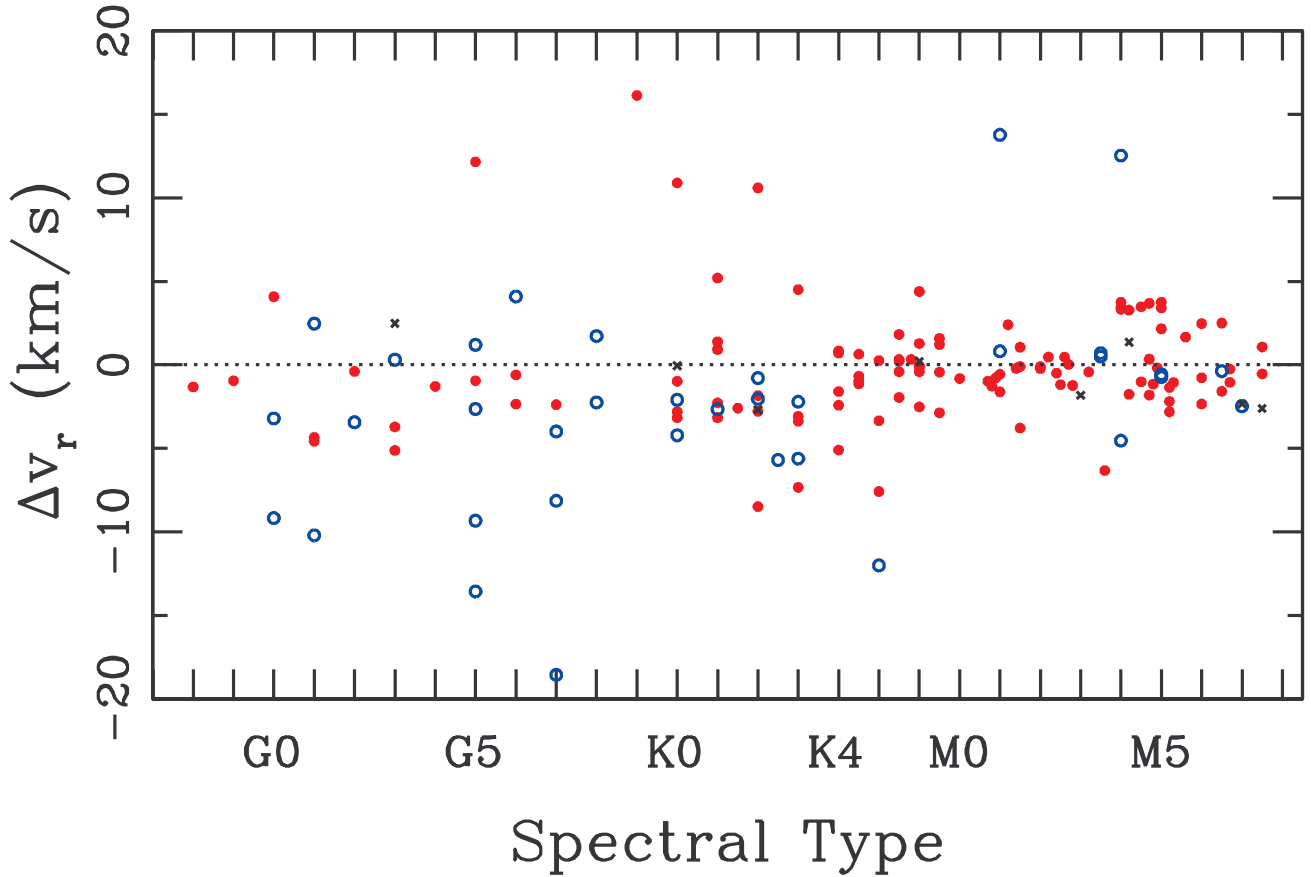}
\figcaption{\label{fig:rv} The discrepancy with respect to the expected RV for a Taurus member, as a function of spectral type, for our sample of candidate disk-free Taurus members. For each star, we compute the difference between the observed RV and the value predicted for that position on the sky (assuming the known space velocity of Taurus, $v_{UVW} = (-15.7, -11.3, -10.1)$ km/s; \citealt{Luhman:2009wd}). The points are color coded as in Figure~\ref{fig:map}, but based on the membership assessment we {\it would have made} based only on the other tests that we use, without using the radial velocity. The spectral types are taken from Table 1, and the RVs are taken from Tables 2 or 7. Based on the RV distribution of lithium-rich stars that must be young (Section 3.3), we assess objects to be likely members if the RV discrepancy is $<$3 km/s. Bona fide members that are single-line spectroscopic binaries might not meet this criterion due to orbital motion, so we only reject objects with a larger RV discrepancy if the RV is shown to be constant in a multi-year time series.}
\end{figure*}


\begin{figure*}
\epsscale{1.0}
\plottwo{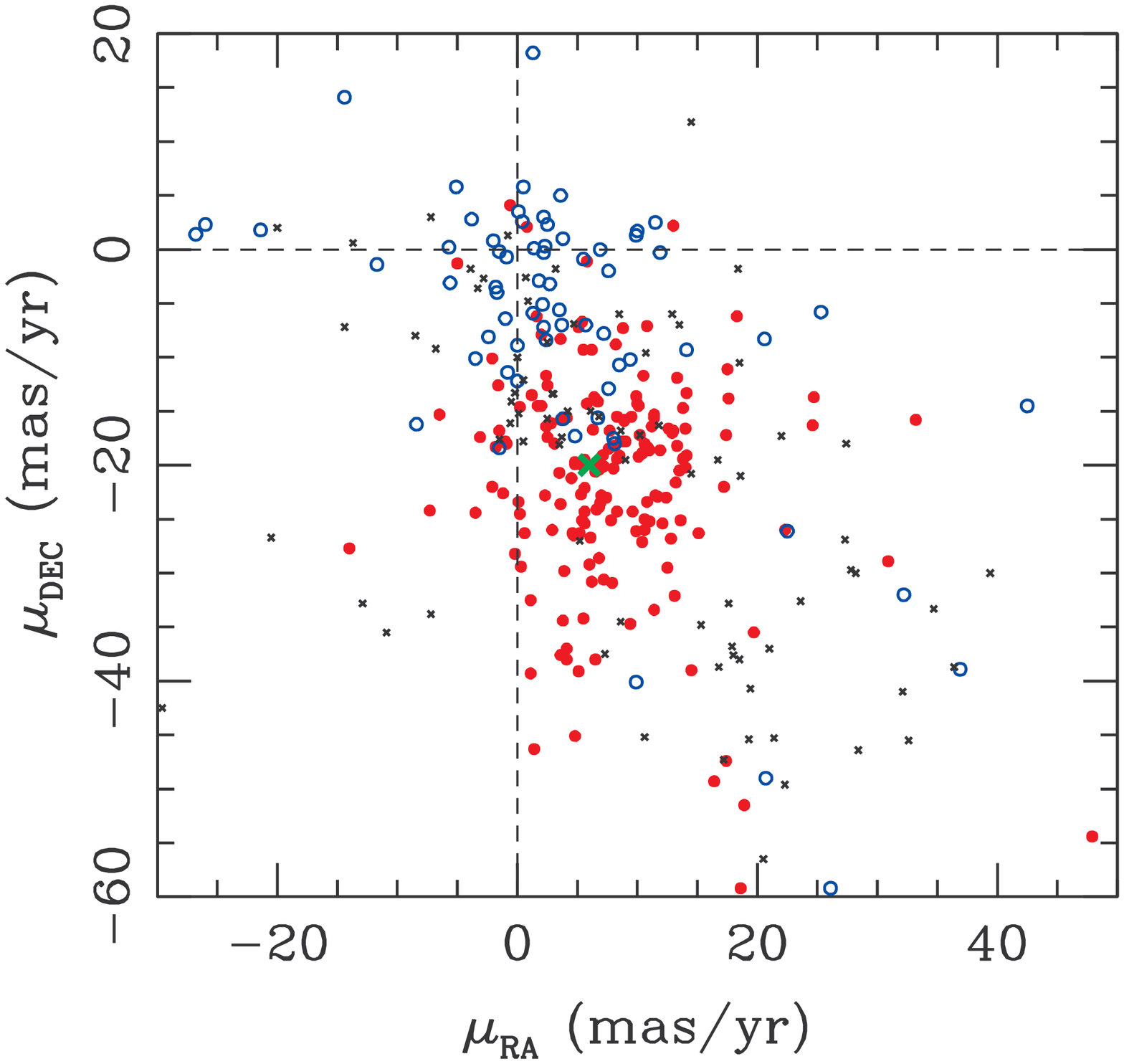}{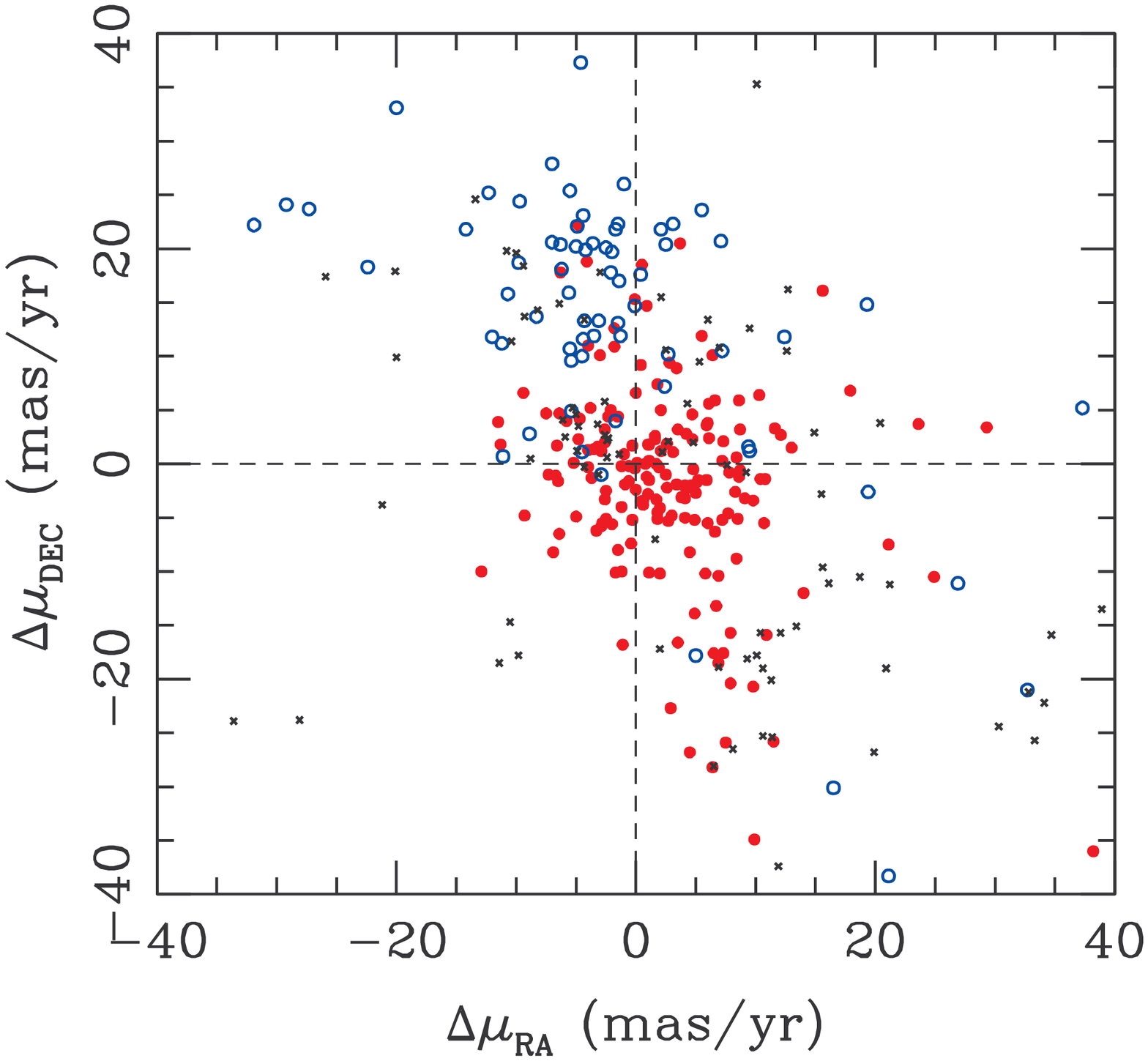}
\figcaption{\label{fig:pm} Left: Proper motion diagram for our sample of candidate disk-free Taurus members. The points are color coded as in Figure~\ref{fig:map}, but based on the membership assessment we {\it would have made} based only on the other tests that we use, without using the proper motion diagram position. The mean proper motion of Taurus is shown with a green X, but the distribution is expected to be significantly broadened by projection effects across the 15$\degr$ extent of Taurus. Right: The corresponding differential proper motion diagram, showing the residual for each target after subtracted the expected proper motion for that object's position on the sky (and hence without projection effects). This analysis also must assume $d = 145$ pc (since almost all candidates lack parallaxes), but given the observed distance dispersion of $\pm$15 pc for Taurus members (Torres et al. 2009), the assumption incurs an additional uncertainty of only $\pm 2$ mas/yr. There clearly is a tighter locus than in the left panel, centered on the origin, which demonstrates that a substantial number of Taurus members are present. Based on the proper motion distribution of lithium-rich stars that must be young (Section 3.3), we assess objects to be likely members if the discrepancy is $\le$10 mas/yr and likely nonmembers if the discrepancy is $>$15 mas/yr.}
\end{figure*}

Spectroscopic signatures of youth become increasingly subtle for stars of earlier spectral type, making such judgements more difficult, and late-type sources can be expensive to observationally confirm. In such cases, an object's membership in a stellar population can also be judged on a probabilistic sense from its position and kinematics. These samples can be polluted by false positives, such as field stars that are comoving by chance. The resulting census also can be incomplete due to false negatives, such as spectroscopic binaries with instantaneous apparent RVs that differ from their systematic RV, or wide binary systems where stellar variability can cause photocenter motion that does not track center-of-mass motion. Outliers on the wings of the distribution due to observational uncertainties also can only be accepted (improving completeness) by accepting more chance contamination of field stars (degrading purity).

In the following subsections, we apply kinematic tests with RVs and proper motions to identify additional likely Taurus members and reject likely field stars. We do not apply any positional tests (i.e., requiring likely members to be clustered with known members) because any such criterion would bias against the detection of new, potentially older subpopulations. In all cases, we test membership against a simple kinematic model where all members are assumed to move with the mean space velocity of Taurus ($v_{UVW} = (-15.7, -11.3, -10.1)$ km/s; \citealt{Luhman:2009wd}), and that velocity is then projected onto the line of sight (for RVs) or the plane of the sky (for proper motions) at the candidate's sky position.

\subsection{Radial Velocities}

Between the Keck/HIRES observations and measurements from the literature, there are 129 candidates with RV measurements available. In Figure~\ref{fig:rv}, we plot the difference as a function of spectral type between the observed RV of each candidate and the expected RV if it moved at the mean Taurus $\vec{v_{UVW}}$. We expect the RV distribution of bona fide, single Taurus members to be inflated by several factors of $\ga$1 km/s, including observational uncertainties, RV jitter from stellar activity (e.g., \citealt{Prato:2008zr,Donati:2014lr}), the large-scale velocity structure of Taurus \citep{Bertout:2006gf}, and the velocity power spectrum around the Taurus mean \citep{Larson:1981qe}. Indeed, we find that of the 86 objects that agree with comovement to within $\le$3 km/s, 48 are flagged as conclusive members while only 6 are flagged as conclusive nonmembers, with 33 remaining ambiguous. For the 41 objects with larger RV offsets, 11 are flagged as conclusive members while 9 are flagged as conclusive nonmembers. We therefore conservatively assess objects to be likely members if their RV agrees with the expected value by $\le$3 km/s.

Binary systems are more difficult to confirm, and this criterion would reject many SB1s or spectrally unresolved SB2s if it were used to reject all objects with discrepant RVs. We therefore suggest that objects with larger offsets should not be rejected without evidence that the velocity is constant and inconsistent with membership. Time-series RVs spanning multiple years are already available for TAP 51 ($\Delta \tau = 6.8$ years; \citealt{Sartoretti:1998bd}) and HD 283759 ($\Delta \tau = 3.9$ years; \citealt{Massarotti:2005ag}), but otherwise is not available in the literature. SB3s also can be assessed more concretely for membership based on the RV of the wide tertiary. \citet{Cuong-Nguyen:2012ul} found that for one SB3 (RX J0412.8+2442) the systemic velocity of the short-period SB2 agrees with the constant velocity of the wide tertiary ($v = +32.3$ km/s) and disagrees significantly with Taurus membership. Similarly, both components of the wide binary pair [LH98] 192 have RVs that are mutually consistent and disagree with membership.




We find an RMS dispersion of $\sigma_{v} = 1.4$ km/s for the candidates that agree with the expected value by $\le$3 km/s, increasing to $\sigma_{v} = 2.1$ km/s for candidates that agree by $\le$6 km/s. Figure~\ref{fig:rv} demonstrates that there is a clear excess of objects of questionable membership ($\delta v > 3$ km/s) with $\Delta v_r < 0$ km/s. The predicted RVs of Taurus members are $v \sim +5$--10 km/s with respect to the LSR, so field stars should have a broader distribution centered at the LSR. The detection of this broader distribution suggests that many of the objects with questionable membership are indeed nonmembers, and instead are distributed about the expected mean for an activity-selected sample of young thin-disk stars.

\subsection{Proper Motions}

In Figure~\ref{fig:pm} (left), we plot the proper motions of the 355 targets with measurements from UCAC4 or our compilation of all-sky surveys. A clear overdensity around the expected proper motion for the center of Taurus ($\mu = $(+6,-20) mas/yr) indicates that there is a large number of comoving Taurus members. However, there is also a clear cluster of targets around the origin (denoting distant objects with negligible proper motions) and an extended distribution of targets with non-zero proper motions that are not consistent with Taurus membership. Distinguishing these populations can provide another test to identify likely background stars. 

This test is complicated by the large survey area. Projection effects lead to different proper motions (by $\pm$5 mas/yr) for true Taurus members with the same $UVW$ velocities, but different positions on the sky. To correct for projection effects, in Figure~\ref{fig:pm} (right) we plot the residuals for each target after subtracting the expected proper motion at its position on the sky (assuming the mean Taurus $UVW$ velocity and a distance of $d = 145$ pc). The Taurus overdensity is tighter in this plot, indicating that projection effects do broaden the distribution.

Among targets comoving within $<$5 mas/yr, there are 46 confirmed members (from Section 4 alone), 4 confirmed nonmembers (Section 4 or Section 5.1), and 18 objects which we were not able to assess in Section 4 or Section 5.1. For targets comoving within 5--10 mas/yr, there are 46 confirmed members, 10 confirmed nonmembers, and 25 objects that lacked membership indicators. For targets comoving within 10--15 mas/yr, there are 22 confirmed members, 16 confirmed nonmembers, and 9 objects that lacked conclusive indicators. For all larger levels of discrepancy, there are 14 confirmed members, 66 confirmed nonmembers, and 63 objects that lacked conclusive indicators.

We therefore assess objects to be likely nonmembers if they disagree with comovement by $>$15 mas/yr (where only $\sim$20\% of the objects with an assessment are bona fide members) and likely members if they agree with comovement by $<$10 mas/yr (where $\sim$90\% of objects with an assessment are bona fide members). In the intermediate range of 10--15 mas/yr, we find that $\sim$50\% of assessed objects are bona fide members. Given that only 9 objects with discrepancies of 10--15 mas/yr lack such an assessment, we suggest that these targets should be assessed in more detail in the future, but they can not currently assessed as members or nonmembers using proper motions.

\section{An Updated Census of Taurus-Auriga}

The conclusive tests of youth described in Section 3 allow us to assess 160 candidates as definitely young (and hence almost certainly Taurus members) based on the presence of low surface gravity (104 candidates) and/or strong lithium absorption (60 candidates). We also can assess another 69 candidates to be definitely old based on their low CMD position (44 candidates) and/or low H$\alpha$ emission (27 candidates). We summarize the outcomes of these assessments in the first four columns of Table~\ref{tab:diag}, denoting objects as ``Y'', ``N'', ``?'' (if data exists and the test is inconclusive) or ``...'' (if the test is inapplicable or the data does not exist).

There are four additional cases where the evidence casts doubt that one of these nominally conclusive indicators of youth can be taken as evidence of Taurus membership, so they must be judged individually. HBC 351 has strong lithium absorption that denotes youth, but it also has a proper motion that is discrepant with Taurus and agrees well with the Pleiades, so we reject it as a Taurus member. The HBC 356/357 binary system also has lithium denoting youth, but it sits low in the CMD and has a discrepant proper motion that better agrees with the more distant Perseus star-forming region, so we also reject it. Finally, two other mid-M candidates ([SCH2006b] J0416272+2053093 and [SCH2006b] J0523020+2428087) also have competing indicators from lithium and/or gravity and H$\alpha$. Given the very fast timescale for lithium depletion in this temperature range, consistent proper motions for both objects, and consistent RV and low surface gravity for the first object, then we accept both as Taurus members.


The kinematic tests similarly allow us to assess many candidates as likely members or nonmembers. Using the proper motion test, we designate 143 objects with ``Y?'' if they agree with comovement within $\le$10 mas/yr, 46 objects as ``?'' if they disagree at 10--15 mas/yr, and 158 objects as ``N?'' if they disagree at $>$15 mas/yr. Using the RV test, we designate 103 candidate members as ``Y?'' if they agree with comovement within $\le$3 km/s, 51 candidates as ``?'' if there is single-epoch data that disagrees with comovement at $>$3 km/s, and 5 candidates as ``N?'' if there are RV time series or measurements from tertiaries that indicate a constantly discrepant systemic velocity. Among the objects without conclusive indicators of age, we identify a total of 58 candidates as likely members and 91 candidates as likely nonmembers, based on the sum of the kinematic evidence. We find that 14 candidates have conflicting indicators from proper motions and radial velocities. We have inspected the proper motion and RV data for those objects and do not find any reasons to disregard the negative results, and therefore we assess them to be likely nonmembers.

In summary, we find that there are 218 confirmed or likely Taurus members, 160 confirmed or likely nonmembers, and 18 candidates that still lack sufficient evidence to draw any conclusions regarding their membership. Of the confirmed or likely Taurus members, 81 sources (37\%) have been largely omitted from the canonical Taurus census (e.g., Esplin et al. 2014). Most of these missing sources are G--K stars drawn from the ROSAT-based surveys of \citet{Wichmann:1996fk} and \citet{Li:1998fj}, suggesting that a substantial reservoir of unrecognized lower-mass members remains to be identified. However, we note that these new Taurus members are still the minority of all ROSAT-selected stars. Our results therefore are not inconsistent with those of Briceno et al. (1997) and Guilliot et al. (1998), who used models of the Milky Way's recent star formation history to argue that the distributed population might be a mixture of young stars that have formed in many regions over the past $10^8$ yr. The majority of these X-ray selected young stars are indeed part of the Milky Way field, but the density is not so large as to overwhelm the distributed population of bona fide Taurus members.

\section{The Age, Kinematics, and IMF of Taurus-Auriga}

\figsetstart
\figsetnum{8}
\figsettitle{Disk-Host and Disk-Free Map}


\begin{figure*}
\epsscale{1.0}
\includegraphics[scale=0.6,trim={0 0 0 0},clip]{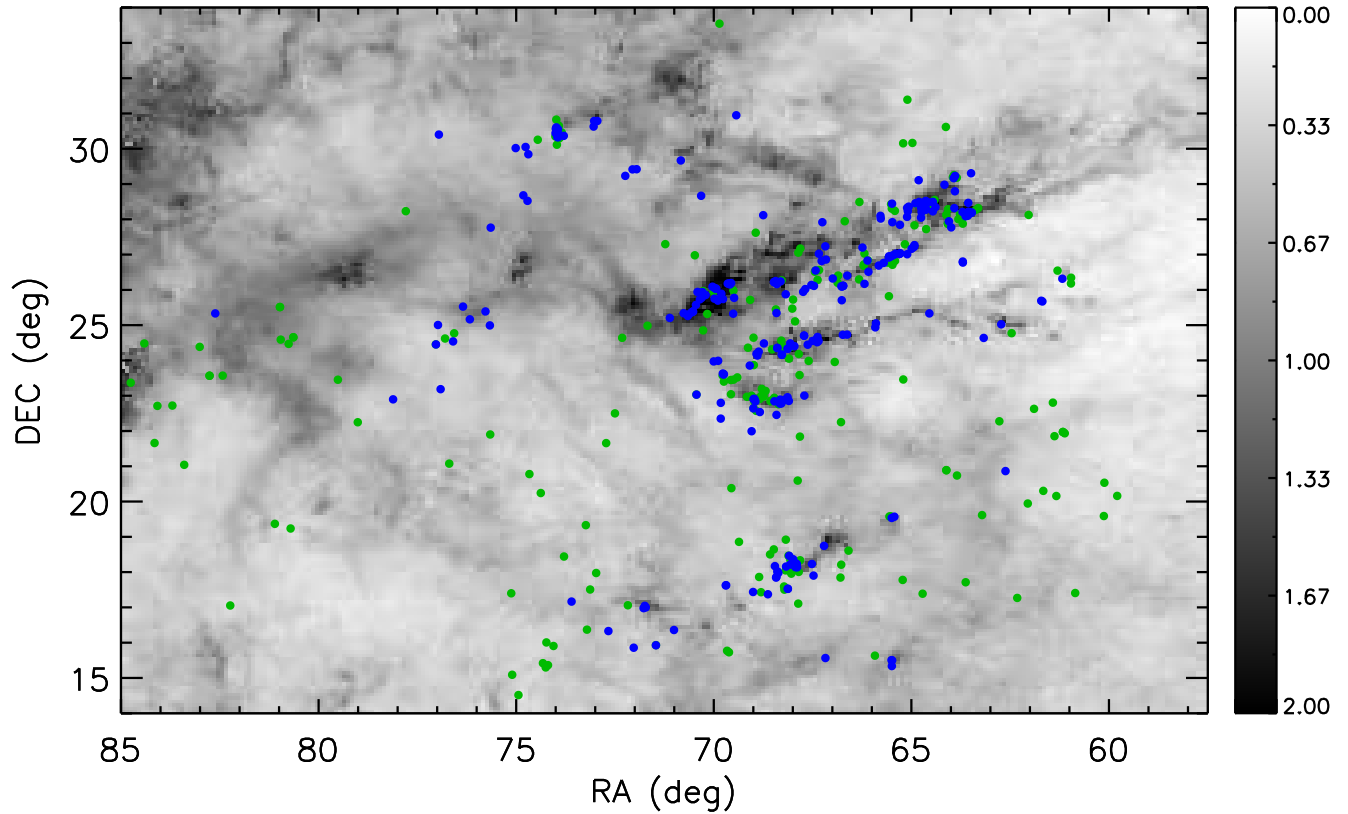}
\figcaption{\label{fig:diskmap} Spatial distribution for all members of Taurus. Disk-free members confirmed in our census are shown with filled green circles. Disk-hosting members of Taurus (Rebull et al. 2010; Luhman et al. 2010; Esplin et al. 2014) are shown with filled blue circles. The background image is an extinction map compiled by \citet{Schlafly:2014lr}. Most members of the disk-hosting population are clearly concentrated around the ongoing sites of star formation in Taurus, whereas the disk-free population also has a more widely distributed component. To improve readability, other realizations of this color scheme (red-green and red-blue) are included as a figure set in the electronic version of the journal.}
\end{figure*}

\figsetgrpstart
\figsetgrpnum{8.1}
\figsetgrptitle{Blue-Green Map}
\figsetplot{diskbluegreen.eps}
\figsetgrpnote{Spatial distribution for all members of Taurus. Disk-free members confirmed in our census are shown with filled green circles. Disk-hosting members of Taurus (Rebull et al. 2010; Luhman et al. 2010; Esplin et al. 2014) are shown with filled blue circles. The background image is an extinction map compiled by \citet{Schlafly:2014lr}. Most members of the disk-hosting population are clearly concentrated around the ongoing sites of star formation in Taurus, whereas the disk-free population also has a more widely distributed component.}
\figsetgrpend

\figsetgrpstart
\figsetgrpnum{8.2}
\figsetgrptitle{Red-Blue Map}
\figsetplot{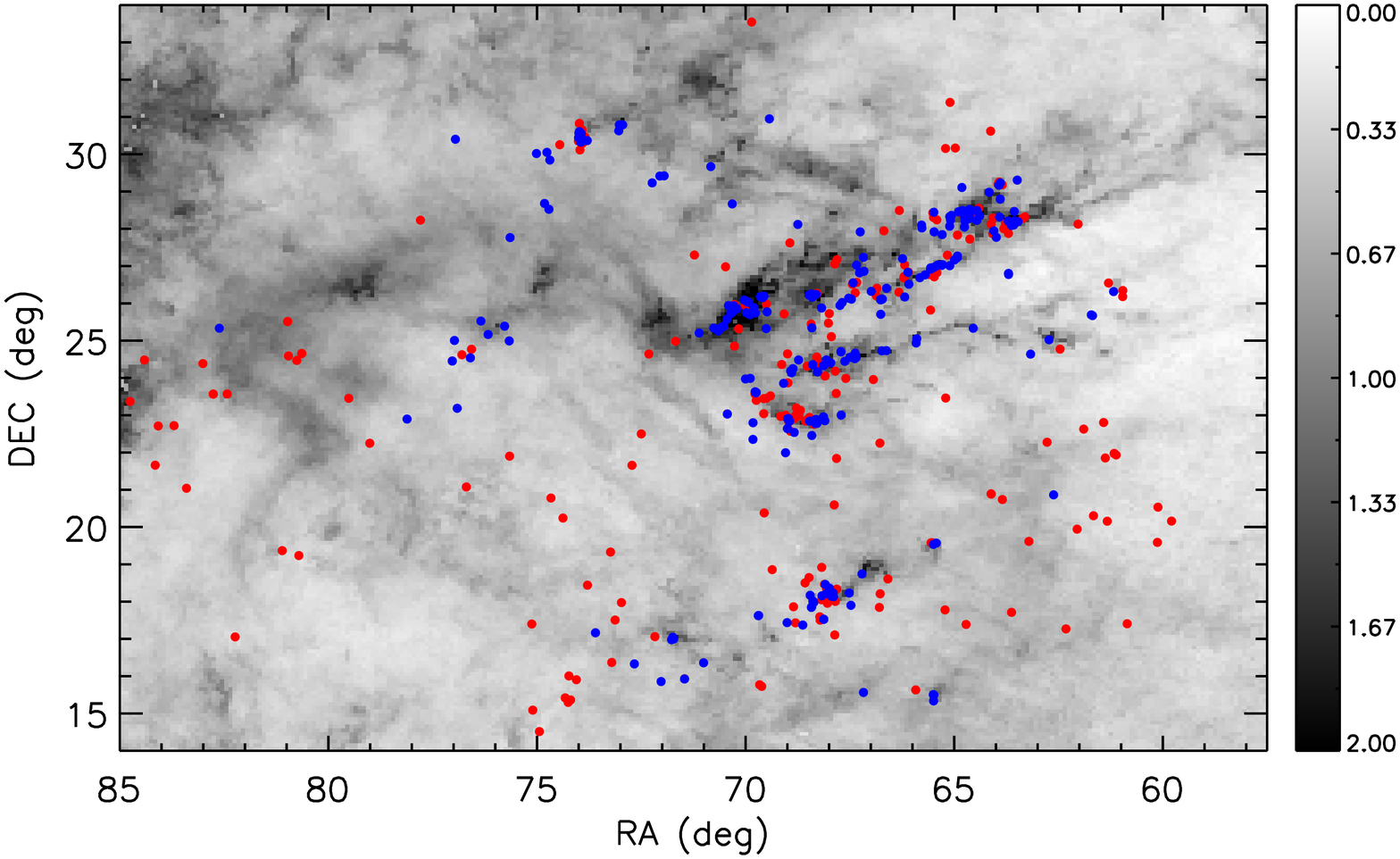}
\figsetgrpnote{Spatial distribution for all members of Taurus. Disk-free members confirmed in our census are shown with filled red circles. Disk-hosting members of Taurus (Rebull et al. 2010; Luhman et al. 2010; Esplin et al. 2014) are shown with filled blue circles. The background image is an extinction map compiled by \citet{Schlafly:2014lr}. Most members of the disk-hosting population are clearly concentrated around the ongoing sites of star formation in Taurus, whereas the disk-free population also has a more widely distributed component.}
\figsetgrpend

\figsetgrpstart
\figsetgrpnum{8.3}
\figsetgrptitle{Blue-Green Map}
\figsetplot{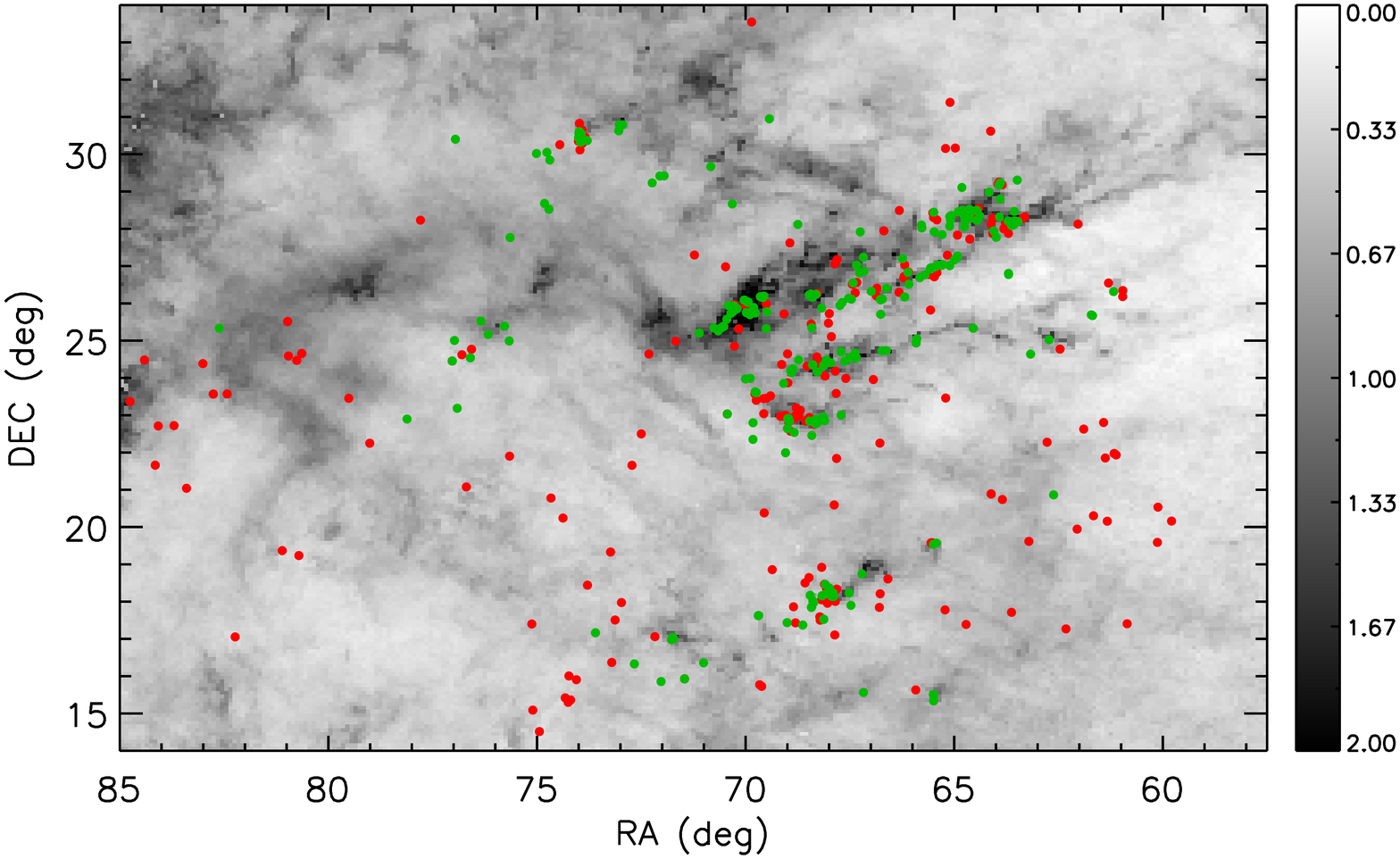}
\figsetgrpnote{Spatial distribution for all members of Taurus. Disk-free members confirmed in our census are shown with filled red circles. Disk-hosting members of Taurus (Rebull et al. 2010; Luhman et al. 2010; Esplin et al. 2014) are shown with filled green circles. The background image is an extinction map compiled by \citet{Schlafly:2014lr}. Most members of the disk-hosting population are clearly concentrated around the ongoing sites of star formation in Taurus, whereas the disk-free population also has a more widely distributed component.}
\figsetgrpend

\figsetend

\figsetstart
\figsetnum{9}
\figsettitle{Disk Fraction and Density Maps}

\begin{figure*}
\epsscale{1.0}
\hspace{-0.3in}\includegraphics[scale=0.46,trim={0 0 0 0},clip]{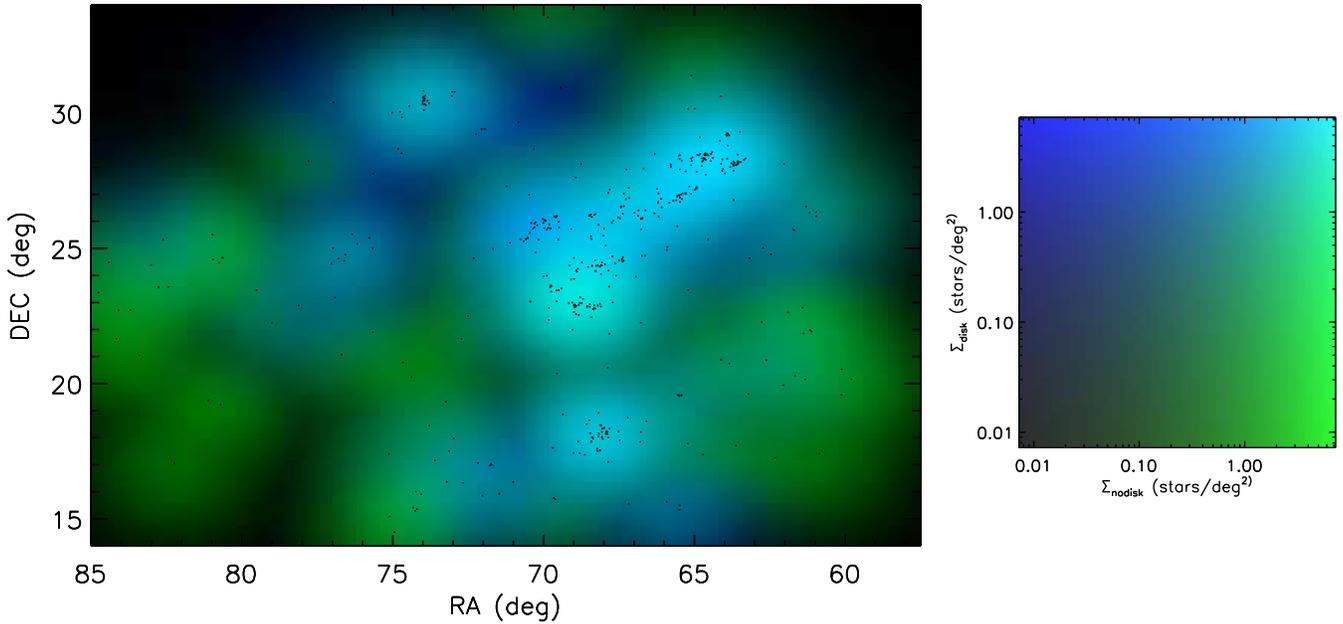}
\figcaption{\label{fig:rgbmap} Stellar density and disk fraction in Taurus, as encompassed in an RGB image. The blue and green channels were computed by convolving the point distributions in Figure~\ref{fig:diskmap} with a Gaussian blur kernel of width $\sigma = 1\degr$, such that the image conveys both the disk fraction (via the hue) and the stellar density (via the total intensity). The mapping of intensity and hue to $\Sigma_{disk}$ and $\Sigma_{nodisk}$ are encompassed in the 2D key shown on the right. The Taurus population appears to be composed of high-density regions with a high disk fraction (i.e., bright and cyan), surrounded by a distributed low-density component with low disk fraction (i.e., faint and red). To improve readability, other realizations of this color scheme (red-green-yellow and red-blue-magenta) are included as a figure set in the electronic version of the journal and at the end of this manuscript.}
\end{figure*}

\figsetgrpstart
\figsetgrpnum{9.1}
\figsetgrptitle{Blue-Green Map}
\figsetplot{DFGbluegreenmap.eps}
\figsetgrpnote{Stellar density and disk fraction in Taurus, as encompassed in an RGB image. The blue and green channels were computed by convolving the point distributions of disk-hosting and disk-free stars in Figure~\ref{fig:diskmap} with a Gaussian blur kernel of width $\sigma = 1\degr$, such that the image conveys both the disk fraction (via the hue) and the stellar density (via the total intensity). The mapping of intensity and hue to $\Sigma_{disk}$ and $\Sigma_{nodisk}$ are encompassed in the 2D key shown on the right. The Taurus population appears to be composed of high-density regions with a high disk fraction (i.e., bright and cyan), surrounded by a distributed low-density component with low disk fraction (i.e., faint and green).}
\figsetgrpend

\figsetgrpstart
\figsetgrpnum{9.2}
\figsetgrptitle{Red-Blue Map}
\figsetplot{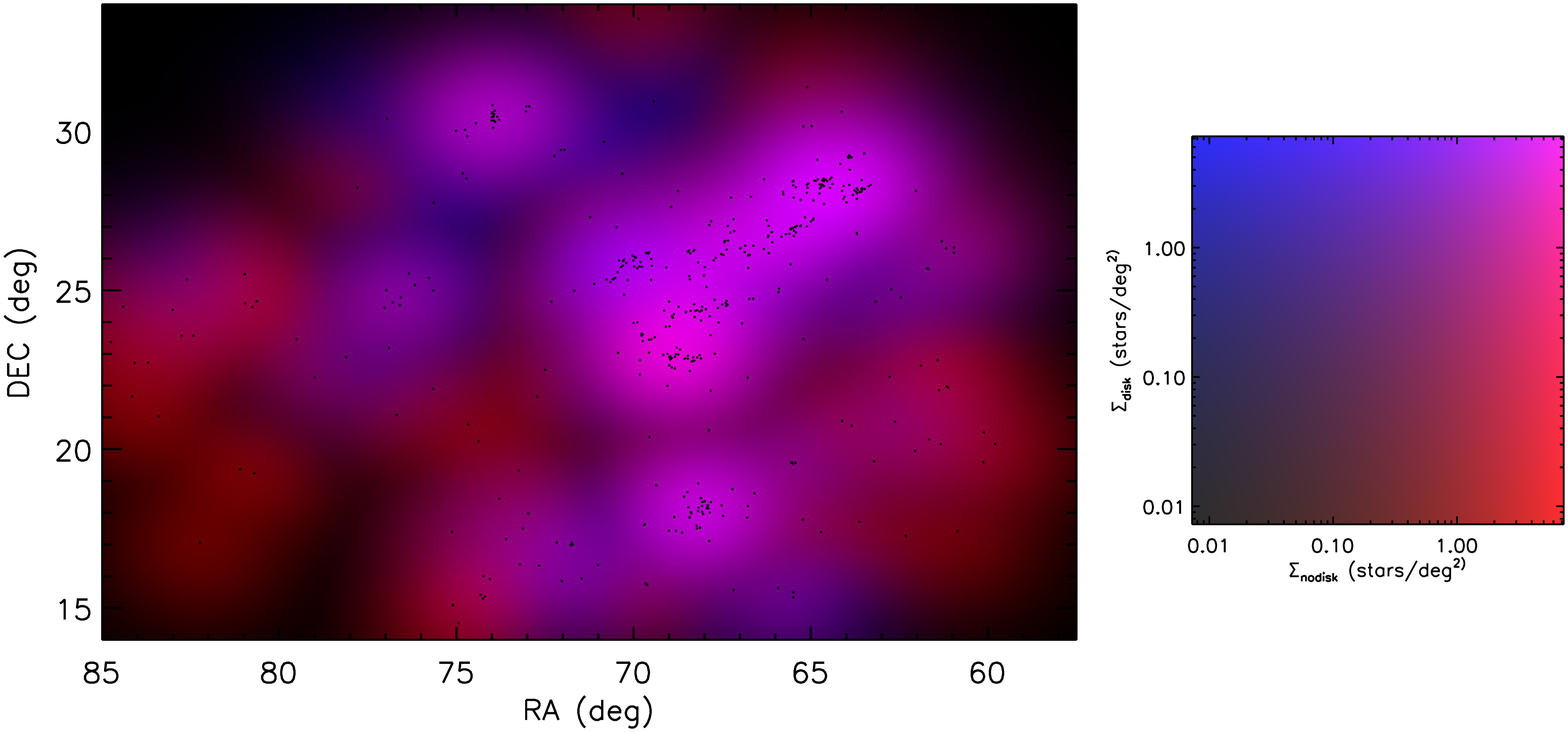}
\figsetgrpnote{Stellar density and disk fraction in Taurus, as encompassed in an RGB image. The blue and green channels were computed by convolving the point distributions of disk-hosting and disk-free stars in Figure~\ref{fig:diskmap} with a Gaussian blur kernel of width $\sigma = 1\degr$, such that the image conveys both the disk fraction (via the hue) and the stellar density (via the total intensity). The mapping of intensity and hue to $\Sigma_{disk}$ and $\Sigma_{nodisk}$ are encompassed in the 2D key shown on the right. The Taurus population appears to be composed of high-density regions with a high disk fraction (i.e., bright and magenta), surrounded by a distributed low-density component with low disk fraction (i.e., faint and red).}
\figsetgrpend

\figsetgrpstart
\figsetgrpnum{9.1}
\figsetgrptitle{Red-Green Map}
\figsetplot{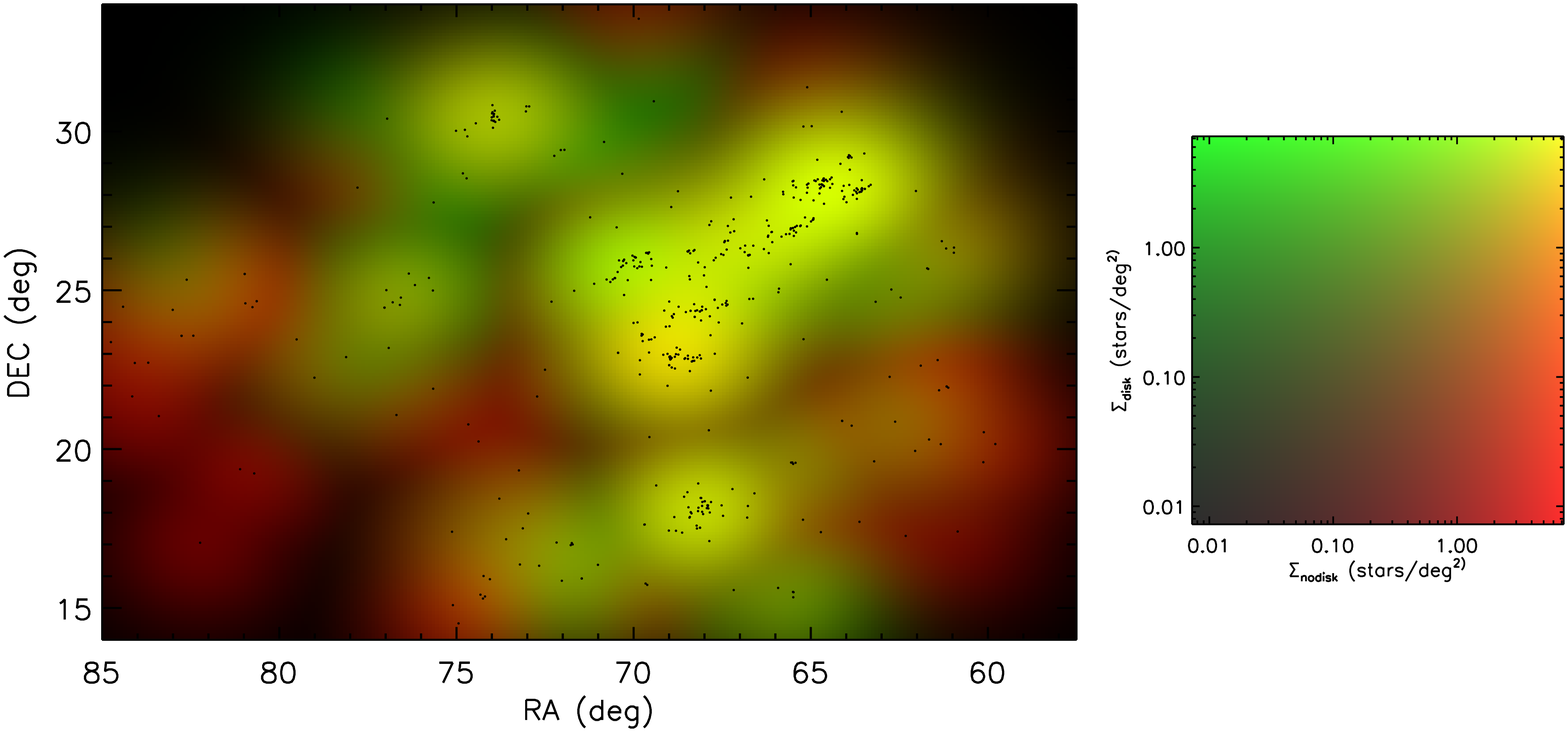}
\figsetgrpnote{Stellar density and disk fraction in Taurus, as encompassed in an RGB image. The blue and green channels were computed by convolving the point distributions of disk-hosting and disk-free stars in Figure~\ref{fig:diskmap} with a Gaussian blur kernel of width $\sigma = 1\degr$, such that the image conveys both the disk fraction (via the hue) and the stellar density (via the total intensity). The mapping of intensity and hue to $\Sigma_{disk}$ and $\Sigma_{nodisk}$ are encompassed in the 2D key shown on the right. The Taurus population appears to be composed of high-density regions with a high disk fraction (i.e., bright and yellow), surrounded by a distributed low-density component with low disk fraction (i.e., faint and red).}
\figsetgrpend

\figsetend

\begin{figure*}
\epsscale{1.0}
\includegraphics[scale=0.6,trim={0 0 0 0},clip]{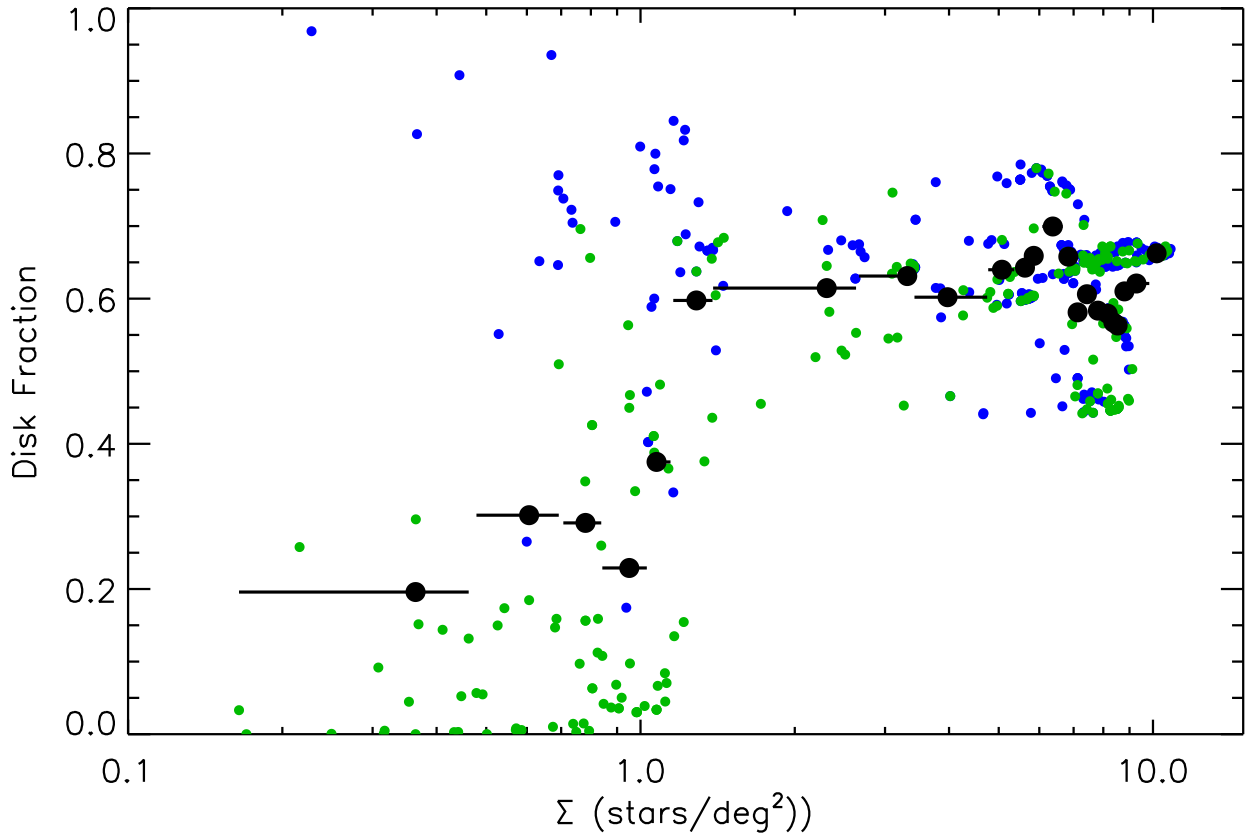}
\figcaption{\label{fig:densityfrac} Disk fraction as a function of stellar density in Taurus. The points represent samplings of Figure~\ref{fig:rgbmap} at the location of disk-free (green) and disk-hosting (blue) members of Taurus, where the total stellar intensity is taken from the sum of the green and blue channels and the disk fraction is taken from their ratio. We also show the average value for bins of 20 Taurus members with black points, where the horizontal line shows the extent of each bin; the poisson uncertainty on these averages is $\pm$0.1, but the scatter is smaller due to covariance between spatially adjacent stars. The correlation between disk fraction and stellar density is demonstrated more quantitatively by the absence of any samplings with disk fraction $F \la 40\%$ for stellar densities $\Sigma \ga 1.5$ deg$^{-2}$.}
\end{figure*}

\begin{figure*}
\epsscale{1.0}
\includegraphics[scale=0.6,trim={0 0 0 0},clip]{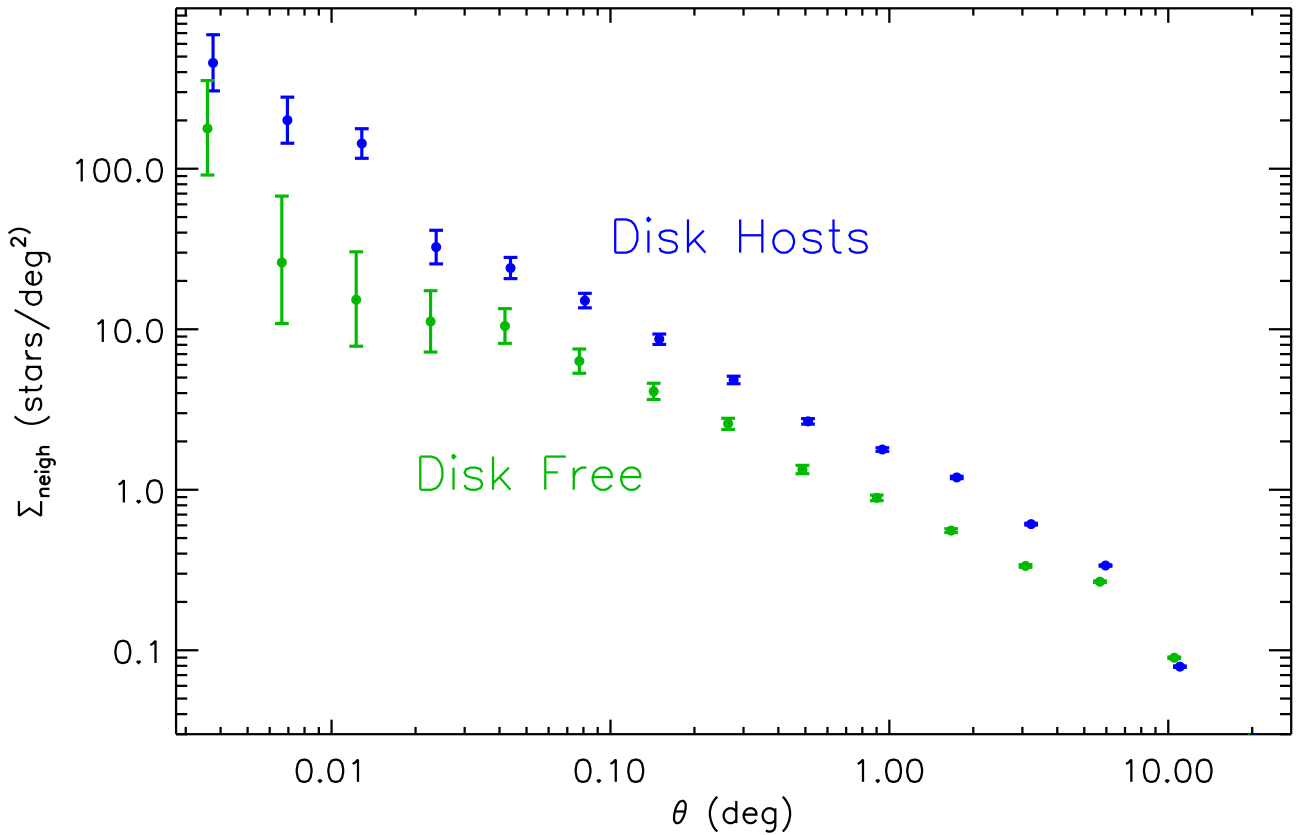}
\figcaption{\label{fig:tpcf} Two-point correlation function for disk-free (green) and disk-hosting (blue) members of Taurus. The two populations are nearly equal, so if the populations were distributed with the same clustering properties, the two TPFCs should overlap. However, the disk-hosting TPCF shows more power across most spatial scales ($\theta \la 6\degr$), while the disk-free population only has more power at the largest scale ($\theta \sim 10\degr$). The disk-hosting members therefore are more tightly clustered than the disk-free members.}
\end{figure*}

\figsetstart
\figsetnum{12}
\figsettitle{HR Diagrams}

\begin{figure*}
\epsscale{1.0}
\includegraphics[scale=0.55,trim={0 0 0 0},clip]{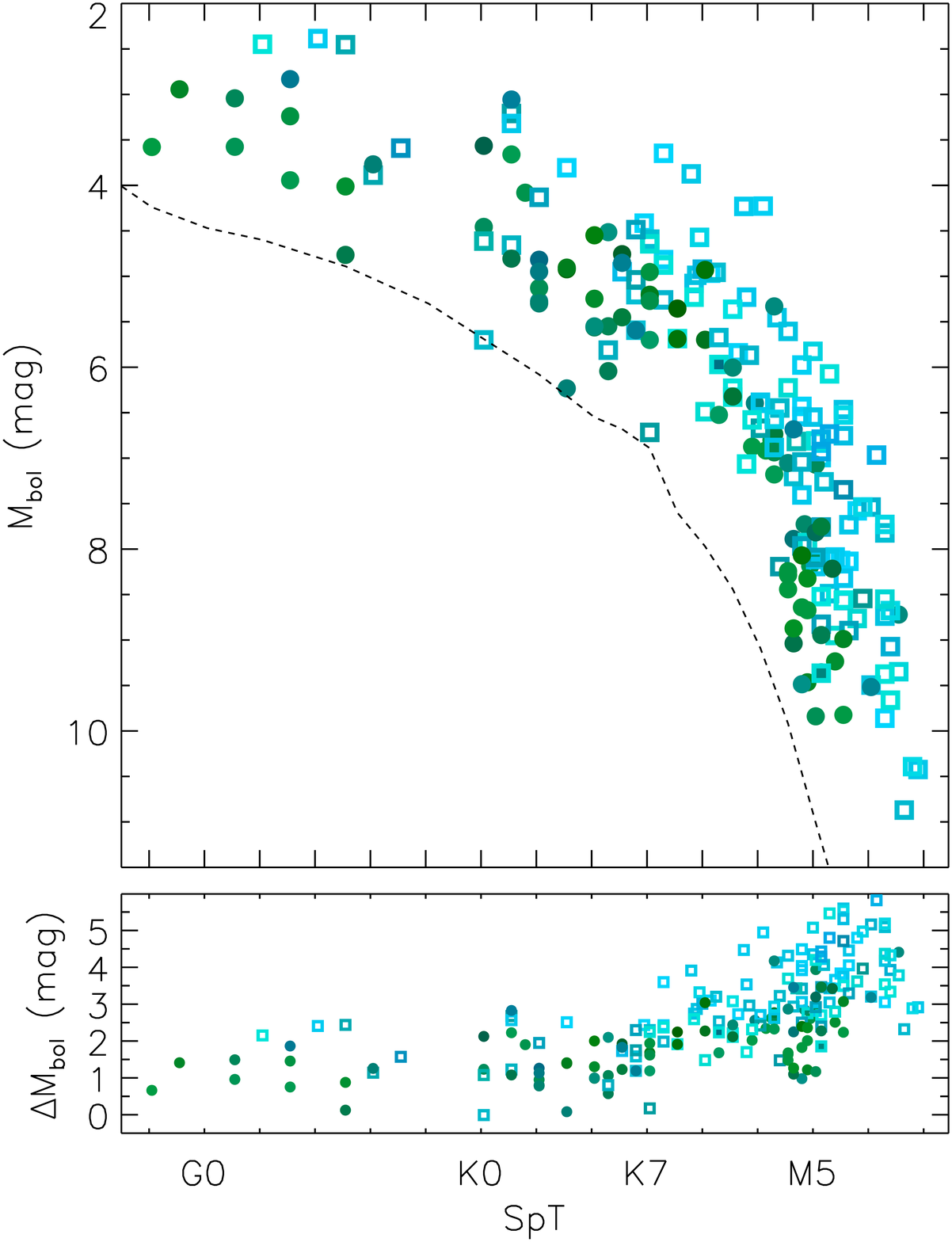}
\figcaption{\label{fig:HRDmemdisk} HR diagram for the disk-free Taurus members of our sample (upper panel) and the height above the main sequence of each member (lower panel). Each point is color coded to match the local stellar density and disk fraction shown in Figure~\ref{fig:rgbmap}, such that the disk-rich clustered population is bright cyan and the disk-free distributed population is dark green. To emphasize the difference between the clustered and distributed populations, objects with a local stellar density of $\Sigma > 1.5$ stars/deg$^2$ are shown with open squares, while objects with $\Sigma \le 1.5$ stars/deg$^2$ are shown with filled circles. The distributed population clearly sits below the clustered population in the HR diagram, indicating either that those objects are either older (by a factor of 2--3) or more distant (by $\sim$80 pc). As in Figure~\ref{fig:hrd}, we also show the field main sequence ($\tau \sim 600$ Myr; $[Fe/H] \sim 0$) as defined in Kraus \& Hillenbrand (2007).}
\end{figure*}

\figsetgrpstart
\figsetgrpnum{12.1}
\figsetgrptitle{Blue-Green HRD}
\figsetplot{HRDmemdiskbluegreen.eps}
\figsetgrpnote{HR diagram for the disk-free Taurus members of our sample (upper panel) and the height above the main sequence of each member (lower panel). Each point is color coded to match the local stellar density and disk fraction shown in the color-matched version of Figure~\ref{fig:rgbmap}, such that the disk-rich clustered population is bright cyan and the disk-free distributed population is dark green. To emphasize the difference between the clustered and distributed populations, objects with a local stellar density of $\Sigma > 1.5$ stars/deg$^2$ are shown with open squares, while objects with $\Sigma \le 1.5$ stars/deg$^2$ are shown with filled circles. The distributed population clearly sits below the clustered population in the HR diagram, indicating either that those objects are either older (by a factor of 2--3) or more distant (by $\sim$80 pc). As in Figure~\ref{fig:hrd}, we also show the field main sequence ($\tau \sim 600$ Myr; $[Fe/H] \sim 0$) as defined in Kraus \& Hillenbrand (2007).}
\figsetgrpend

\figsetgrpstart
\figsetgrpnum{12.2}
\figsetgrptitle{Red-Blue HRD}
\figsetplot{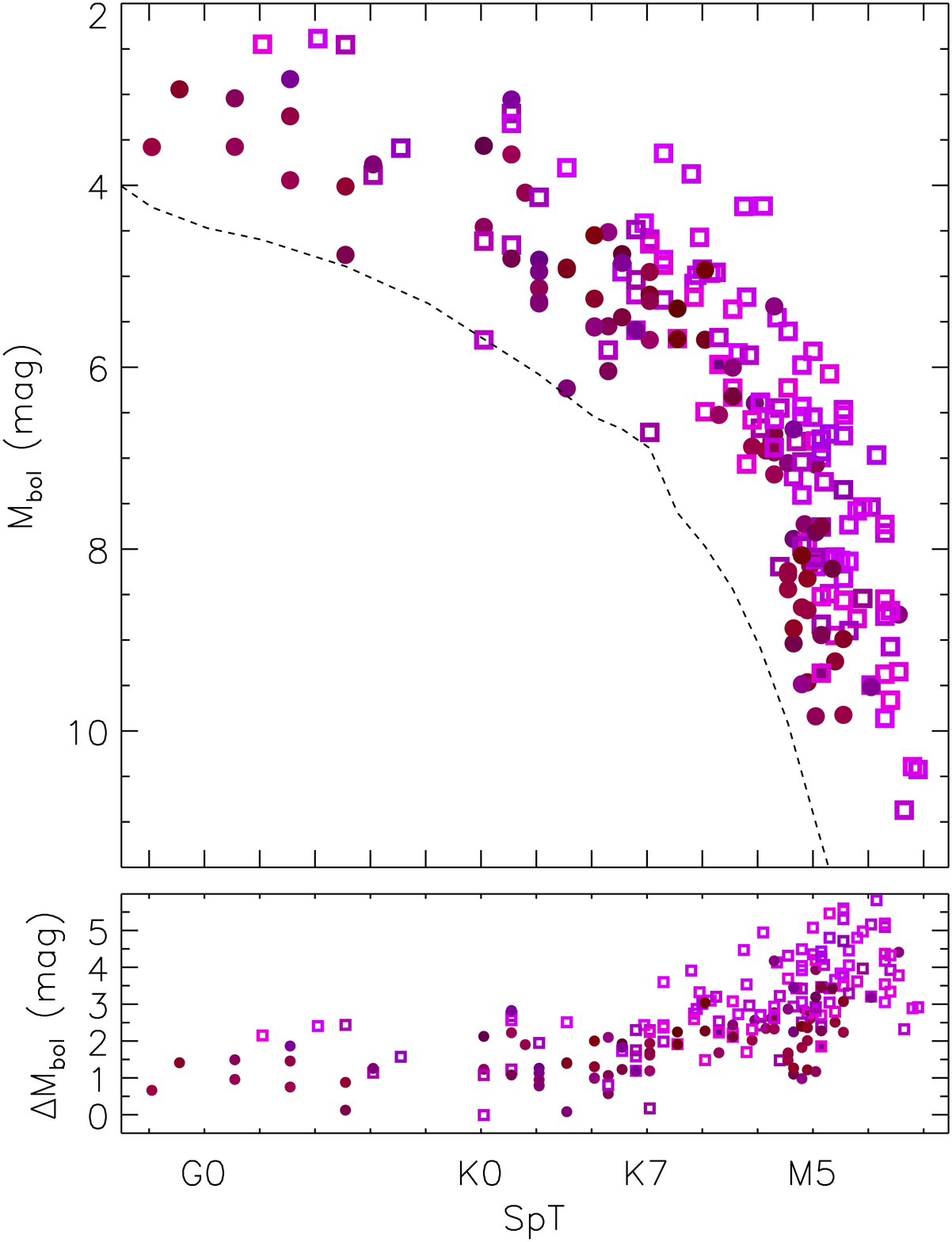}
\figsetgrpnote{HR diagram for the disk-free Taurus members of our sample (upper panel) and the height above the main sequence of each member (lower panel). Each point is color coded to match the local stellar density and disk fraction shown in the color-matched version of Figure~\ref{fig:rgbmap}, such that the disk-rich clustered population is bright magenta and the disk-free distributed population is dark red. To emphasize the difference between the clustered and distributed populations, objects with a local stellar density of $\Sigma > 1.5$ stars/deg$^2$ are shown with open squares, while objects with $\Sigma \le 1.5$ stars/deg$^2$ are shown with filled circles. The distributed population clearly sits below the clustered population in the HR diagram, indicating either that those objects are either older (by a factor of 2--3) or more distant (by $\sim$80 pc). As in Figure~\ref{fig:hrd}, we also show the field main sequence ($\tau \sim 600$ Myr; $[Fe/H] \sim 0$) as defined in Kraus \& Hillenbrand (2007).}
\figsetgrpend

\figsetgrpstart
\figsetgrpnum{12.3}
\figsetgrptitle{Blue-Green HRD}
\figsetplot{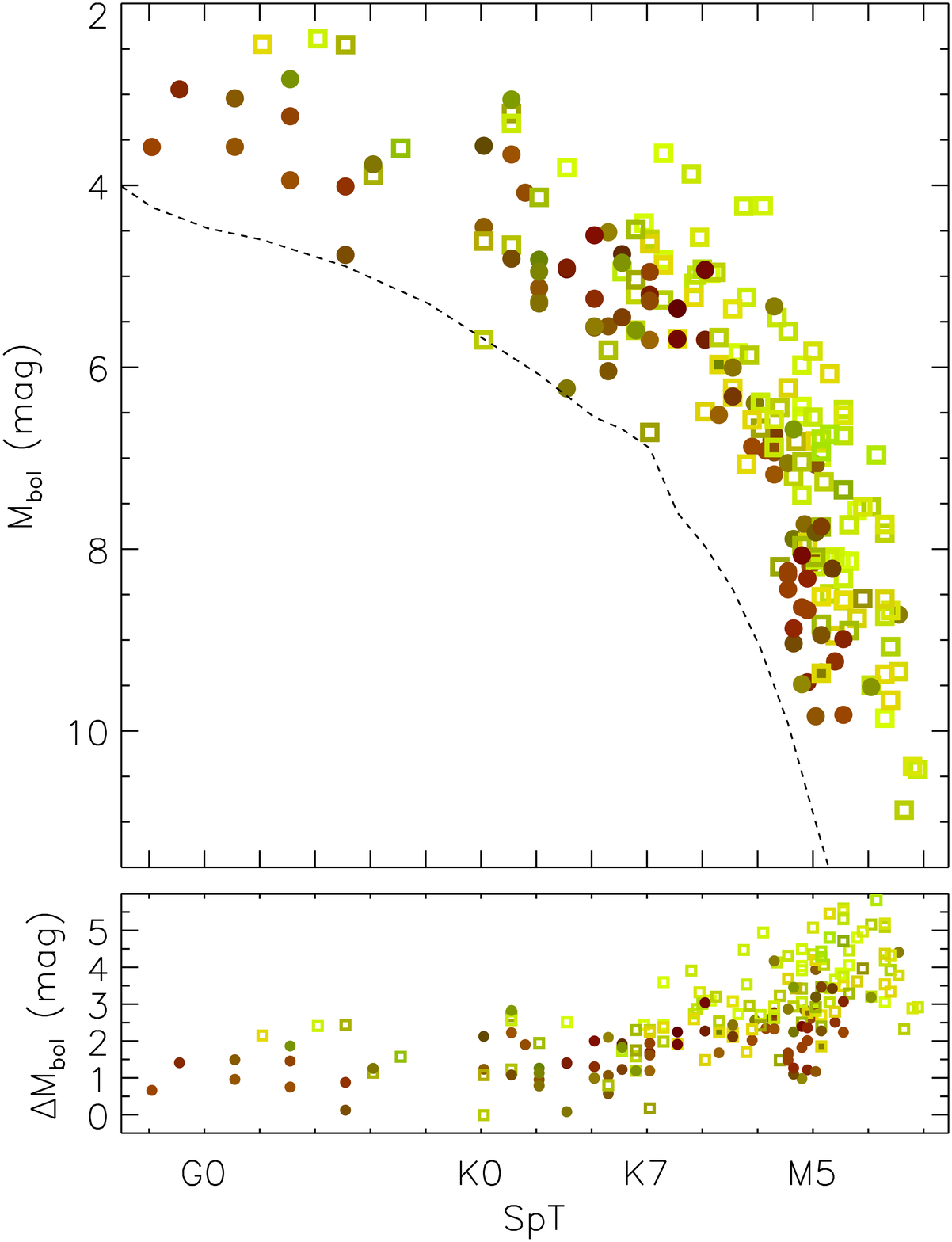}
\figsetgrpnote{HR diagram for the disk-free Taurus members of our sample (upper panel) and the height above the main sequence of each member (lower panel). Each point is color coded to match the local stellar density and disk fraction shown in the color-matched version of Figure~\ref{fig:rgbmap}, such that the disk-rich clustered population is bright yellow and the disk-free distributed population is dark red. To emphasize the difference between the clustered and distributed populations, objects with a local stellar density of $\Sigma > 1.5$ stars/deg$^2$ are shown with open squares, while objects with $\Sigma \le 1.5$ stars/deg$^2$ are shown with filled circles. The distributed population clearly sits below the clustered population in the HR diagram, indicating either that those objects are either older (by a factor of 2--3) or more distant (by $\sim$80 pc). As in Figure~\ref{fig:hrd}, we also show the field main sequence ($\tau \sim 600$ Myr; $[Fe/H] \sim 0$) as defined in Kraus \& Hillenbrand (2007).}
\figsetgrpend

\figsetend

\figsetstart
\figsetnum{13}
\figsettitle{HR Diagrams with Gaia}

\begin{figure*}
\epsscale{1.0}
\includegraphics[scale=0.55,trim={0 0 0 0},clip]{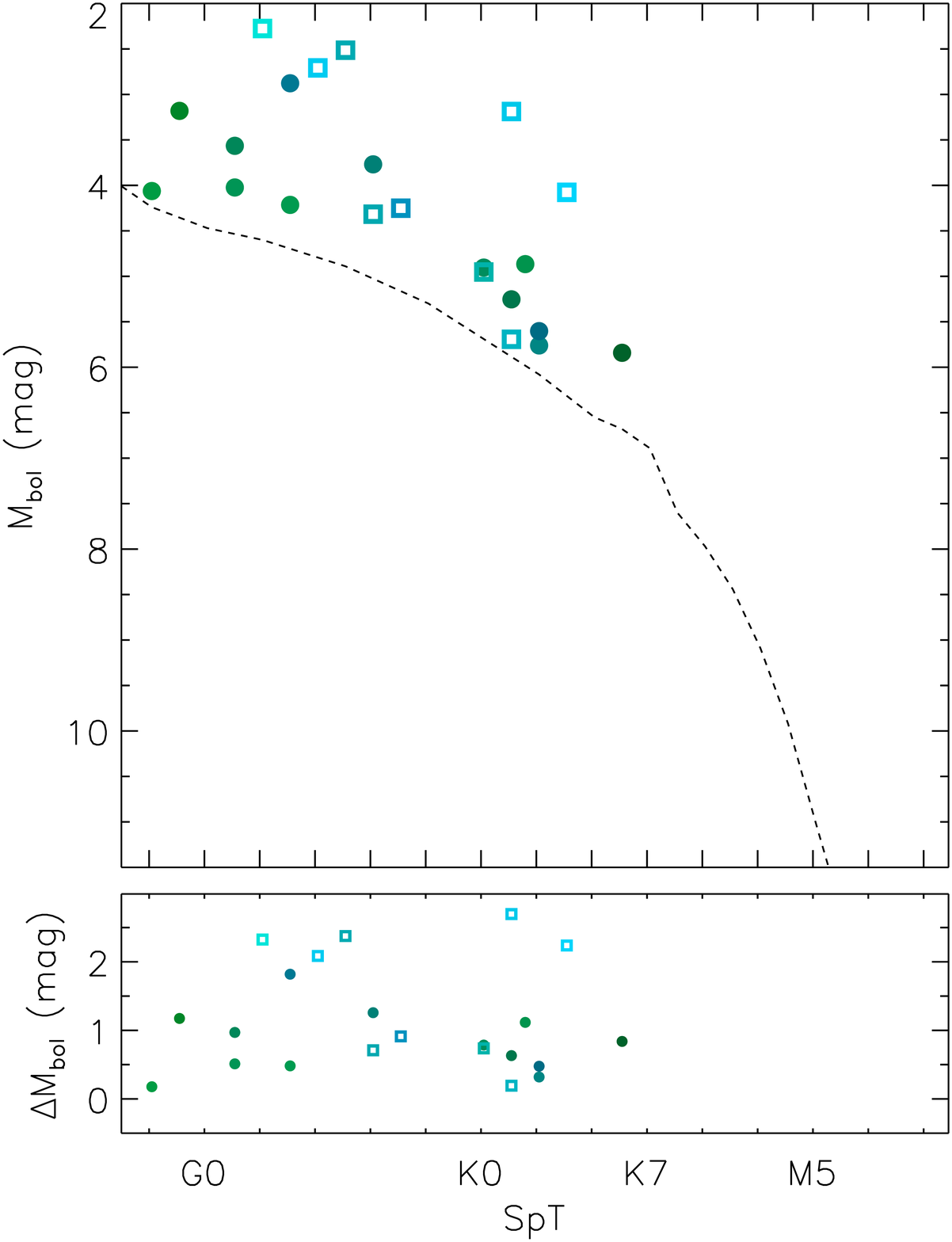}
\figcaption{\label{fig:HRDmemdiskGAIA} HR diagram and residuals as in Figure~\ref{fig:HRDmemdisk}, but with luminosities estimated from Gaia distances (Gaia Collaboration 2016) for the 22 sample members with parallaxes in the Tycho-Gaia Astrometric Solution, rather than using our assumed mean distance of 145 pc. The distributed population sits unambiguously above the main sequence, confirming its pre-main sequence nature. However, most of the distributed population is located at a moderately closer distance ($d \sim $110--130 pc) while the objects in the clustered population are indeed at the distance of Taurus, accentuating the difference in ages.}
\end{figure*}

\figsetgrpstart
\figsetgrpnum{13.1}
\figsetgrptitle{Blue-Green Gaia HRD}
\figsetplot{HRDmemdiskGAIAbluegreen.eps}
\figsetgrpnote{HR diagram and residuals as in Figure~\ref{fig:HRDmemdisk}, but with luminosities estimated from Gaia distances (Gaia Collaboration 2016) for the 22 sample members with parallaxes in the Tycho-Gaia Astrometric Solution, rather than using our assumed mean distance of 145 pc. The distributed population sits unambiguously above the main sequence, confirming its pre-main sequence nature. However, most of the distributed population is located at a moderately closer distance ($d \sim $110--130 pc) while the objects in the clustered population are indeed at the distance of Taurus, accentuating the difference in ages.}
\figsetgrpend

\figsetgrpstart
\figsetgrpnum{13.1}
\figsetgrptitle{Red-Blue Gaia HRD}
\figsetplot{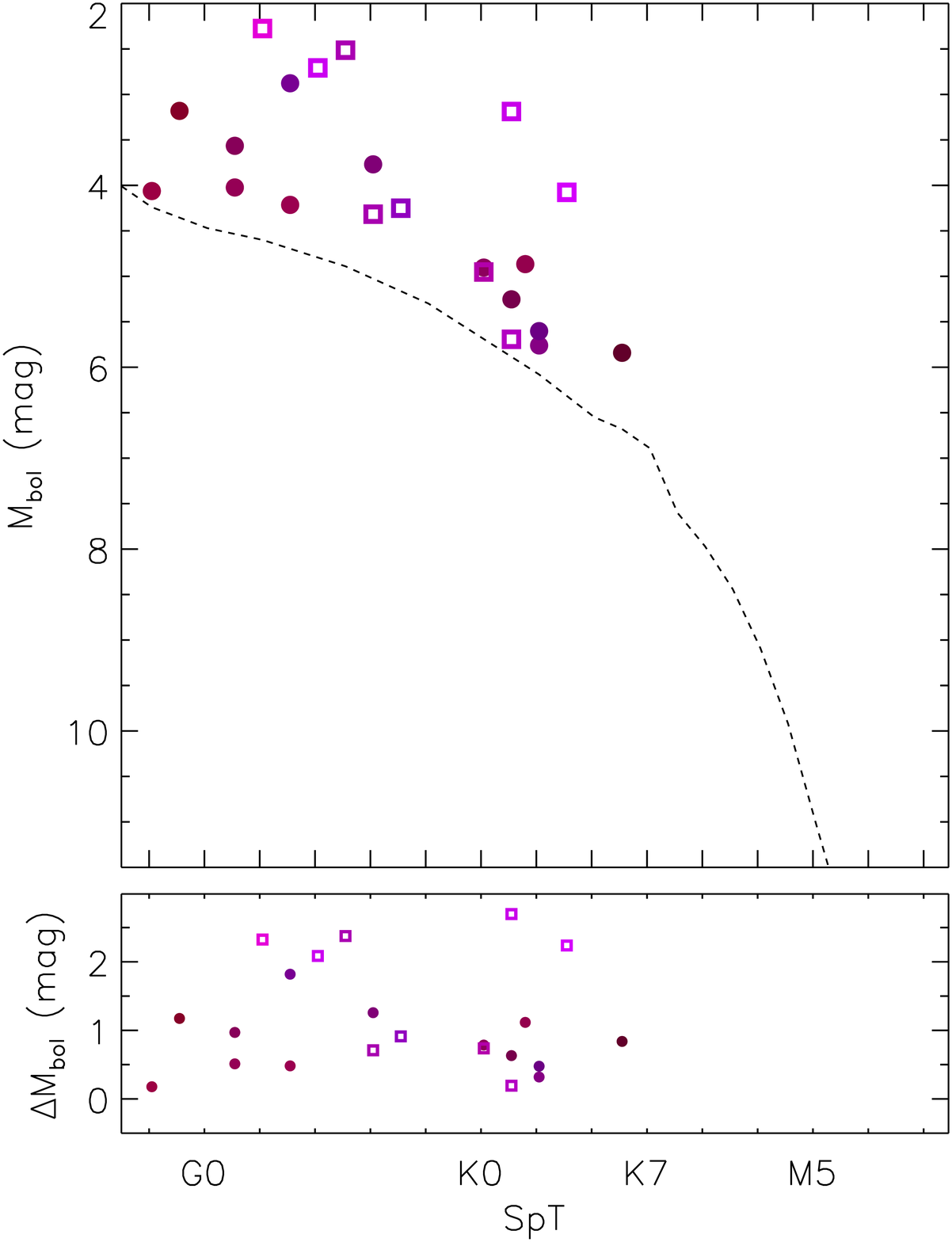}
\figsetgrpnote{HR diagram and residuals as in Figure~\ref{fig:HRDmemdisk}, but with luminosities estimated from Gaia distances (Gaia Collaboration 2016) for the 22 sample members with parallaxes in the Tycho-Gaia Astrometric Solution, rather than using our assumed mean distance of 145 pc. The distributed population sits unambiguously above the main sequence, confirming its pre-main sequence nature. However, most of the distributed population is located at a moderately closer distance ($d \sim $110--130 pc) while the objects in the clustered population are indeed at the distance of Taurus, accentuating the difference in ages.}
\figsetgrpend

\figsetgrpstart
\figsetgrpnum{13.1}
\figsetgrptitle{Red-Green Gaia HRD}
\figsetplot{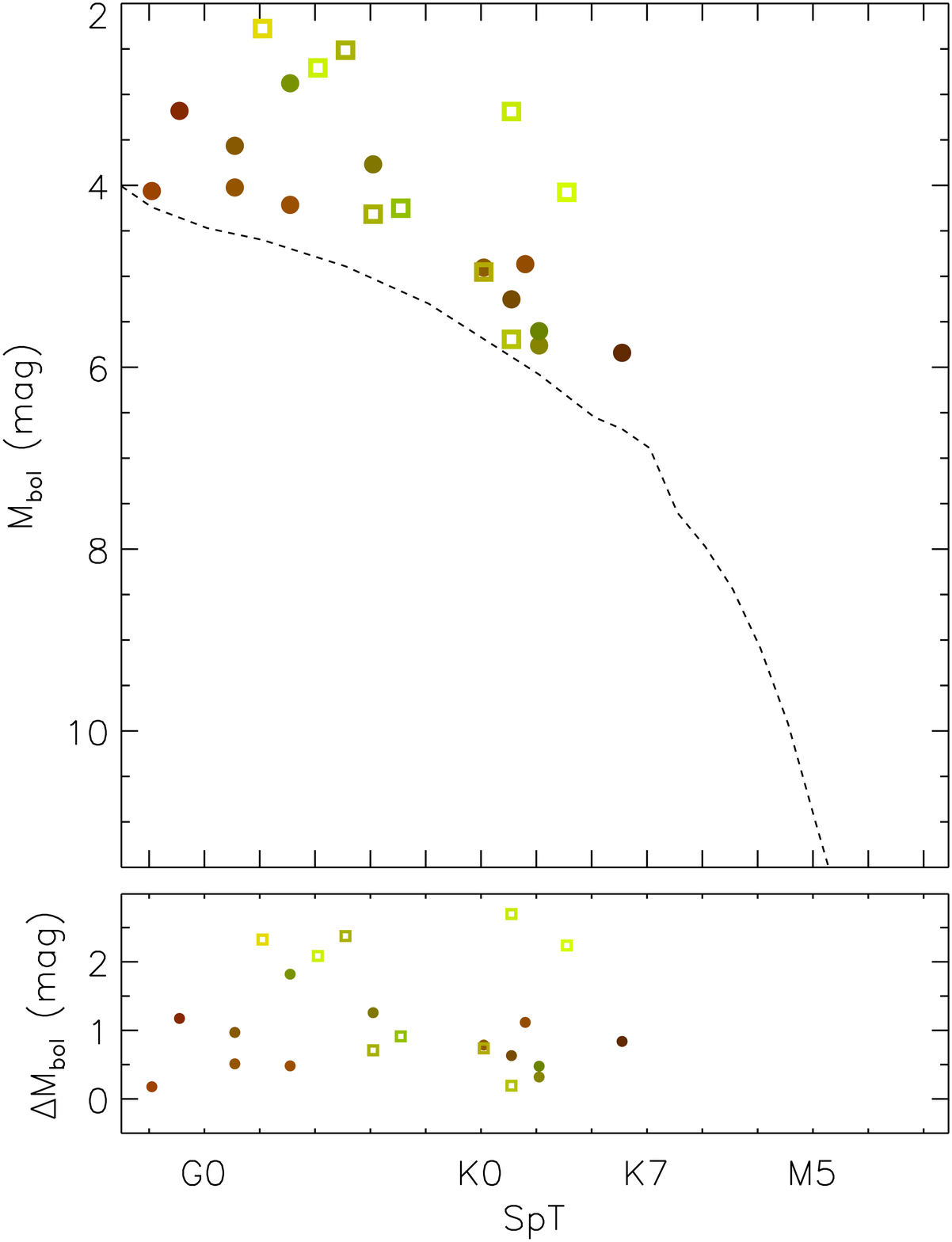}
\figsetgrpnote{HR diagram and residuals as in Figure~\ref{fig:HRDmemdisk}, but with luminosities estimated from Gaia distances (Gaia Collaboration 2016) for the 22 sample members with parallaxes in the Tycho-Gaia Astrometric Solution, rather than using our assumed mean distance of 145 pc. The distributed population sits unambiguously above the main sequence, confirming its pre-main sequence nature. However, most of the distributed population is located at a moderately closer distance ($d \sim $110--130 pc) while the objects in the clustered population are indeed at the distance of Taurus, accentuating the difference in ages.}
\figsetgrpend

\figsetend

\begin{figure*}
\epsscale{1.0}
\includegraphics[scale=0.6,trim={0 0 0 0},clip]{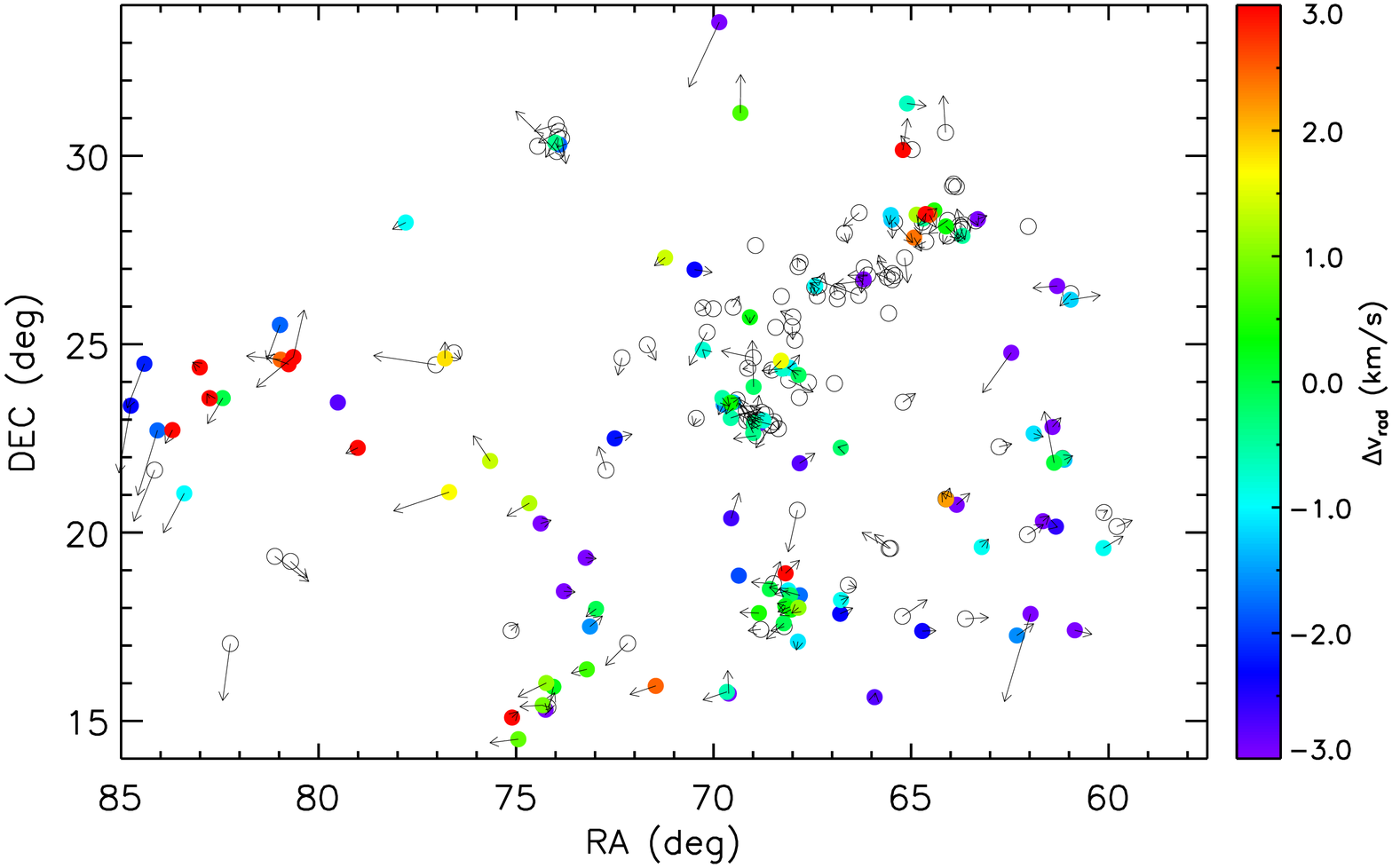}
\figcaption{\label{fig:alloff} RV and proper motion residuals for disk-free Taurus members as a function of sky position. Arrows indicate the magnitude and direction of the residual after subtracting the projection of the mean Taurus $\vec{v_{UVW}}$ on the plane of the sky at the given position; an arrow with length of 1$\degr$ corresponds to a proper motion residual of 15 mas/yr. Point colors indicate the residual in $v_{rad}$ after subtracting the projection of the mean Taurus $\vec{v_{UVW}}$ onto the line of sight at the given position. Some sample members lack a proper motion (denoted by lack of an arrow), a radial velocity (denoted by an open circle), or both (not plotted). A full clustering analysis is very complicated due to incompleteness of the input sample and the availabile data, but two trends do emerge. In the main body of Taurus, the distributed population tends to have a negative velocity residual while objects in the sites of ongoing star formation (by definition) are moving at the mean RV. On the eastern edge of Taurus, there are two populations with RVs that differ by 5 km/s, and the population with a negative residual RV also shows coherent proper motion residuals that could indicate either a large relative $\vec{v_{tan}}$ or a smaller distance.}
\end{figure*}

\begin{figure*}
\epsscale{1.0}
\includegraphics[scale=0.33,trim={0 0 0 0},clip]{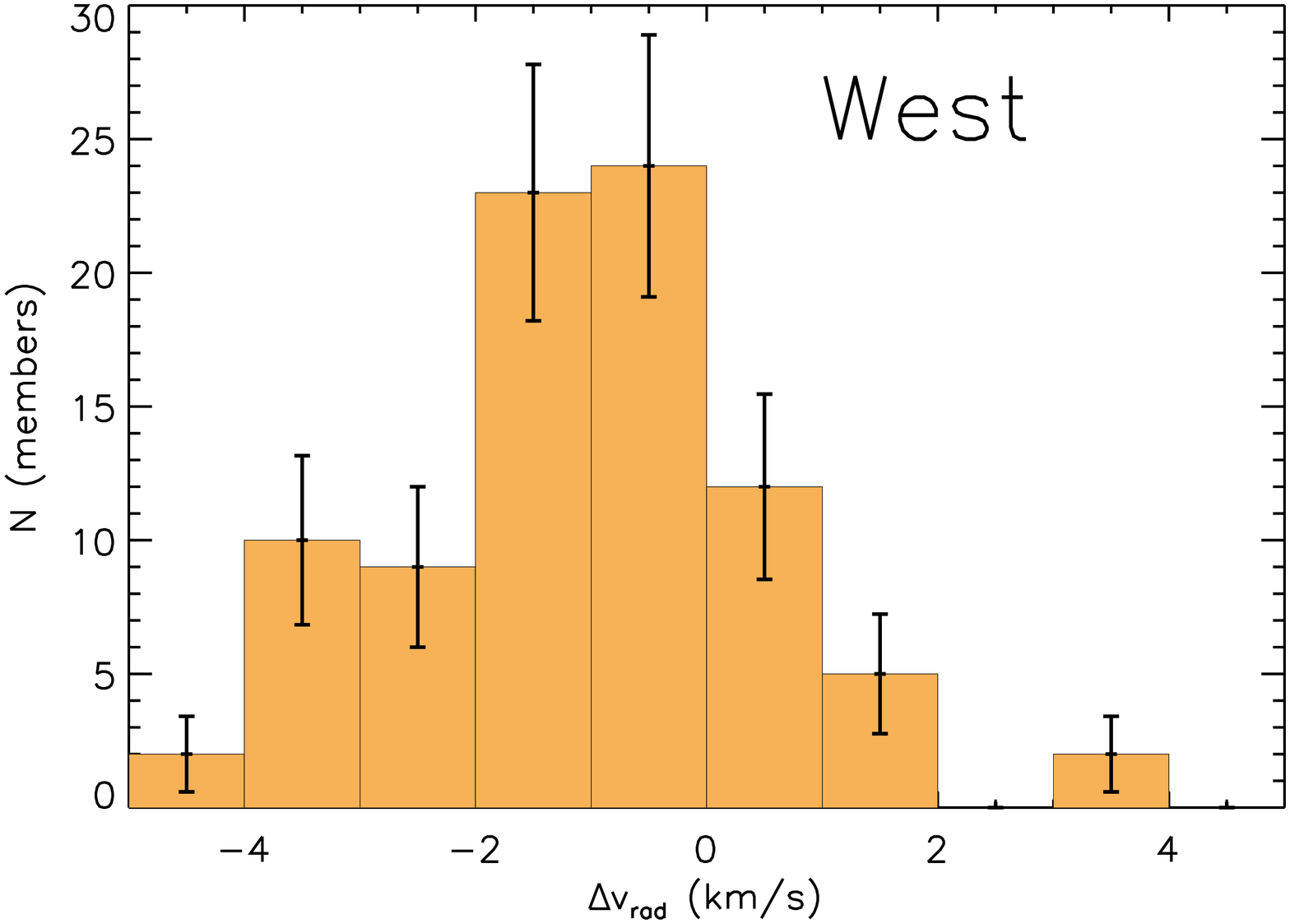}
\hspace{-0.3in}
\includegraphics[scale=0.33,trim={0 0 0 0},clip]{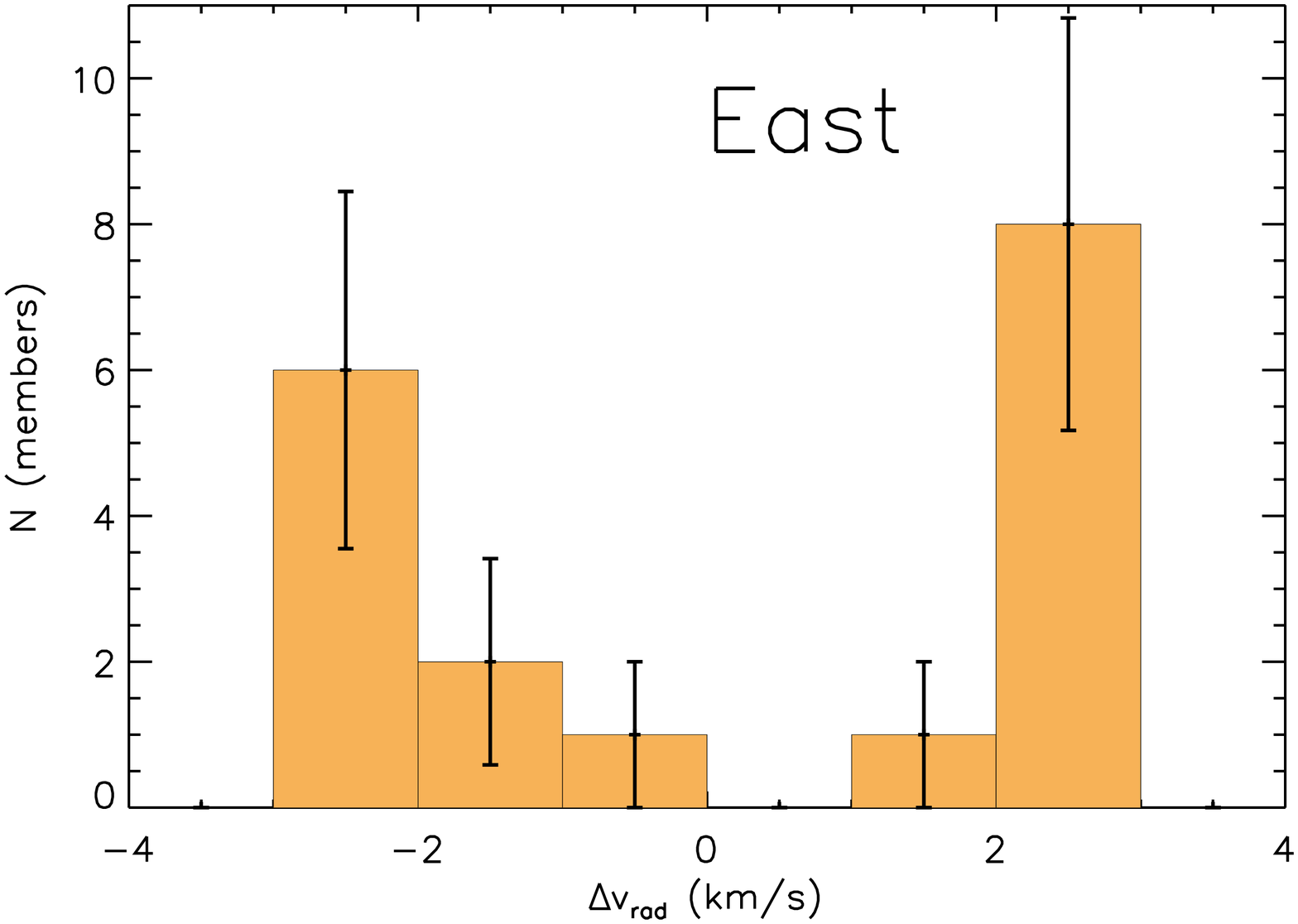}
\figcaption{\label{fig:vraddist} RV residual distributions for objects in the main area of Taurus (Left; $\alpha < 77\degr$) and the eastern edge (Right; $\alpha > 77\degr$). The western distribution is consistent with either a unimodal value offset by 1 km/s from canonical values, or the presence of a substantial negative tail. Based on the spatial distribution of objects with different RVs in Figure~\ref{fig:alloff}, the negative tail is preferred and specifically represents the distributed population. The eastern distribution appears to consist of two populations that differ in radial velocity by 5 km/s; the negative population also has a coherent residual proper motion indicating either a tangential velocity of $\Delta v_{tan} = 12$ km/s or a kinematic distance of $d \sim 100$ pc.}
\end{figure*}

The balance of disk-hosting and disk-free stars in Taurus is a result of several formative and evolutionary processes. Due to conservation of angular momentum, protostars must form with protostellar disks that act as the channel for mass accretion from the envelope to the central protostar. Once the envelope is exhausted, the remaining mass in the disk should evolve on the viscous timescale ($\tau \sim 10^6$--$10^7$ yr) to accrete downward onto the star. Finally, the disk should eventually be dispersed by photoevaporation or outflows, or incorporated into planetary companions. The dispersal timescale is observationally demonstrated to lengthen with decreasing stellar mass (e.g., \citealt{Carpenter:2006hf,Carpenter:2009qe}). There also should be a dependence on the initial disk mass (which stochastically results from the initial specific angular momentum of the envelope). The presence of a close binary companion ($a \la 50$ AU) also leads to rapid disk dispersal in the majority of cases (e.g., \citealt{Cieza:2009fr,Duchene:2010cj,Kraus:2012qe,Cheetham:2015qy}), perhaps via efficient photoevaporation \citep{Alexander:2012la}. The disk-free population therefore might include Taurus members that are preferentially older, more massive, and more highly multiple.

In the following subsections, we analyze and discuss the spatial distribution, ages, mass function, and kinematics for our broadened census of Taurus-Auriga. We find via multiple diagnostics that there is a widely distributed population of comoving young stars surrounding the well-studied molecular clouds. This distributed population has a substantially lower disk fraction, numerous members with depleted lithium, and lower typical position in the HR diagram, suggesting that the members are older than the population clustered around the molecular clouds. The enlarged census does not substantially change the inferred IMF, but instead indicates that previous generations of star formation have not been fully recognized in most extant studies of Taurus. Kinematics demonstrate that all of these young objects might ultimately be part of a much wider star-forming event.

We note here that to avoid potential issues with color perception among readers or gamut differences between print and screen, Figures 9, 10, 12, and 13 are presented as figure sets in the online version of the journal. We also include the alternative figures at the end of this document.

\subsection{The Distributed Population of Disk-Free Stars}

If there is a finite age spread in Taurus, then the disk-free population should preferentially contain the older stars that either have left their formative molecular cloud, or have had the natal cloud dispersed through internal or external feedback from radiations, winds, or supernovae. The natural expectation is therefore that any star-forming region should see a broader spatial dispersion among the disk-free stars than among disk-hosting stars, a trend that is broadly seen in many star-forming regions (e.g., \citealt{Evans:2009fk}). However, the existence and extent of any dispersed population in Taurus has not been addressed in most of the canonical member surveys. As we discussed with respect to our sample construction, numerous candidates have been suggested to fall in this distributed population (\citealt{Wichmann:1996fk,Li:1998fj,Slesnick:2006pi,Gomez-de-Castro:2015kx}). Many of these claims have been subsequently considered and taken as evidence of contamination by interloper field stars \citep{Briceno:1997qy}, but some surveys have included those stars after demonstrating that at least some must be Taurus members \citep{Cuong-Nguyen:2012ul,Daemgen:2015uq}. Our results suggest that approximately half of those stars should indeed be considered as part of the broader Taurus-Auriga ecosystem.

In Figure~\ref{fig:diskmap}, we plot the spatial distributions of disk-free stars from this work and disk-hosting protostars as compiled by \citet{Rebull:2010xf}, \citet{Luhman:2010cr}, and \citet{Esplin:2014lr}; in the background we show the extinction map of \citet{Schlafly:2014lr}. A large fraction of both the disk-hosting and disk-free stars clearly track the high-extinction filaments and clumps of the main Taurus clouds, demonstrating a direct link between those clouds and the recent formation of young stars. There are also clumps or filaments of mixed stars located well away from the high-extinction regions, which could indicate locations where star formation occurred recently but has ended with the dispersal of the local cloud material. However, there also appears to be a distributed population of disk-free stars that are not clearly associated with disk hosts or with cloud material.

In Figure~\ref{fig:rgbmap}, we emphasize the relative distributions of disk-free and disk-hosting stars by smoothing the spatial distributions of Figure~\ref{fig:diskmap} and presenting them as the appropriate channels of an equivalent RGB image. This image therefore conveys both local stellar density (via intensity) and local disk fraction (via hue) in a way that can be interpreted visually. This figure demonstrates a fundamental dichotomy of the Taurus population. The central molecular clouds have a high stellar density and a mixed disk population, and hence are bright and cyan. The outlying areas have a low stellar density and preferentially disk-free population, and hence are faint and green. There are no bright green areas (i.e., dense clusters of disk-free stars) or faint cyan areas (i.e., distributed mixed populations), and the only faint blue areas surround the cyan (and hence indicate potential incompleteness of the disk-free census near the clouds).

In Figure~\ref{fig:densityfrac}, we quantify this relation at the location of each Taurus member by plotting the total surface density (the sum of both channels) and the disk fraction (the ratio of the red channel to the sum of both channels), both of which we extract out of Figure~\ref{fig:rgbmap}. We also show the binned average disk fraction across ranges of surface density. At high density, the disk fraction clearly converges to 60\%, which is consistent with past studies of Taurus (e.g., \citet{Luhman:2010cr}). However, the disk fraction falls at surface density of $\la$1 star deg$^{-2}$. At $0.3 < \Sigma < 1.0$ stars deg$^{-2}$, only 25\% of Taurus members have disks, and all of the most isolated Taurus members are disk free. The lower disk fraction implies that the distributed Taurus members are older, on average, than the Taurus members near the sites of current star formation.

An alternate way to display the different clustering properties is via the two-point correlation function or TPCF (as embodied in the mean surface density of neighbors), which has previously been used to explore clustering in young populations (e.g., \citealt{Gomez:1993zi,Simon:1997er,Kraus:2008fr}). In Figure~\ref{fig:tpcf}, we show the TPCFs for disk-hosting stars around other disk-hosting stars and for disk-free stars around other disk-free stars. Both distributions can be characterized with parallel power law distributions across the range of scales from 0.04\degr\, to 4\degr, demonstrating that clustering does occur. However, the surface density of disk hosts is significantly higher than for disk-free stars on all scales $\theta \la 6\degr$. In contrast, the relations converge and cross at the largest scale ($\theta \sim 10\degr$). This behavior demonstrates that disk-host stars are more tightly clustered on small scales, whereas disk-free stars are less likely to be clustered on small scales and are more likely to be smoothly distributed across the entire field. 

The TPCFs are parallel across two orders of magnitude ($\theta = $0.03--3\degr), which indicates that many of the disk-free stars are clustered in the same way as disk-host stars and have only dispersed $\la 0.03\degr$ or $\la 0.1$ pc. The slope of the power law ($\alpha = -1$) is the value expected for filamentary substructure across this range of scales \citep{Gomez:1993zi,Simon:1997er,Kraus:2008fr}. However, the disk-free TPCF is displaced a factor of 2 lower than the disk-host TPCF, despite nearly equal numbers of stars contributing to each relation. This offset suggests that approximately half of the disk-free stars belong to the clustered population that traces the disk-host stars, while the other half are in the distributed population that contributes a negligible amount of power to the TPCF at smaller scales.

Detailed spatial and kinematic information and a dynamical traceback will be needed to conclusively demonstrate the origin of the distributed population of disk-free stars, as they could have either dispersed from the current sites of ongoing star formation or formed in situ from molecular cloud material that has since evaporated. Sparse star-forming populations have small-scale velocity dispersions as low as 200 m/s \citep{Kraus:2008fr}, but the RV distribution in Figure~\ref{fig:rv} suggests that large-scale features could have relative velocities as large as $v_{tan} = $1--2 km/s or $\mu \sim 1.5$ mas/yr. At this velocity, subpopulations within Taurus could travel 30 pc or 12\degr\, within 20 Myr after stellar birth, while also internally dispersing by 4 pc or 1.5\degr . Distinguishing dispersion from in-situ formation is therefore beyond the limits of currently available data, but will be feasible with early data releases from Gaia.

\subsection{Lithium and the Age(s) of the Distributed Population}

The extant measurements can not distinguish whether the distributed population formed in situ or by dispersal. However, both explanations still require the distributed population to be older than the concentrated disk-hosting population, indicating that Taurus is a long-lived structure that has produced multiple generations of comoving, cospatial star formation, in analogy to the Sco-Cen OB association (Pecaut \& Mamajek 2016). The measurement of stellar ages is fraught with uncertainty, due to a combination of measurement errors, unrecognized systematic uncertainties like multiplicity and excesses, and possibly even fundamental scatter in the properties of stars as a result of different assembly histories, rotation rates, or magnetic field strengths (e.g., \citealt{Hartmann:1997rf,Baraffe:2010lr,Somers:2015yq,Feiden:2016fk}. However, relative ages could potentially be inferred from the disk fraction itself (e.g., \citealt{Haisch:2001om,Hernandez:2008ai,Hillenbrand:2009uz}), and some Taurus members are old enough for the lithium depletion boundary to have appeared.

As we demonstrate in Figure~\ref{fig:densityfrac}, the typical disk fraction at low stellar densities ($0.3 < \Sigma < 1.0$ stars/deg$^2$) is only $F \sim 25\%$, similar to the mass-averaged disk fraction of the Upper Scorpius star-forming region \citep{Carpenter:2006hf,Luhman:2012wd} at an age of $\tau \sim 11$ Myr \citep{Pecaut:2012dp}. We therefore conclude that the typical age for the distributed population is $\sim$10 Myr, or $\sim$5 times larger than the median age of the canonical population \citep{Kraus:2009fk}. However, even the disk hosts in low-density environments appear to be clustered in Figures~\ref{fig:diskmap} and \ref{fig:rgbmap}, suggesting that there are spatially dependent age variations in the distributed population. The census of distributed disk-free stars is almost certainly incomplete (i.e., almost totally lacking early-M dwarfs that represent the peak of the IMF), so the true disk fraction could be overestimated and hence the age could be underestimated.

Lithium depletion provides a more robust age indicator, but only if at least some stars are old enough to have depleted their surface lithium ($\tau \ga 15$ Myr; \citealt{Baraffe:2015qy}). We find that six likely Taurus members with early-M spectral types appear to be nearly or fully depleted of lithium: the disk-free stars [LH98] 185 (M0), [LH98] 188 B (M3), RX J0456.7+1521 (M3.5), 2MASS J04110570+2216313 (M3.5), and [SCH2006b] J0502377+2154050 (M4.25), as well as the disk-hosting star StHa 34 (M3). Based on the lithium curves of growth from \citet{Palla:2007kx}, absence of lithium ($EW[Li] < 100$ m\AA) corresponds to a depletion from the interstellar value by at least three orders of magnitude ($A[Li]\le0$, depleted from $A[Li]=3.3$; \citealt{Palla:2007kx}). We find that [LH98] 185 still barely exceeds this threshold ($EW[Li] = 130$ m\AA), suggesting it might be on the cusp of lithium depletion.

If we map these spectral types to $T_{eff}$ using the temperature scale of \citet{Herczeg:2014oq}, then the models of \citet{Baraffe:2015qy} imply lithium depletion ages of $\ge15$ Myr for the M0 star, $\ge20$ Myr for the M3--M3.5 stars, and $\ge$30 Myr for M4.25 stars. However, there is evidence that the temperatures predicted by models for a given mass are systematically too warm by $\sim$200 K \citep{Kraus:2015mz,Rizzuto:2016lr,Feiden:2016fk}. If the temperatures of the models are shifted to cooler values (i.e., the mass at a given spectral type is increased), then the corresponding ages would be $\sim$25 Myr at M0, $\sim$10--15 Myr at M3 and M3.5, and $\sim$20--25 Myr at M4.25. These values are more consistent with empirical measures of lithium depletion, which find that no stars are fully depleted in Upper Sco at $\tau \sim 10$ Myr \citet{Rizzuto:2015fj}, while the M3--M4 stars are depleted in UCL/LCC at 15--20 Myr (Pecaut \& Mamajek 2016), and the M2--M4 stars are depleted in the Beta Pic Moving Group (Messina et al. 2016). Given the substantial uncertainty in $T_{eff}$ for these stars, the ages should be regarded as uncertain; while the M0 and M4.25 stars could be older, the uncertainty in the spectral types makes it equally plausible that they are simply a different temperature. However, no star should deplete lithium within $\tau \la 15$ Myr, suggesting a firm lower limit on their age. 

As was pointed out by \citet{White:2005gz} and \citet{Hartmann:2005ve}, StHa 34 is particularly interesting in having a circumstellar disk (suggesting $\tau \la 20$ Myr) but no lithium (suggesting $\tau \ga 20$ Myr for an M3 star), features that are in tension unless the system falls within a very narrow age range or some disks are exceptionally long-lived. Gas-rich disks do still exist in the TW Hya and Beta Pic moving groups ($\tau \sim $15--25 Myr; Binks \& Jeffries 2016), but are dispersed by the age of Tuc-Hor (Kraus et al. 2014). The space velocity of StHa 34 is $v_{UVW} = (-15.3 \pm 0.6, -8.6 \pm 2.3, -9.7 \pm 2.1)$ km/s, assuming $v_{rad} = 17.9 \pm 0.6$ km/s (White \& Hillenbrand 2007), $\mu = (+0.9, -13.3) \pm 3.1$ mas/yr (UCAC4; Zacharias et al. 2012), and a distance of $d = 145 \pm 15$ pc \citep{Torres:2009ct}. The system therefore does indeed appear comoving with the mean Taurus velocity to within $1 \sigma$. The existence of StHa34 indicates that comoving, cospatial star formation has been occurring in the Taurus region for $\ga$20 Myr.

While parallactic distances are not available for most of our sample members, it is still informative to consider their HR diagram position if we assume all objects have the mean distance of Taurus. A similar analysis of Sco-Cen by Pecaut \& Mamajek (2016) illustrated a clear signature of spatially dependent stellar ages, demonstrating the long star-formation history of that region. In Figure~\ref{fig:HRDmemdisk}, we plot the HR diagram positions of all the disk-free Taurus members, color-coding each point by the local density of members (intensity) and disk fraction (hue) from the RGB image of Figure~\ref{fig:rgbmap}, and also show the residual height above the main sequence. Points from the high-density, disk-rich (young) regions appear to sit preferentially $\sim$1 magnitude higher in the HR diagram, with the dichotomy most clearly present in the K stars and M4--M6 stars that comprise the majority of the distributed members in our study. This offset indicates either that the distributed population is a factor of $\sim$2--3 older, or that it sits $\sim$80 pc behind the clustered population while also possessing a substantially different space velocity that coincidentally results in the same bulk proper motion. 

Finally, many of the early-type stars in our sample were included in the recent release of the Tycho-Gaia Astrometric Solution (Gaia Collaboration 2016), which includes parallaxes with typical precisions of $\sigma_{\pi} = $0.2--0.3 mas. With distance measurements, we can compute refined luminosities and consider the stars' HR diagram positions more robustly. In Figure~\ref{fig:HRDmemdiskGAIA}, we plot the HR diagram for 22 TGAS stars that we assess to be likely or confirmed Taurus members, demonstrating that the stars do indeed sit above the field main sequence (and hence are young pre-main sequence stars). However, we find that the distributed population is located at a moderately closer distance ($d \sim $110--130 pc) while the objects in the clustered population are indeed at the distance of Taurus, accentuating the difference in ages.

\subsection{Implications for the Initial Mass Function}

Taurus has long been suggested to have an unusual IMF \citep{Luhman:2000dk,Luhman:2004bs,Luhman:2009wd,Luhman:2017lr}, hosting an excess of stars with $M = 0.7-1.0 M_{\odot}$ and a deficit of stars with $M > 1 M_{\odot}$. Stars with $M > 1 M_{\odot}$ have spectral types earlier than M, and hence many of the traditional diagnostics of youth (such as lithium and low surface gravity) are not useful in assessing candidate young stars. The only meaningful indicators of potential Taurus membership are high X-ray or UV emission (which are elevated for young FGK stars, and persist for $> 10^8$ yr; \citealt{Shkolnik:2009lr}) or comovement with the association's proper motion and radial velocity. The missing stars therefore could have been identified as candidates by the ROSAT surveys of \citet{Wichmann:1996fk} and \citet{Li:1998fj}, without being confirmed at sufficient confidence to join the established census of Taurus that was promulgated by subsequent works. This trend could have been further exacerbated by the mass-dependent lifetimes of protoplanetary disks (e.g., \citealt{Carpenter:2006hf}). High-mass stars have preferentially shorter disk lifetimes, so in any population with some members older than $\tau \ga 1$--2 Myr, the disk-free population (which is harder to identify as young) would contain a preferentially higher number of high-mass stars. To test this possibility, we cross-referenced our updated list of likely disk-free Taurus members with the fields for which \citet{Luhman:2004bs} and \citet{Luhman:2009wd} compiled their unusual IMFs. 

The fields studied by \citet{Luhman:2004bs}, which summarized the efforts of several earlier surveys, used optical color-magnitude diagrams to select candidates that were subsequently observed with optical or near-infrared spectroscopy to confirm signatures of youth. The subsequent releases of Spitzer mid-infrared photometry and XMM/Newton X-ray fluxes revealed many additional low-mass members (e.g., \citealt{Guieu:2006tn,Luhman:2006tg,Slesnick:2006xr}), and hence the early IMFs clearly require revision. However, all of these additional members are of spectral type M (with $M < 0.7 M_{\odot}$), and we can not add any additional members that were not already included in the census by \citet{Esplin:2014lr}. We therefore conclude that the absence of higher-mass members ($M > 1 M_{\odot}$) has only been exacerbated, and none of the stars from \citet{Wichmann:1996fk} and \citet{Li:1998fj} can remedy the shortfall. The more modern census of \citet{Luhman:2009wd} considered all available member searches for the fields studied by the XEST survey \citep{Gudel:2007uq,Scelsi:2007sd}, and again found too many solar-type members and not enough high-mass members. No members with $M > 1 M_{\odot}$ have been found in those fields, and we only add a single new member to that sample (RXJ0422.1+1934; SpT = M3.5), and hence we again find that the measured IMF remains discrepant after our updated census. While this paper was under review, \citet{Luhman:2017lr} published the discovery of additional late-type members of Taurus with spectral types of mid-M or later. They found that the addition of these low-mass objects lessened the deficit of brown dwarfs in comparison to the proposed excess of 0.7--1.0 $M_{\odot}$ stars. However, they did not add any high-mass members, and hence that discrepancy remains.

Our study only considers objects that have previously been suggested as Taurus members, so it remains plausible that the measured Taurus IMF is biased by an incomplete census. It would be premature to conclude that the true Taurus IMF is different until a comprehensive study of F--K stars has been conducted. Furthermore, \citet{Mooley:2013fj} have identified a number of B--A stars in Taurus that might significantly modify the inferred IMF, though their membership has been disputed (e.g., Esplin et al. 2014), and \citet{Herczeg:2014oq} have found that some Taurus members were significantly misclassified when originally observed in the 1960s and 1970s. We therefore suggest that the shape of the Taurus IMF is not yet settled.

\subsection{Kinematic Sub- and Super-Structure}

Stellar populations are typically assumed to have a common motion through space. However, this assumption begins to break down on large scales or with sufficiently precise velocity information. Taurus-Auriga appears to be much younger than a crossing time on all scales $\ga0.1\degr$ \citep{Kraus:2008fr}, so it is not yet (and likely never will be) virialized. Even the primordial velocities are not uniform though, and instead reflect the turbulent power spectrum of the natal molecular cloud. The empirical determinations in Larson's Laws \citep{Larson:1981qe} find that the velocity dispersion scales with the square root of physical size; more widely separated regions have statistically larger velocity differences. Based on past determinations of the $UVW$ space velocity, the dispersion between the core subgroups of Taurus (separated by $\la 5 \degr$) is $\sim$1--2 km/s (e.g., Luhman et al. 2009). However, there may be evidence of a larger gradient for individual stars with precise distances and kinematics from VLBI (e.g., Torres et al. 2009). We should therefore expect velocity differences on $\sim 20 \degr$ scales of at least $\sim$4--5 km/s. Moreover, the observed velocities should vary as projection angle changes, and the observed proper motion should change in amplitude (but not direction) depending on the distance of a given star.

In Figure~\ref{fig:alloff}, we show five dimensions of the kinematics (two spatial dimensions and three velocity dimensions) for our sample of disk-free likely Taurus members. To encompass this data in the two-dimensional figure, we plot the positions of all objects on the sky, using filled circles that are color-coded to the RV residual about the mean Taurus $UVW$ and arrows showing the magnitude and direction of the proper motion residuals about the mean Taurus $UVW$. Not all sample members have both proper motions and RVs, so in some cases we show only open circles. This figure therefore visually conveys the correlations between sky position, $\Delta v_{rad}$, and $\Delta \vec{\mu}$, based on coherence of colors and velocity vectors.

The ideal analysis for such a dataset would use clustering algorithms to identify substructures within the sample. However, the sample is likely to be spatially incomplete (especially on the outskirts of Taurus) and not all dimensions are populated for all objects, so a clustering analysis seems unlikely to yield significant new insights until Gaia provides a uniform set of distances and precise proper motions. However, some trends do appear to emerge for carefully selected subsets. We specifically consider a main-group subsample ($\alpha < 77 \degr$, encompassing most objects) and an eastern subsample ($\alpha > 77 \degr$, encompassing objects east of the traditional boundary of Taurus).

In Figure~\ref{fig:vraddist} (left), we show the distribution of radial velocities for all sample members in the main-group subsample. The distribution clearly conveys the result seen visually in Figure~\ref{fig:alloff}: there is an excess of objects with negative RV residuals, denoting motion toward the Sun. This excess can be quantified using either the skewness of the distribution ($skew[\Delta v_{rad}] = -72.2$) or by the relative weights of the tails (where there are 21 sample members with $\Delta v_{rad} < -2$ km/s and 2 sample members with $\Delta v_{rad} > +2$ km/s. The symmetry of the distribution would be restored if the mean radial velocity were instead changed by 1 km/s, such that the distribution was again centered on zero. However, the spatial distinction between objects argues against this interpretation; disk-free objects located near the ongoing sites of star formation indeed have velocities near the canonical value, and it is only the distributed population that shows a strong excess of negative residual velocities. We therefore conclude that the distributed population of disk-free stars is indeed non-comoving in the radial direction by $\sim$2--3 km/s. There is not yet significant evidence of an offset in $v_{tan}$; among sample members with $\Delta v_{rad} < -2.0$ km/s, the average proper motion residuals are ($\Delta \mu_{\alpha}$ , $\Delta \mu_{\delta}$) = ( -2.1$\pm$1.1 , +1.6$\pm$1.3 ) mas/yr. As we discuss in Section 7.2 and Figure~\ref{fig:HRDmemdiskGAIA}, the distributed population also is located on the near side of the molecular clouds, indicating a potential spatial offset. However, a more complete census is needed to confirm that the distance difference does not result from incompleteness for more distant objects that might have fallen below the ROSAT detection limits.

In Figure~\ref{fig:vraddist} (right), we show the distribution of radial velocities for all sample members in the eastern subsample. Given that the radial velocity uncertainties are typically $\la 1$ km/s (e.g., Table~\ref{tab:hires}), the presence of two peaks separated by $\sim$5 km/s is suggestive of substructure as well. There plausibly could be an intrinsic velocity dispersion widening a unimodal distribution, though, and the Kolmogorov-Smirnov statistic with respect to a unimodal distribution with mean of 0.8 km/s and standard deviation of 2.7 km/s is inconclusive ($D = 0.26$, $P = 0.14$). However, Figure~\ref{fig:alloff} demonstrates that the objects with negative $\Delta v_{rad}$ also have large and coherent proper motion residuals, while the objects with positive $\Delta v_{rad}$ do not. If we divide the eastern subsample at $\Delta v_{rad} = 0$ km/s, the objects with $\Delta v_{rad} < 0$ km/s have mean proper motion residuals of ($\Delta \mu_{\alpha}$ , $\Delta \mu_{\delta}$) = ( +5.8$\pm$0.5 , -15.7$\pm$3.2 ) mas/yr, while the objects with $\Delta v_{rad} > 0$ km/s have mean proper motion residuals of ($\Delta \mu_{\alpha}$ , $\Delta \mu_{\delta}$) = ( +4.5$\pm$2.2 , +1.8$\pm$3.1 ) mas/yr. These values are discrepant at 4$\sigma$, suggesting that the two sets are indeed distinct. If the objects with $\Delta v_{rad} > 0$ km/s (the yellow and red points) are taken as distinct from the other eastern objects, then they visually trace a possible extension of an RV gradient across the main population, as suggested for the main body of Taurus by Torres et al. (2009). However, the presence of objects with negative RVs at $\alpha \sim 73 \degr$ suggests that this might not be the case. 

Intriguingly, the proper motions for objects with $\Delta v_{rad} < 0$ km/s have the same direction as would be expected for Taurus members, they are simply 50\% larger. We therefore can not distinguish whether those objects are at the same distance and differ by $\Delta v_{tan} = 12$ km/s, whether they are comoving in $v_{tan}$ and are located at a smaller distance ($d \sim 100$ pc), or whether distance and $v_{tan}$ both differ. Intriguingly, these objects are also consistent with the space velocity of 118 Tau ($d \sim 130$ pc; van Leeuwen 2007), which has been suggested to host its own young moving group of stars by \citet{Mamajek2016}\footnote{\citet{Mamajek2016} reports the 118 Tau group to have mean position $\alpha$, $\delta$ = 83$^{\circ}$.1, 24$^{\circ}$.0, diameter $\sim$4$^{\circ}$, proper motion $\mu_{\alpha}$, $\mu_{\delta}$ $\simeq$ +4, -39 ($\pm$1, $\pm$1) mas\,yr$^{-1}$, radial velocity 18\,$\pm$\,2 km\,s$^{-1}$, mean distance 121\,$\pm$\,6 pc, and age $\sim$10 Myr. The group's original membership consisted of the 118 Tau binary, HD 36546, [SCH2006b] J0539009+2322081, [SCH2006b] J0537385+2428518, [LH98] 184, [LH98] 188, [LH98] 201, [LH98] 204, [LH98] 211, [LH98] 213, and [LH98] 219.  An updated distance can be calculated by combining the revised Hipparcos parallaxes for 118 Tau and HD 36546 with the Gaia DR1 TGAS parallaxes for [LH98] 204, 213, and 219. The median parallax of 8.78 mas is consistent with distance $\sim$114 pc.}. The 118 Tau group therefore might include these low-mass objects and be kinematically related to Taurus.

Finally, the eastern objects with $\Delta v_{rad} < 0$ and kinematics similar to 118 Tau have proper motions and radial velocities that are intriguingly intermediate between those of Taurus and those of the proposed young group of stars associated with 32 Orionis (Mamajek 2007; Mamajek 2016; Burgasser et al. 2016), which is located immediately to the south ($\alpha = 77\degr$, $\delta = +6\degr$). The 32 Orionis group is located at $d = 90$ pc, with a proper motion of $\mu = (+8,-33)$ mas/yr and a radial velocity of $v_{rad} = +18$ km/s. Indeed, the corresponding space velocity for 32 Ori ($UVW = (-12, -19, -9)$ km/s) only differs from that of Taurus by (-4,-8,-1) km/s, a difference that is fully compatible with Larson's Law and the large distance between them. We therefore suggest that the ongoing star-formation event that is today producing Taurus members might have previously stretched beyond even the boundary of our current search, encompassing a volume of space and length of time rivaling that of Orion (Bally 2008) or Sco-Cen (e.g., Pecaut \& Mamajek 2016).

\section{Summary}

We have systematically reconsidered the status of 396 objects that were previously suggested to be candidate members of the Taurus-Auriga star-forming complex. We find that 218 of these objects are likely Taurus members, of which $\sim$1/3 have not been included in the canonical census used for most Taurus studies and are prime targets for future detailed study. Most of these additional members are not located near the sites of ongoing star formation, but instead are distributed more uniformly around and between those sites; our updated census therefore does not change previous suggestions of an unusual IMF in Taurus. Intriguingly, an analysis of spatial distributions for disk-hosting and disk-free stars demonstrates that young stars are found in either high-density areas with a higher disk fraction, or low-density areas with a low disk fraction. Based on the disk fraction and a handful of lithium-depleted members, our results show that Taurus is host to a distributed older population ($\tau \sim 10$--20 Myr) that formed in previous episodes of comoving, cospatial star formation.

\acknowledgements

We thank T. Dupuy, B. Bowler, S. Andrews, and D. Jaffe for helpful discussions on the nature of Taurus-Auriga and how to best present its complexity. We also thank the referee for providing a helpful critique of the work. A.W.M. was supported through Hubble Fellowship grant 51364 awarded by STScI, which is operated by AURA for NASA, under contract NAS 5-26555.

This research has made use of the Keck Observatory Archive (KOA), which is operated by the W. M. Keck Observatory and the NASA Exoplanet Science Institute (NExScI), under contract with the National Aeronautics and Space Administration. Some of the data presented herein were obtained at the W.M. Keck Observatory, which is operated as a scientific partnership among the California Institute of Technology, the University of California and the National Aeronautics and Space Administration. The Observatory was made possible by the generous financial support of the W.M. Keck Foundation. 

The authors wish to recognize and acknowledge the very significant cultural role and reverence that the summit of Mauna Kea has always had within the indigenous Hawaiian community.  We are most fortunate to have the opportunity to conduct observations from this mountain.

\bibliographystyle{aasjournal.bst}
\bibliography{ms.bbl}

\clearpage

\startlongtable
\begin{deluxetable}{llllrlrlrrll}
\tabletypesize{\tiny}
 \tablewidth{0pt}
 \tablecaption{Candidate Class III Members of Taurus-Auriga \label{tab:sample}}
 \tablehead{
 \colhead{2MASS} & \colhead{Other Name} & \colhead{SpT} & \colhead{Ref} & \colhead{$A_V$} & \colhead{Ref} & \colhead{$r'$} & \colhead{Ref} & \colhead{$K_s$} & \colhead{$\mu$} & \colhead{Ref} & \colhead{Binary}
\\ 
\colhead{} & \colhead{} & \colhead{}& \colhead{}  & \colhead{(mag)} & \colhead{} & \colhead{(mag)} & \colhead{} & \colhead{(mag)} & \colhead{(mas/yr)} & \colhead{} & \colhead{Ref} }
 \startdata
J03500343+2235300 & [LH98] 106 & G0 & 26 & 2.4 & 11 & 8.77 & 8 & 6.04 & (+156.9,-60.6)$\pm$1.0 & 1 & \\
J03502840+1631151 & [LH98] 108 & G5 & 26 & 0.5 & 11 & 10.38 & 7,8 & 8.63 & (+27.3,-26.9)$\pm$0.6 & 1 & \\
J03505432+2350056 & [LH98] 110 & G7 & 26 & 0.8 & 11 & 11.08 & 7,8 & 9.05 & (+19.3,-45.4)$\pm$0.7 & 1 & \\
J03520223+2439479 & HBC351 & K5 & 14 & 0.0 & 14 & 11.63 & 7,8 & 9.07 & (+17.4,-47.4)$\pm$1.7 & 1 & 37\\
J03533134+2631411 & [LH98] 117 & G7 & 26 & 1.4 & 11 & 11.62 & 7,8 & 9.24 & (+21.4,-45.3)$\pm$1.3 & 1 & \\
J03542523+2421363 & [LH98] 119 & G5 & 26 & 0.4 & 11 & 10.79 & 7,8 & 9.09 & (+17.2,-47.3)$\pm$1.0 & 1 & \\
J03542950+3203013 & HBC352 & G0 & 14 & 0.9 & 14 & 11.61 & 7 & 9.58 & (+2.4,-8.4)$\pm$2.5 & 1 & 38\\
J03543017+3203043 & HBC353 & G5 & 14 & 1.0 & 14 & 12.09 & 7 & 9.86 & (+9.4,-10.2)$\pm$1.7 & 1 & \\
\enddata
\tablecomments{All $K_s$ magnitudes are from 2MASS \citep{Skrutskie:2006nb}. Other references:
1) \citet{Zacharias:2012lr}
; 2) \citet{Kraus:2007mz}
; 3) \citet{Zacharias:2015zr}
; 4) \citet{Bouvier:2008mz}
; 5) \citet{Harris:1999ly}
; 6) \citet{Ahn:2012ys}
; 7) \citet{Evans:2002sp}
; 8) \citet{Henden:2012dq}
; 9) \citet{Slesnick:2006xr}
; 10) \citet{Zacharias:2004zi}
; 11) This Work (Photometric); 
; 12) This Work (Spectroscopic); 
; 13) \citet{Herczeg:2014oq}
; 14) \citet{Kenyon:1995dg}
; 15) \citet{Briceno:1999nm}
; 16) \citet{Rebull:2010xf}
; 17) \citet{Luhman:2009wd}
; 18) \citet{Luhman:2010cr}
; 19) \citet{Scelsi:2008my}
; 20) \citet{Strom:1994dq}
; 21) \citet{Briceno:2002jo}
; 22) \citet{Esplin:2014lr}
; 23) \citet{Findeisen:2010mj}
; 24) \citet{Gizis:1999yq}
; 25) \citet{Guieu:2006tn}
; 26) \citet{Li:1998fj}
; 27) \citet{Luhman:2004bs}
; 28) \citet{Luhman:2006dp}
; 29) \citet{Luhman:1998qa}
; 30) \citet{Luhman:2003fu}
; 31) \citet{Luhman:2009wd}
; 32) \citet{Martin:2001rp}
; 33) \citet{Reid:1999nq}
; 34) \citet{Slesnick:2006xr}
; 35) \citet{Walter:1988ae}
; 36) \citet{Wichmann:1996fk}
; 37) \citet{Leinert:1993gd}
; 38) \citet{Daemgen:2015uq}
; 39) \citet{Kraus:2011tg}
; 40) \citet{Kraus:2006ul}
; 41) \citet{Ghez:1993xh}
; 42) \citet{Todorov:2014fk}
; 43) \citet{Correia:2006pi}
; 44) \citet{Chen:1990an}
; 45) \citet{Kraus:2012bh}
; 46) \citet{Simon:1995yi}
; 47) \citet{Mundt:1983lr}
; 48) R. White, priv. comm.
; 49) \citet{Sartoretti:1998bd}
; 50) \citet{Simon:1992qy}
; 51) \citet{Konopacky:2007pd}
; 52) \citet{Todorov:2010nl}
; 53) \citet{Mathieu:1989lr}
; 54) \citet{Cuong-Nguyen:2012ul}
.}
 \end{deluxetable}
 
\clearpage

\startlongtable
\begin{deluxetable*}{lllrl}
 \tabletypesize{\tiny}
 \tablewidth{0pt}
 \tablecaption{Literature H$\alpha$ Measurements for Candidate Class III Taurus Members \label{tab:litha}}
 \tablehead{
 \colhead{2MASS} & \colhead{Other Name} & \colhead{SpT} & \colhead{$EW[H\alpha]$} & \colhead{Ref}
\\ 
\colhead{} & \colhead{} & \colhead{} & \colhead{(\AA)} 
 }
 \startdata
J03502840+1631151 & [LH98] 108 & G5 & -1.3 & 29\\
J03505432+2350056 & [LH98] 110 & G7 & -1.4 & 14\\
J03520223+2439479 & HBC351 & K5 & -2.8 & 13\\
J03533134+2631411 & [LH98] 117 & G7 & -3.0 & 14\\
J03542523+2421363 & [LH98] 119 & G5 & -1.6 & 29\\
J03542950+3203013 & HBC352 & G0 & 2.2 & 13\\
J03543017+3203043 & HBC353 & G5 & 2.1 & 13\\
J03543556+2537111 & HBC354 & K3 & 1.2 & 13\\
\enddata
\tablecomments{References:
1) \citealt{Nguyen:2009vj}
; 2) \citet{Briceno:1993hc}
; 3) \citet{Cohen:1979qm}
; 4) \citet{Duchene:1999ux}
; 5) \citet{Fernandez:1998gf}
; 6) \citet{Findeisen:2010mj}
; 7) \citet{Gizis:1999yq}
; 8) \citet{Gomez:1992dt}
; 9) \citet{Guieu:2006tn}
; 10) \citet{Hartmann:1991gf}
; 11) \citet{Herczeg:2014oq}
; 12) Herczeg, priv. comm.
; 13) \citet{Kenyon:1998ol}
; 14) \citet{Li:1998fj}
; 15) \citet{Luhman:2003fu}
; 16) \citet{Luhman:2009wd}
; 17) \citet{Martin:1998ve}
; 18) \citet{Martin:1994ul}
; 19) \citet{Mohanty:2005sh}
; 20) \citet{Muzerolle:2003xd}
; 21) \citet{Rebull:2010xf}
; 22) \citet{Reid:1999nq}
; 23) \citet{Scelsi:2008my}
; 24) \citet{Shkolnik:2009lr}
; 25) \citet{Slesnick:2006xr}
; 26) \citet{Walter:1986qf}
; 27) \citet{Walter:1988ae}
; 28) \citet{White:2003xz}
; 29) \citet{White:2007pd}
; 30) \citet{Wichmann:1996fk}
.}
 \end{deluxetable*}
 
 \clearpage

\startlongtable
\begin{deluxetable*}{lllrl}
 \tabletypesize{\tiny}
 \tablewidth{0pt}
 \tablecaption{Literature Lithium Measurements for Candidate Class III Taurus Members \label{tab:litli}}
 \tablehead{
 \colhead{2MASS} & \colhead{Other Name} & \colhead{SpT} & \colhead{$EW[Li]$} & \colhead{Ref} 
\\ 
\colhead{} & \colhead{} & \colhead{} & \colhead{(m\AA)}
 }
 \startdata
 J03502840+1631151 & [LH98] 108 & G5 & 205 & 19\\
J03505432+2350056 & [LH98] 110 & G7 & 220 & 11\\
J03520223+2439479 & HBC351 & K5 & 350 & 17\\
J03533134+2631411 & [LH98] 117 & G7 & 310 & 11\\
J03542523+2421363 & [LH98] 119 & G5 & 196 & 19\\
J03542950+3203013 & HBC352 & G0 & 130 & 13\\
J03543017+3203043 & HBC353 & G5 & 223 & 13\\
J03543556+2537111 & HBC354 & K3 & 240 & 13\\
\enddata
\tablecomments{References:
1) \citet{Xing:2010rr}
; 2) \citet{Barrado-y-Navascues:2004nx}
; 3) \citet{Briceno:1993hc}
; 4) \citet{Briceno:1999nm}
; 5) \citet{Fernandez:1998gf}
; 6) \citet{Findeisen:2010mj}
; 7) \citet{Gomez:1992dt}
; 8) \citet{Hartmann:1991gf}
; 9) \citet{Herbig:1988no}
; 10) \citet{Herbig:1986bh}
; 11) \citet{Li:1998fj}
; 12) \citet{Martin:1998ve}
; 13) \citet{Martin:1994ul}
; 14) \citet{Muzerolle:2003xd}
; 15) \citet{Scelsi:2008my}
; 16) \citet{Strom:1989lq}
; 17) \citet{Walter:1988ae}
; 18) \citet{White:2003xz}
; 19) \citet{White:2007pd}
; 20) \citet{Wichmann:2000dq}
.}
 \end{deluxetable*}
 
 \clearpage

 \startlongtable
 \begin{deluxetable*}{lllrl}
 \tabletypesize{\tiny}
 \tablewidth{0pt}
 \tablecaption{Literature Radial Velocity Measurements for Candidate Class III Taurus Members \label{tab:litrv}}
 \tablehead{
 \colhead{2MASS} & \colhead{Other Name} & \colhead{SpT} & \colhead{$v_{rad}$} & \colhead{Ref}
\\ 
\colhead{} & \colhead{} & \colhead{} & \colhead{(km/s)}
 }
 \startdata
J03500343+2235300 & [LH98] 106 & G0 & 6.9 & 7\\
J03502840+1631151 & [LH98] 108 & G5 & 8.00 & 18\\

J03520223+2439479 & HBC351 & K5 & 3.66 & 6\\
J03533134+2631411 & [LH98] 117 & G7 & 7.13 & 10\\
J03542523+2421363 & [LH98] 119 & G5 & 2.24 & 18\\
J03542950+3203013 & HBC352 & G0 & 10.7 & 6\\
J03543017+3203043 & HBC353 & G5 & 15.1 & 16\\
J03543556+2537111 & HBC354 & K3 & 9.9 & 16\\
\enddata
\tablecomments{References:
1) \citet{Wichmann:2000dq}
; 2) \citet{Biazzo:2012ly}
; 3) \citet{Hartmann:1986cr}
; 4) \citet{Massarotti:2005ag}
; 5) \citet{Muzerolle:2003xd}
; 6) \citet{Cuong-Nguyen:2012ul}
; 7) \citet{Nordstrom:2004wd}
; 8) \citet{Reid:1999nq}
; 9) \citet{Rice:2010lk}
; 10) \citet{Rosvick:1992nx}
; 11) \citet{Sartoretti:1998bd}
; 12) \citet{Scelsi:2008my}
; 13) \citet{Shkolnik:2010de}
; 14) \citet{Shkolnik:2012ee}
; 15) \citet{Torres:2013fk}
; 16) \citet{Walter:1988ae}
; 17) \citet{White:2003xz}
; 18) \citet{White:2007pd}
.}
 \end{deluxetable*}

 \clearpage

\startlongtable
\begin{deluxetable*}{llrrrrrrr}
 \tabletypesize{\tiny}
 \tablewidth{0pt}
 \tablecaption{Keck/HIRES Observations of Candidate Taurus Members \label{tab:hires}}
 \tablehead{
 \colhead{2MASS J} & \colhead{Other Name} & \colhead{Epoch} & \colhead{$t_{int}$} & \colhead{SNR} & \colhead{$EW[H\alpha]$} & \colhead{$EW[Li]$} & \colhead{$v_{rad}$} & \colhead{$v\sin(i)$} 
\\ 
\colhead{} & \colhead{} & \colhead{(UT Date)} & \colhead{(sec)} & \colhead{(@6600\AA)} & \colhead{(\AA)} & \colhead{(m\AA)} & \colhead{(km/s)} & \colhead{(km/s)}
 }
 \startdata
04072456+2332553 & [SCH2006b] J0407246+2332554 & 20061212 & 1800 & 24 & -5.5 & 0 & 11.8 $\pm$ 0.5 & 14.2 $\pm$ 0.6 \\
04073502+2237394 & [SCH2006b] J0407350+2237396 & 20061213 & 1800 & 21 & -16.4 & 612 & 15.4 $\pm$ 2.5 & 97.3 $\pm$ 3.5 \\
04162725+2053091 & [SCH2006b] J0416272+2053093 & 20061213 & 1800 & 20 & -4.0 & 561 & 19.3 $\pm$ 0.5 & 14.3 $\pm$ 0.6 \\
04270739+2215037 & [SCH2006b] J0427074+2215039 & 20061213 & 1800 & 9 & -15.5 & 443 & 16.8 $\pm$ 0.6 & 14.9 $\pm$ 0.7 \\
04295950+2433078 & [SCH2006b] J0429595+2433080 & 20061213 & 1800 & 6 & -83.2 & 307 & 19.3 $\pm$ 0.8 & 18.4 $\pm$ 1.0 \\
04380007+2327167 & [SCH2006b] J0438001+2327167 & 20061212 & 1800 & 11 & -8.3 & 576 & 15.6 $\pm$ 0.5 & 16.8 $\pm$ 0.7 \\
04381630+2326402 & [SCH2006b] J0438163+2326404 & 20061213 & 1500 & 20 & -9.1 & 556 & 17.3 $\pm$ 0.6 & 20.6 $\pm$ 0.8 \\
04385859+2336351 & [SCH2006b] J0438586+2336352 & 20061213 & 1500 & 11 & -25.8 & 532 & 16.7 $\pm$ 0.6 & 16.3 $\pm$ 0.8 \\
\enddata
 \end{deluxetable*}
 
 \clearpage

\startlongtable
 \begin{deluxetable*}{llrrlrrr}
 \tabletypesize{\tiny}
 \tablewidth{0pt}
 \tablecaption{UH88/SNIFS Observations of Candidate Taurus Members \label{tab:snifs}}
 \tablehead{
 \colhead{2MASS J} & \colhead{Other Name} & \colhead{Epoch} & \colhead{$t_{int}$} & \colhead{SpT} & \colhead{$A_V$} & \colhead{$EW[H\alpha]$} & \colhead{Na$_{8189}$ \tablenotemark{a}}
 \\ 
\colhead{} & \colhead{} & \colhead{(UT Date)} & \colhead{(sec)} & \colhead{} & \colhead{(mag)} & \colhead{(\AA)} 
 }
 \startdata
04003106+1935207 & HD 285281       & 20141210 & 200 & F9 & 1 & 0.3 & .. \\
04031395+2552597 & HBC356/357 & 20141210 & 110 & K4.5 & 0.1 & -0.6 & .. \\
04051959+2009256 & RXJ0405.3+2009  & 20141130 & 80 & K1.5 & 0.1 & 0.7 & .. \\
04065134+2541282 & RXJ0406.8+2541  & 20141208 & 200 & K6.5 & 1.3 & -3.9 & 0.937 \\
04090973+2901306 & RXJ0409.1+2901  & 20141211 & 80 & K2.5 & 0.1 & 0 & .. \\
04091700+1716081 & RXJ0409.2+1716  & 20141210 & 110 & M1 & 0.5 & -5.3 & 0.926 \\
04095113+2446211 & RXJ0409.8+2446  & 20150111 & 170 & M1.5 & 0.1 & -1.9 & 0.923 \\
04110570+2216313 & .. & 20150112 & 520 & M3.5 & 0.2 & -6.7 & 0.882 \\
\enddata
\tablenotetext{a}{The $Na_{8189}$ measurement is the spectral index defined by Slesnick et al. (2007).}
 \end{deluxetable*}

 \clearpage
 
 \startlongtable
 \begin{deluxetable*}{lllr}
 \tabletypesize{\tiny}
 \tablewidth{0pt}
 \tablecaption{$Na_{8189}$ Indices for Candidate Class III Taurus Members of Herczeg \& Hillenbrand (2014) \label{tab:litna}}
 \tablehead{
 \colhead{2MASS} & \colhead{Other Name} & \colhead{SpT} & \colhead{$Na_{8189}$}
\\ 
\colhead{} & \colhead{} & \colhead{} & \colhead{(\AA)} 
 }
 \startdata
J04035084+2610531 & HBC359 & M2.8 & 0.89 \\
J04043936+2158186 & HBC360 & M3.4 & 0.88 \\
J04043984+2158215 & HBC361 & M3.2 & 0.89 \\
J04053087+2151106 & HBC362 & M2.7 & 0.90 \\
J04131414+2819108 & LkCa1 & M3.6 & 0.94 \\
J04132722+2816247 & Anon1 & M0.5 & 0.95 \\
J04144739+2803055 & XEST 20-066 & M5.2 & 0.96 \\
J04144797+2752346 & LkCa3 & M2.4 & 0.93 \\
\enddata
 \end{deluxetable*}

\clearpage

\startlongtable
 \begin{deluxetable*}{llccccccc}
 \tabletypesize{\tiny}
 \tablewidth{0pt}
 \tablecaption{Membership Diagnostics \label{tab:diag}}
 \tablehead{
 \colhead{2MASS} & \colhead{Other Name} & \colhead{Lithium} & \colhead{Gravity} & \colhead{HRD} & \colhead{H$\alpha$} & \colhead{PM} & \colhead{RV} & \colhead{Final}
 }
 \startdata
J03500343+2235300 & [LH98] 106 & .. & .. & ? & .. & N? & ? & N? \\	
J03502840+1631151 & [LH98] 108 & ? & .. & ? & ? & N? & ? & N? \\	
J03505432+2350056 & [LH98] 110 & ? & .. & ? & ? & N? & .. & N? \\	
J03520223+2439479 & HBC351 & Y & .. & ? & ? & N? & ? & N \\	
J03533134+2631411 & [LH98] 117 & ? & .. & ? & ? & N? & ? & N? \\	
J03542523+2421363 & [LH98] 119 & ? & .. & ? & ? & N? & ? & N? \\	
J03542950+3203013 & HBC352 & ? & .. & N & ? & N? & ? & N \\	
J03543017+3203043 & HBC353 & ? & .. & N & ? & ? & Y? & N \\
\enddata
\tablecomments{}
 \end{deluxetable*}

\begin{figure*}
\epsscale{1.0}
\includegraphics[scale=0.6,trim={0 0 0 0},clip]{diskredblue.eps}
\figcaption{Spatial distribution for all members of Taurus. Disk-free members confirmed in our census are shown with filled red circles. Disk-hosting members of Taurus (Rebull et al. 2010; Luhman et al. 2010; Esplin et al. 2014) are shown with filled blue circles. The background image is an extinction map compiled by \citet{Schlafly:2014lr}. Most members of the disk-hosting population are clearly concentrated around the ongoing sites of star formation in Taurus, whereas the disk-free population also has a more widely distributed component.}
\end{figure*}

\begin{figure*}
\epsscale{1.0}
\hspace{-0.3in}\includegraphics[scale=0.46,trim={0 0 0 0},clip]{DFGredbluemap.eps}
\figcaption{Stellar density and disk fraction in Taurus, as encompassed in an RGB image. The blue and green channels were computed by convolving the point distributions of disk-hosting and disk-free stars in Figure~\ref{fig:diskmap} with a Gaussian blur kernel of width $\sigma = 1\degr$, such that the image conveys both the disk fraction (via the hue) and the stellar density (via the total intensity). The mapping of intensity and hue to $\Sigma_{disk}$ and $\Sigma_{nodisk}$ are encompassed in the 2D key shown on the right. The Taurus population appears to be composed of high-density regions with a high disk fraction (i.e., bright and magenta), surrounded by a distributed low-density component with low disk fraction (i.e., faint and red).}
\end{figure*}

\begin{figure*}
\epsscale{1.0}
\includegraphics[scale=0.55,trim={0 0 0 0},clip]{HRDmemdiskredblue.eps}
\figcaption{\label{fig:HRDmemdisk} HR diagram for the disk-free Taurus members of our sample (upper panel) and the height above the main sequence of each member (lower panel). Each point is color coded to match the local stellar density and disk fraction shown in Figure~\ref{fig:rgbmap}, such that the disk-rich clustered population is bright magenta and the disk-free distributed population is dark red. To emphasize the difference between the clustered and distributed populations, objects with a local stellar density of $\Sigma > 1.5$ stars/deg$^2$ are shown with open squares, while objects with $\Sigma \le 1.5$ stars/deg$^2$ are shown with filled circles. The distributed population clearly sits below the clustered population in the HR diagram, indicating either that those objects are either older (by a factor of 2--3) or more distant (by $\sim$80 pc). As in Figure~\ref{fig:hrd}, we also show the field main sequence ($\tau \sim 600$ Myr; $[Fe/H] \sim 0$) as defined in Kraus \& Hillenbrand (2007).}
\end{figure*}

\begin{figure*}
\epsscale{1.0}
\includegraphics[scale=0.55,trim={0 0 0 0},clip]{HRDmemdiskGAIAredblue.eps}
\figcaption{\label{fig:HRDmemdiskGAIA} HR diagram and residuals as in Figure~\ref{fig:HRDmemdisk}, but with luminosities estimated from Gaia distances (Gaia Collaboration 2016) for the 22 sample members with parallaxes in the Tycho-Gaia Astrometric Solution, rather than using our assumed mean distance of 145 pc. The distributed population sits unambiguously above the main sequence, confirming its pre-main sequence nature. However, most of the distributed population is located at a moderately closer distance ($d \sim $110--130 pc) while the objects in the clustered population are indeed at the distance of Taurus, accentuating the difference in ages.}
\end{figure*}

\begin{figure*}
\epsscale{1.0}
\includegraphics[scale=0.6,trim={0 0 0 0},clip]{diskredgreen.eps}
\figcaption{Spatial distribution for all members of Taurus. Disk-free members confirmed in our census are shown with filled red circles. Disk-hosting members of Taurus (Rebull et al. 2010; Luhman et al. 2010; Esplin et al. 2014) are shown with filled green circles. The background image is an extinction map compiled by \citet{Schlafly:2014lr}. Most members of the disk-hosting population are clearly concentrated around the ongoing sites of star formation in Taurus, whereas the disk-free population also has a more widely distributed component.}
\end{figure*}

\begin{figure*}
\epsscale{1.0}
\hspace{-0.3in}\includegraphics[scale=0.46,trim={0 0 0 0},clip]{DFGredgreenmap.eps}
\figcaption{Stellar density and disk fraction in Taurus, as encompassed in an RGB image. The blue and green channels were computed by convolving the point distributions of disk-hosting and disk-free stars in Figure~\ref{fig:diskmap} with a Gaussian blur kernel of width $\sigma = 1\degr$, such that the image conveys both the disk fraction (via the hue) and the stellar density (via the total intensity). The mapping of intensity and hue to $\Sigma_{disk}$ and $\Sigma_{nodisk}$ are encompassed in the 2D key shown on the right. The Taurus population appears to be composed of high-density regions with a high disk fraction (i.e., bright and yellow), surrounded by a distributed low-density component with low disk fraction (i.e., faint and red).}
\end{figure*}

\begin{figure*}
\epsscale{1.0}
\includegraphics[scale=0.55,trim={0 0 0 0},clip]{HRDmemdiskredgreen.eps}
\figcaption{\label{fig:HRDmemdisk} HR diagram for the disk-free Taurus members of our sample (upper panel) and the height above the main sequence of each member (lower panel). Each point is color coded to match the local stellar density and disk fraction shown in Figure~\ref{fig:rgbmap}, such that the disk-rich clustered population is bright yellow and the disk-free distributed population is dark red. To emphasize the difference between the clustered and distributed populations, objects with a local stellar density of $\Sigma > 1.5$ stars/deg$^2$ are shown with open squares, while objects with $\Sigma \le 1.5$ stars/deg$^2$ are shown with filled circles. The distributed population clearly sits below the clustered population in the HR diagram, indicating either that those objects are either older (by a factor of 2--3) or more distant (by $\sim$80 pc). As in Figure~\ref{fig:hrd}, we also show the field main sequence ($\tau \sim 600$ Myr; $[Fe/H] \sim 0$) as defined in Kraus \& Hillenbrand (2007).}
\end{figure*}

\begin{figure*}
\epsscale{1.0}
\includegraphics[scale=0.55,trim={0 0 0 0},clip]{HRDmemdiskGAIAredgreen.eps}
\figcaption{\label{fig:HRDmemdiskGAIA} HR diagram and residuals as in Figure~\ref{fig:HRDmemdisk}, but with luminosities estimated from Gaia distances (Gaia Collaboration 2016) for the 22 sample members with parallaxes in the Tycho-Gaia Astrometric Solution, rather than using our assumed mean distance of 145 pc. The distributed population sits unambiguously above the main sequence, confirming its pre-main sequence nature. However, most of the distributed population is located at a moderately closer distance ($d \sim $110--130 pc) while the objects in the clustered population are indeed at the distance of Taurus, accentuating the difference in ages.}
\end{figure*}

\end{document}